\documentclass[11pt,a4paper]{article}
\pdfoutput=1
\usepackage{jheppub} 
\usepackage{url}
\usepackage{multirow}
\usepackage{array,booktabs,colortbl,multirow}
\usepackage{colortbl,xcolor,colordvi,color}
\usepackage{rotating}
\usepackage{graphicx}
\usepackage{physics}
\usepackage{longtable}
\usepackage{amssymb}
\usepackage{amsmath}
\usepackage{ulem}
\allowdisplaybreaks
\usepackage{placeins}

\makeatletter
\def\@fpheader{\relax}
\makeatother

\definecolor{colorDG}{HTML}{008000}

\title{Constraining new physics effective interactions via a
  global fit of electroweak, Drell-Yan, Higgs, top, and flavour observables}

\author[a]{J. de Blas}
\author[b]{, A. Goncalves}
\author[c,d]{, V. Miralles}
\author[b]{, L. Reina}
\author[e]{, L. Silvestrini}
\author[e]{and M. Valli}

\affiliation[a]{Departamento de F\'isica Te\'orica y del Cosmos, Universidad de  Granada, Campus de Fuentenueva, E-18071 Granada, Spain}
\affiliation[b]{Physics Department, Florida State University,\\ Tallahassee, FL 32306-4350, USA}
\affiliation[c]{
Departament de F\'isica, Universitat d'Alacant,
Campus de Sant Vicent del Raspeig, E-03690 \mbox{Alacant}, Spain
}
\affiliation[d]{School of Physics and Astronomy, University of Manchester, Oxford Road, Manchester M13 9PL, UK}
\affiliation[e]{INFN, Sezione di Roma,\\ Piazzale A. Moro 2, I-00185 Rome, Italy}

\emailAdd{deblasm@ugr.es}
\emailAdd{agoncalvesdossantos@fsu.edu}
\emailAdd{victor.miralles@ua.es}
\emailAdd{reina@hep.fsu.edu}
\emailAdd{luca.silvestrini@roma1.infn.it}
\emailAdd{mauro.valli@roma1.infn.it}

\abstract{We present results from a global fit of Standard Model parameters and dimension-6 SMEFT Wilson coefficients that includes electroweak, Drell-Yan, Higgs-boson, top-quark, and flavour observables. Fits obtained by floating individual coefficients are also discussed. The leading-order scale dependence of the SMEFT Wilson coefficients is consistently included in the evolution from the UV scale to the electroweak scale and the low-energy scale of flavour observables. In defining the SMEFT set of active operators we consider both the $U(3)^5$ and the $U(2)^5$ flavour symmetric limits. All fits are obtained within the \texttt{HEPfit} framework and are based on the most recent experimental results and state-of-the-art theoretical predictions for all the observables considered.}

\begin{document}
\maketitle
\flushbottom

\section{Introduction}
\label{sec:introduction}

The consistency and predictivity of the Standard Model (SM) of
particle physics up to energy scales of a few hundred GeV is
impressive and has reached an unprecedented level of accuracy. At the
same time its limits are well known: from missing a viable candidate
of dark matter to not explaining the baryon asymmetry of the universe
or the nature of dark energy. Furthermore, constituents of the SM
itself, namely the Higgs boson and neutrinos, are still quite
mysterious. Indeed, neutrino masses cannot be described within the SM and require a new physics (NP) explanation. On the other hand, the properties of the Higgs boson as well as the origin of the electroweak (EW) symmetry breaking are encoded but not explained by the SM, 
and the Higgs particle could be a potential messenger of NP.  Although weakly-interacting light NP may have escaped detection so far, most likely the explanation of these phenomena is to be sought at energies beyond the ones directly probed so far, in a still unknown \textit{ultraviolet} (UV) theory that reduces to the SM at energy
scales of a few hundred GeV.

Low-energy experiments as well as high-energy lepton and hadron colliders such as LEP, the Tevatron, past and present B-factories, and the Large Hadron Collider (LHC) have set lower bounds
on UV physics beyond the SM (BSM) both directly and indirectly. Given the energies explored and the absence of deviations from SM predictions, often probed with percent or permille level sensitivity, we expect potential
NP to live at energies quite above the EW scale, namely at or above the TeV scale. Indeed the region between 1-10 TeV will be the focus of
the high-luminosity LHC (HL-LHC) and of future-collider explorations, both via higher precision and higher energy.

From a theoretical point of view, the problem of investigating the
effects of unknown heavy NP lends itself naturally to an
effective field theory (EFT) approach, where the effects of NP at
the EW scale are encoded in non-renormalizable local
interactions of the SM fields. The SM is then to be seen as the
4-dimensional part of a more general EFT whose
Lagrangian also includes operators of higher dimension obtained from the
projection of a given UV theory onto the space of fields of the SM.  The SM
Effective Field Theory (SMEFT) realizes this idea by extending the
Lagrangian of the SM by a complete set of Lorentz and (SM-)gauge
invariant effective interactions of increasing mass dimensionality,
built in terms of SM fields.
These interactions do, in general, break the SM accidental symmetries 
but could also satisfy additional discrete and
global symmetry properties, depending on the kind of NP they encode
(see~\cite{Falkowski:2023hsg,Isidori:2023pyp} for a general introduction and~\cite{Aebischer:2025qhh} for a state-of-the-art review of the SMEFT and its studies).

Different UV extensions of the SM project differently on
the effective interactions of the SMEFT Lagrangian below the UV scale $\Lambda$~\cite{deBlas:2017xtg,Guedes:2023azv,Guedes:2024vuf},
such that constraining the effect of SMEFT interactions with available
data could discriminate between classes of UV models and guide future
BSM explorations. Most notably, different flavour symmetries assumed
at the UV scale $\Lambda$ modify the number and structure of SMEFT
interactions present below $\Lambda$ and directly affect a broad
spectrum of observables. Several studies~\cite{Aoude:2020dwv,Faroughy:2020ina,Bissmann:2020mfi,Bruggisser:2021duo,Bruggisser:2022rhb,Bellafronte:2023amz,Grunwald:2023nli,Greljo:2023adz,Allwicher:2023shc,Bartocci:2023nvp,Mantani:2025bqu,Kala:2025srq} have investigated the impact of different UV flavour assumptions and emphasised the need to
move beyond the initial flavour-blind approach. A popular first step beyond assuming complete invariance
under family rotation for the five kinds of fermions present in the SM
(the so-called $U(3)^5$ flavour symmetry of the two left-handed doublets and
three right-handed singlets of the SM gauge group) consists in
singling out the third generation and still assuming invariance under
rotation within the first two families (the so-called $U(2)^5$
flavour symmetry assumption). This choice is motivated by the scale hierarchy
present among fermions and naturally appears in several BSM models. In
our study we will consider both cases and discuss pros and
cons of both assumptions in current fits. We emphasize that when specifying the flavour symmetry we only refer to the NP contribution encoded in the Wilson coefficients of operators of $d>4$. Typical examples of such NP models are given by heavy gauge bosons coupling universally to all flavours ($U(3)^5$, see e.g.~ref.~\cite{Salvioni:2009mt}) or coupling differently to the third generation ($U(2)^5$). Since the $U(2)^5$ symmetry describes to a good approximation the quark flavour structure of the SM, one could also assume that the flavour invariance involves also the renormalizable part of the Lagrangian, with suitable breaking terms arising in both the $d\leq4$ and $d>4$ terms in the SMEFT, as discussed in detail in ref.~\cite{Faroughy:2020ina} and typically happening in models seeking to explain the SM flavour hierarchy (see for example ref.~\cite{Barbieri:2023qpf}).

The EFT framework also allows to systematically account for quantum
effects. In particular, the SMEFT Wilson coefficients satisfy well-defined
renormalization group (RG) equations that allow to determine their
scale dependence and evolve them from the UV scale $\Lambda$ to the
lower scales where they enter the calculation of physical
observables. In this way, not only the precision of theoretical
predictions of such observables is improved by a more efficient
resummation of their perturbative expansion, but the potentially large
interplay between different operators that mix under renormalization
group evolution (RGE) is taken into account.

Overall, the great advantage of employing an effective field theory framework is that we can implement in a very general and systematic way a
bottom-up approach to the exploration of some unknown UV theory using
our knowledge of the physics we can currently test.
Had patterns to emerge (e.g. enhanced or suppressed SMEFT coefficients,
correlation and flat directions in SMEFT space, etc.)  they could help
in selecting classes of models whose investigation could then become
more specific. 
The effective field theory framework is also useful in this regard.
Using a top-down approach, one can easily re-interpret the SMEFT results
in terms of specific models by using the matching between a given UV scenario
and the effective theory. 
In the absence of unambiguous signs of BSM physics, the
SMEFT general approach, although less informative, can indeed
provide a very powerful exploratory tool.  With this respect, we now have
at our disposal for the first time a very broad spectrum of
sufficiently accurate experimental measurements that can be
simultaneously used to constrain the effective interactions of the
SMEFT. This has indeed become a major effort in particle physics
phenomenology and various comprehensive studies based on fits of the
SMEFT that include a broad spectrum of observables have appeared in
the recent
literature~\cite{Ellis:2020unq,Ethier:2021bye,Garosi:2023yxg,Allwicher:2023shc,Bartocci:2023nvp,Celada:2024mcf,ATLAS:2024lyh,Bartocci:2024fmm,CMS:2025ugn,terHoeve:2025gey}.

In this paper we present the results of both individual and global
fits of the SMEFT coefficients that include EW, Drell-Yan,
Higgs-boson, top-quark, and flavour-physics observables. For the first time in the literature, in each fit we also float all SM parameters and all the relevant hadronic parameters in the flavour sector, thus taking fully into account the SM parametric and theoretical uncertainties, which are mostly relevant in the flavour sector. This turns the global fit into a highly challenging endeavour, particularly for the $U(2)^5$ case where it involves a total of about two hundred parameters. 

We will be
working in the SMEFT between the UV scale $\Lambda$
and the EW scale ($\mu_{W}$), while at $\mu_{W}$ we
will transition to the Low Energy Effective Theory (LEFT)~\cite{Jenkins:2017jig}
where all the heavy SM degrees of freedom ($W,Z,H,t$) are
integrated out. To fully benefit of the EFT formalism and in order to
account for impactful quantum effects, we will consider the scale
dependence of the SMEFT and LEFT Wilson coefficients using
leading-order RGE, with tree-level matching conditions at the EW scale $\mu_{W}$. All fits are
performed in the open-source \texttt{HEPfit} framework~\cite{DeBlas:2019ehy,hepfitsite} using
for each observable state-of-the-art theoretical predictions and the
most recent experimental measurements.

The outcome of the fits presented in this paper will determine the ranges
of SM parameters and SMEFT Wilson coefficients at the UV scale
$\Lambda$ that are compatible with the selected broad set of
experimental measurements. We will present both fits in which only one
coefficient at a time is active and fits in which all SMEFT
coefficients are simultaneously active at the scale
$\Lambda$. Single-coefficient fits can be used to extract
information on the set of observables that most constrain a given
coefficient and can also give preliminary indications on the main RGE
effects, while the global fit fully represents the combined effect of
using a broad spectrum of observables and considering the operator
mixing induced by the RGE. At the same time, a global fit of SM
parameters and SMEFT coefficients can be very sensitive to the
parametric and theoretical uncertainties and thus less effective,
given current accuracies, to constrain the theory at hand. This dilution of the sensitivity is less evident in the $U(3)^5$ scenario, but has a strong impact in the $U(2)^5$ case, calling for the inclusion of more observables and/or for higher theoretical and experimental precision on the currently included ones.

Furthermore, in order to determine to which extent with current data
we can constrain NP living at different scales, we will
consider two representative benchmark values for the NP scale: $\Lambda = 3$, and $10$~TeV including full RGE, and also compare them with results obtained without RGE (and $\Lambda=1$~TeV). Not knowing the scale of NP, this is quite relevant to answer the fundamental question about what kind of observables are
going to play the most important role and to what level of precision
we need to determine them both theoretically and experimentally in
order to explore NP up to a scale of 10 TeV.

The paper consists of this introductory section followed by three main core
sections and two appendices. 
A final section summarises the main conclusions of our study and presents an  outlook on future developments. Concerning the core sections, in
Section~\ref{sec:smeft-to-left} we review elements of the SMEFT relevant
to our study with particular emphasis on the implementation of the two
aforementioned flavour scenarios, namely $U(3)^5$ and $U(2)^2$, and a detailed
discussion of how the RGE of the SMEFT 
coefficients and SM parameters is included in our study. 
Technical details on how
these two scenarios are implemented in the SMEFT basis of operators considered in our study
are provided in Appendix~\ref{sec:app-SMEFT-flav}. 
Section~\ref{sec:eft-coeff-constraints} is dedicated to a
discussion of the fitting framework and procedure, including how the
various sets of observables have been implemented in \texttt{HEPfit}. 
A complete set of results is illustrated and discussed in
Section~\ref{sec:results} while the full outcome of both individual and
global fits, for all three UV scales $\Lambda=1,\,3,$ and 10~TeV, with and
without RGE, are collected in a series of tables in Appendix~\ref{sec:app-results}.

\section{Framework: effective interactions from new physics}
 \label{sec:smeft-to-left}

 \subsection{The Standard Model Effective Field Theory: formalism and assumptions}
 \label{sec:smeft}

Assuming the presence of a significant mass gap between the scale of NP $\Lambda$ and the energies explored in current experiments, we can describe NP effects via the SMEFT Lagrangian:
\begin{equation}
\label{eq:smeft-lagrangian}
\mathcal{L}_{\mathrm{SMEFT}}=\mathcal{L}_{\mathrm{SM}}+\sum_{d>4}\frac{1}{\Lambda^{d-4}}\sum_{i}C_i^{(d)}\mathcal{O}_i^{(d)},
\end{equation}
where ${\cal O}_i^{(d)}$ are Lorentz and (SM)-gauge invariant operators of canonical mass dimension $d$ built only of SM fields, $i$ collectively labels each operator according to its
field composition and quantum numbers, and $\Lambda$ is the cutoff of the effective theory indicatively identified with the UV scale of NP. If lepton number is conserved by the NP, the leading-order effects appear at dimension 6. 
In this work, we truncate the $\mathcal{L}_{\mathrm{SMEFT}}$ to include only dimension-6 operators and adopt the Warsaw basis introduced in ref.~\cite{Grzadkowski:2010es}. 
We also neglect baryon number violating and CP-odd operators. This results in the set of operators listed in Table~\ref{tab:smeft-operators}.
\begin{table}[ht!] 
  {\small
  \centering
  \renewcommand{\arraystretch}{1.5}
  \scalebox{0.87}{
  \begin{tabular}{|c|c|c|} 
  \hline
  $X^3$ & 
  $\phi^6$~ and~ $\phi^4 D^2$ &
  $\psi^2\phi^3$\\
  \hline
  $\mathcal{O}_G=f^{ABC} G_\mu^{A\nu} G_\nu^{B\rho} G_\rho^{C\mu} $ &  
  $\mathcal{O}_\phi=(\phi^\dag \phi)^3$ &
  $\mathcal{O}_{e\phi}^{[pr]}=(\phi^\dag \phi)(\bar l_p \phi e_r )$\\
  $\mathcal{O}_{W}=\varepsilon^{IJK} W_\mu^{I\nu} W_\nu^{J\rho} W_\rho^{K\mu}$ &   
  $\mathcal{O}_{\phi\Box}=(\phi^\dag \phi) \Box(\phi^\dag \phi)$ &
  $\mathcal{O}_{u\phi}^{[pr]}=(\phi^\dag \phi)(\bar q_p \widetilde{\phi} u_r )$\\  
  &
  $\mathcal{O}_{\phi D}=\left(\phi^\dag D^\mu \phi\right)^\star \left(\phi^\dag D_\mu \phi\right)$ &
  $\mathcal{O}_{d\phi}^{[pr]}=(\phi^\dag \phi)(\bar q_p \phi d_r )$\\ 
  \hline
 $X^2 \phi^2$ &
 $\psi^2 X \phi$ &
 $\psi^2 \phi^2 D$\\ 
  \hline
  $\mathcal{O}_{\phi G}=\phi^\dag \phi\, G^A_{\mu\nu} G^{A\mu\nu}$ & 
  $\mathcal{O}_{e W}^{[pr]}=(\bar l_p \sigma^{\mu\nu} e_r) \tau^I \phi W_{\mu\nu}^I$ &
  $\mathcal{O}_{\phi l}^{(1)[pr]}=(\phi^\dag i\!\!\stackrel{\leftrightarrow}{D}_\mu \phi) (\bar l_p \gamma^\mu l_r)$\\
  $\mathcal{O}_{\phi W}=\phi^\dag \phi\, W^I_{\mu\nu} W^{I\mu\nu}$ & 
  $\mathcal{O}_{e B}^{[pr]}=(\bar l_p \sigma^{\mu\nu} e_r) \phi B_{\mu\nu}$ &
  $\mathcal{O}_{\phi l}^{(3)[pr]}=(\phi^\dag i\!\! \stackrel{\leftrightarrow}{D^I_\mu} \phi) (\bar l_p \tau^I \gamma^\mu l_r)$\\
  $\mathcal{O}_{\phi B}=\phi^\dag \phi\, B_{\mu\nu} B^{\mu\nu}$ &
  $\mathcal{O}_{uG}^{[pr]}=(\bar q_p \sigma^{\mu\nu} T^A u_r) \widetilde{\phi}\, G_{\mu\nu}^A$ &
  $\mathcal{O}_{\phi e}^{[pr]}= (\phi^\dag i\!\!\stackrel{\leftrightarrow}{D}_\mu \phi) (\bar e_p \gamma^\mu e_r)$\\  
  $\mathcal{O}_{\phi WB}= \phi^\dag \tau^I \phi\, W^I_{\mu\nu} B^{\mu\nu}$ &
  $\mathcal{O}_{uW}^{[pr]}=(\bar q_p \sigma^{\mu\nu} u_r) \tau^I \widetilde{\phi}\, W_{\mu\nu}^I$ &
  $\mathcal{O}_{\phi q}^{(1)[pr]}=(\phi^\dag i\!\!\stackrel{\leftrightarrow}{D}_\mu \phi) (\bar q_p \gamma^\mu q_r)$\\
      &  
  $\mathcal{O}_{u B}^{[pr]}=(\bar q_p \sigma^{\mu\nu} u_r) \widetilde{\phi}\, B_{\mu\nu}$&
  $\mathcal{O}_{\phi q}^{(3)[pr]}=(\phi^\dag i \!\!\stackrel{\leftrightarrow}{D^I_\mu} \phi) (\bar q_p \tau^I \gamma^\mu q_r)$\\
     &  
  $\mathcal{O}_{dG}^{[pr]}=(\bar q_p \sigma^{\mu\nu} T^A d_r) \phi\, G_{\mu\nu}^A$ & 
  $\mathcal{O}_{\phi u}^{[pr]}=(\phi^\dag i\!\!\stackrel{\leftrightarrow}{D}_\mu \phi) (\bar u_p \gamma^\mu u_r)$\\ 
     &  
  $\mathcal{O}_{dW}^{[pr]}=(\bar q_p \sigma^{\mu\nu} d_r) \tau^I \phi\, W_{\mu\nu}^I$ &
  $\mathcal{O}_{\phi d}^{[pr]}=(\phi^\dag i\!\!\stackrel{\leftrightarrow}{D}_\mu \phi) (\bar d_p \gamma^\mu d_r)$\\
      &  
  $\mathcal{O}_{dB}^{[pr]}=(\bar q_p \sigma^{\mu\nu} d_r) \phi\, B_{\mu\nu}$ &
  $\mathcal{O}_{\phi ud}^{[pr]}=(\widetilde{\phi}^\dag iD_\mu \phi)(\bar u_p \gamma^\mu d_r)$\\  
  \hline 
  $(\bar LL)(\bar LL)$ & 
  $(\bar RR)(\bar RR)$ &
  $(\bar LL)(\bar RR)$\\
  \hline
  $\mathcal{O}_{ll}^{[prst]}=(\bar l_p \gamma_\mu l_r)(\bar l_s \gamma^\mu l_t)$ &
  $\mathcal{O}_{ee}^{[prst]}=(\bar e_p \gamma_\mu e_r)(\bar e_s \gamma^\mu e_t)$ &
  $\mathcal{O}_{le}^{[prst]}=(\bar l_p \gamma_\mu l_r)(\bar e_s \gamma^\mu e_t)$ \\
  $\mathcal{O}_{qq}^{(1)[prst]}=(\bar q_p \gamma_\mu q_r)(\bar q_s \gamma^\mu q_t)$ &
  $\mathcal{O}_{uu}^{[prst]}=(\bar u_p \gamma_\mu u_r)(\bar u_s \gamma^\mu u_t)$ &
  $\mathcal{O}_{lu}^{[prst]}=(\bar l_p \gamma_\mu l_r)(\bar u_s \gamma^\mu u_t)$ \\
  $\mathcal{O}_{qq}^{(3)[prst]}=(\bar q_p \gamma_\mu \tau^I q_r)(\bar q_s \gamma^\mu \tau^I q_t)$ &
  $\mathcal{O}_{dd}^{[prst]}=(\bar d_p \gamma_\mu d_r)(\bar d_s \gamma^\mu d_t)$ &
  $\mathcal{O}_{ld}^{[prst]}=(\bar l_p \gamma_\mu l_r)(\bar d_s \gamma^\mu d_t)$ \\
  $\mathcal{O}_{lq}^{(1)[prst]}=(\bar l_p \gamma_\mu l_r)(\bar q_s \gamma^\mu q_t)$ &
  $\mathcal{O}_{eu}^{[prst]}=(\bar e_p \gamma_\mu e_r)(\bar u_s \gamma^\mu u_t)$ &
  $\mathcal{O}_{qe}^{[prst]}=(\bar q_p \gamma_\mu q_r)(\bar e_s \gamma^\mu e_t)$ \\
  $\mathcal{O}_{lq}^{(3)[prst]}=(\bar l_p \gamma_\mu \tau^I l_r)(\bar q_s \gamma^\mu \tau^I q_t)$ &
  $\mathcal{O}_{ed}^{[prst]}=(\bar e_p \gamma_\mu e_r)(\bar d_s\gamma^\mu d_t)$ &
  $\mathcal{O}_{qu}^{(1)[prst]}=(\bar q_p \gamma_\mu q_r)(\bar u_s \gamma^\mu u_t)$ \\ 
  &
  $\mathcal{O}_{ud}^{(1)[prst]}=(\bar u_p \gamma_\mu u_r)(\bar d_s \gamma^\mu d_t)$ &
  $\mathcal{O}_{qu}^{(8)[prst]}=(\bar q_p \gamma_\mu T^A q_r)(\bar u_s \gamma^\mu T^A u_t)$ \\ 
  & 
  $\mathcal{O}_{ud}^{(8)[prst]}=(\bar u_p \gamma_\mu T^A u_r)(\bar d_s \gamma^\mu T^A d_t)$ &
  $\mathcal{O}_{qd}^{(1)[prst]}=(\bar q_p \gamma_\mu q_r)(\bar d_s \gamma^\mu d_t)$ \\
  &&
  $\mathcal{O}_{qd}^{(8)[prst]}=(\bar q_p \gamma_\mu T^A q_r)(\bar d_s \gamma^\mu T^A d_t)$\\
  \hline 
  $(\bar LR)(\bar LR)$ & 
  $(\bar LR)(\bar RL)$ &
  \\
  \hline
  $\mathcal{O}_{quqd}^{(1)[prst]}=(\bar q_p^i u_r)\epsilon_{ij}(\bar q_s^j d_t)$ &
  $\mathcal{O}_{ledq}^{[prst]}=(\bar l_p^i  e_r)(\bar d_s q_{ti})$ &
  \\
  $\mathcal{O}_{quqd}^{(8)[prst]}=(\bar q_p^i T^A u_r)\epsilon_{ij}(\bar q_s^j T^A d_t)$&
   &
  \\
  $\mathcal{O}_{lequ}^{(1)[prst]}=(\bar l_p^i e_r)\epsilon_{ij}(\bar q_s^j u_t)$ &
 &
  \\
  $\mathcal{O}_{lequ}^{(3)[prst]}=(\bar l_p^i \sigma_{\mu\nu} e_r)\epsilon_{ij}(\bar q_s^j \sigma^{\mu\nu} u_t)$&
   &
  \\
  \hline
  \end{tabular}
  }
  }
  \caption{\it Dimension-6 operators in the Warsaw basis, adapted from ref.~\cite{Grzadkowski:2010es}. We collectively denote by $X$ the field
    strength tensors of the SM gauge group ($X=G^{\mu\nu},W^{\mu\nu},B^{\mu\nu}$), by $\psi$ a generic
    fermion field, and by $\phi$ the scalar field of the
    SM. Among the fermion fields, $q$ and
    $l$ denote $SU(2)_L$ left-handed doublets while $u,d,e$ $SU(2)_L$ right-handed
    singlets, with family indices specified by the superscript in square
    brackets.\label{tab:smeft-operators}}
\end{table}

Furthermore, we do not consider the operator  $\mathcal{O}_{\phi}$ since it 
does not enter at leading order in any of the observables considered in this study. In fact, the main effect of this operator is to modify the Higgs-boson self-interactions, and can only be weakly constrained with current Higgs pair-production data~\cite{CMS:2022dwd,ATLAS:2024ish}. We also notice that while this operator can be generated from other bosonic interactions via renormalization (to be discussed in Section~\ref{sec:wc-running}), it does not enter in the 1-loop anomalous dimension matrix entries for any other SMEFT interaction, i.e. no other operator is generated from a non-zero value of $C_\phi$ at the UV scale.

In this study we consider two possible flavour assumptions at the new-physics scale: the $U(3)^5$ and the $U(2)^5$ flavour symmetric limits. The $U(3)^5$ limit assumes that the NP is flavour-blind,  and leaves a total of 41 independent Wilson
coefficients (40 if not considering $C_\phi$). The notation for these coefficients and the relations with the SMEFT Wilson coefficients in the general case are presented in Appendix~\ref{sec:app-SMEFT-U35}.

To define the $U(2)^5$ limit, it is necessary to specify a direction in flavour space that identifies the third generation. In this work, we choose the direction of the third family to be the one defined by the Yukawa couplings at the NP scale $\Lambda$, and consider the two possibilities where the third family is defined: 1) in the basis of diagonal up-type quark Yukawa couplings (denoted as \textit{UP basis} in the following), or 2) in the basis of diagonal down-type quark Yukawa couplings (denoted as \textit{DOWN basis} in the following). It is important to emphasise that these two classes of scenarios do not correspond to a mere change of basis, but rather reflect different assumptions about the flavour structure of the underlying UV physics. Notice that such a flavour misalignment arises exclusively in the left-handed quark sector --- see the discussion in \cite{Silvestrini:2018dos} --- while right-handed quark Yukawa couplings, as well as the full charged-lepton Yukawa couplings, can always be chosen flavour diagonal.

Having specified the basis, the $U(2)^5$ limit is described by a total of 124 (123 if not considering $C_\phi$) independent Wilson coefficients~\cite{Faroughy:2020ina}. As in the $U(3)^5$ case, the relations between the $U(2)^5$ Wilson coefficients and those of the general SMEFT are presented in Appendix~\ref{sec:app-SMEFT-U25}, which also sets the notation for the presentation of the $U(2)^5$-symmetric results, following ref.~\cite{Allwicher:2023shc}.

\subsection{Renormalization group evolution in the SMEFT}
\label{sec:wc-running}

Assuming that the SMEFT coefficients are generated by BSM physics at the UV scale $\Lambda$, the Wilson coefficients at the EW scale $\mu_W$ are obtained by running them from the UV scale $\Lambda$ to the EW scale $\mu_W$ using the leading order (LO) RG equations derived in refs.~\cite{Jenkins:2013zja,Jenkins:2013wua,Alonso:2013hga}. Lacking the next-to-leading SMEFT anomalous dimension, we can control only log-enhanced higher-order corrections to SMEFT contributions. Hence, we use tree-level matching conditions and matrix elements, and compute all SMEFT Wilson coefficients at a common scale $\mu_W$ that for convenience we choose to be $\mu_W = M_W$. For all high-$p_T$ observables (Higgs, top, and Drell-Yan), we use the same scale $\mu_W$, since the difference between the relevant physical scale of those processes and $M_W$ is such that no large logarithms arise, except for a few regions of phase space which currently still have large experimental uncertainties. 

Working at linear order in the SMEFT coefficients, we can write the Wilson coefficients at the EW scale $\mu_W$ as:
\begin{equation}
C_i(\mu_W)=U(\mu_W,\Lambda)_{ij}C_j(\Lambda),
\label{eq:wc-evolution}
\end{equation}
where $U(\mu_W,\Lambda)$ is the evolutor between the scales $\Lambda$ and $\mu_W$ computed neglecting the effect of SMEFT coefficients in the running of the SM parameters. At this order, one would still need to recompute the evolutor $U(\mu_W,\Lambda)$ for each given value of the SM parameters in the Monte Carlo integration. However, as stated above, at linear order we can neglect the effect of the SMEFT in the extraction of the SM parameters from experimental measurements when computing $U(\mu_W,\Lambda)$. This further justifies neglecting also the small uncertainties on SM parameters in the computation of $U(\mu_W,\Lambda)$, so that the evolutor becomes a constant matrix for fixed $\mu_W$ and $\Lambda$, allowing for a dramatic increase in the speed of the numerical computation of the Wilson coefficients at the scale $\mu_W$. 

Concerning the SM parameters, in principle one could just define them at the scale $\Lambda$, evolve them to the EW scale $\mu_W$ using the full SMEFT, and then use the measurements of masses and couplings as constraints. However, this is extremely inefficient for the parameters that are very precisely determined experimentally, such as $G_F$, $M_Z$, $\alpha$ and the lepton masses. For this reason, we follow the same approach as in SM EW precision physics and choose to express all observables in terms of a set of input parameters, namely $G_F$, $\alpha$ or $M_W$, $M_Z$, $M_h$ and the lepton masses. This amounts to trading implicitly the values of the SM parameters at the EW scale, including the corrections from the SMEFT, for the chosen set of input values. Of course, care must be taken in the EW sector, where the SM relations between masses and couplings are modified by the SMEFT.


One could be tempted to apply the same procedure to quark masses and CKM angles~\cite{Descotes-Genon:2018foz}. However, one of the symmetry assumptions we are using, namely exact $U(2)^5$ for $d=6$ SMEFT operators, requires specifying the structure of the SM Yukawa couplings at the scale $\Lambda$, and the relations between quark masses, CKM angles, and Yukawa couplings are modified by SMEFT contributions. Thus, one cannot just trade the experimental values of quark masses and CKM angles for the corresponding SM parameters at the EW scale, and keeping the SMEFT contributions in the Yukawa matrices RGE is essential.\footnote{To our knowledge, these corrections have been neglected in all the SMEFT fits that have considered the RGE evolution of the SMEFT coefficients so far.} To impose the required flavour structure, we choose as parameters in the quark flavour sector the Yukawa couplings at the scale $\Lambda$, which are then evolved to the EW scale including the effect of the SMEFT in the running. Once evolved to the EW scale, the Yukawa couplings are used to compute the quark masses and CKM angles, including SMEFT contributions, which are then compared to the experimental values. Concerning the leptons, since we are assuming diagonal Yukawa couplings for the charged leptons and individual lepton numbers are preserved, the flavour orientation at the scale $\Lambda$ is irrelevant and, as stated above, we can trade the lepton Yukawa couplings at the EW scale for the corresponding masses.

The RG evolution from the scale $\Lambda$ to $\mu_W$ is performed within the \texttt{HEPfit} framework using the \texttt{RGEsolver} library~\cite{DiNoi:2022ejg}. 
The SMEFT Lagrangian at the scale $\mu_W$ is then expanded around the minimum of the scalar potential, and the quark mass matrices are diagonalised through the biunitary transformations
\begin{equation}
  M_{u,d} = V_{u,d}^R M_{u,d}^\mathrm{diag} V_{u,d}^{L\dagger}\,.
\end{equation}
The SMEFT operators involving quark fields are then transformed to the mass eigenstate basis using the matrices $V_{u,d}^{L,R}$ after removing unphysical phases. The unitary CKM matrix is defined as $V_\mathrm{CKM} \equiv V_u^{L\dagger} V_d^L$.\footnote{Note that the determination of the CKM matrix elements from the experimental data is affected by additional corrections from SMEFT operators contributing to the relevant CKM observables. All these potential new physics effects are included in our calculations.} The SMEFT coefficients in the mass eigenstate basis are then used to compute the observables at the scale $\mu_W$.\footnote{Under our flavour assumptions, the flavour rotation is expected to have a negligible effect for the calculation of flavour-diagonal observables such as EW, Higgs boson, and top quark observables.} Most observables in the flavour sector involve lower energies, and are therefore computed in the LEFT, where the massive gauge bosons, the Higgs field and the top quark are integrated out. The matching between the two EFT's, the SMEFT and the LEFT, is implemented according to ref.~\cite{Jenkins:2017jig}. For each flavour observable, as detailed in Section~\ref{sec:flavour}, the LEFT coefficients are evolved to the relevant energy scale using the LEFT RG equations.

\section{Constraints on new physics from current data}
\label{sec:eft-coeff-constraints}

We perform a global fit of the SMEFT including EW, Drell-Yan, Higgs-boson, top-quark, and flavour observables. The fit has been implemented in the \texttt{NPSMEFTd6General} model of the open-source \texttt{HEPfit} framework~\cite{DeBlas:2019ehy,hepfitsite}. In this section, after specifying the adopted statistical framework and fitting procedure, we describe the methods used for the calculation of the different classes of observables entering the fit and provide references for the corresponding experimental measurements.

\subsection{Introduction to the fitting procedure}
\label{sec:fit-intro}

We perform a Bayesian analysis to extract the posterior probability density for the SMEFT coefficients $C_i(\Lambda)$ given the experimental data, a uniform prior for $C_i(\Lambda)$ in the $[-4\pi,4\pi]\sim[-15,15]$ perturbative range\footnote{This limit can be understood by demanding that a loop contribution from dimension-6 operators gives a contribution not larger than a genuine dimension-eight operator, and so forth for higher and higher dimensions. This is satisfied for $C_i \lesssim 4 \pi$, following naive dimensional analysis~\cite{Manohar:1983md,Manohar:2018aog}.}, and a Gaussian prior for the input parameters $G_F$, $\alpha$ or $M_W$ (depending on the input scheme), $M_Z$, $M_h$, $\alpha_s(M_Z)$ and the lepton masses, as described in Section \ref{sec:wc-running}. Concerning instead quark masses and CKM matrix elements, we consider uniform priors in the ranges reported in Table \ref{tab:smparameters}, and use the experimental information to constrain the relevant combination of SM parameters and SMEFT cofficients. SM parameters are always floating in the fit, while for SMEFT coefficients we perform fits with one operator at a time and fits with all coefficients floating simultaneously. To assess the impact of the different sets of observables, we also perform fits excluding one set at a time, and determine in this way which observables gives the strongest constraint.  
For each fit, the probability density function (pdf) of the SMEFT coefficients $C_i(\Lambda)$ is sampled using its numerical representation through the Markov Chain Monte Carlo (MCMC) implemented in 
\texttt{HEPfit} using the BAT library~\cite{Caldwell:2008fw}.
From the MCMC samples, we compute $68\%$ and $95\%$ Highest Posterior Density Intervals (HPDI) for the SMEFT coefficients. From the largest absolute value of the $95\%$ HPDI we compute the lower bound on the effective NP interaction scale $\Lambda/\sqrt{|C|}$.  

{\begin{table}[htbp]
  \centering
  \begin{tabular}{|c|c|c|c|}
    \hline
    Parameter & Range & Parameter & Range \\
    \hline
    $m_t(m_t)$ & $[155,171]$ GeV & $\lambda_\mathrm{\scriptscriptstyle{CKM}}$ & $[0.223,0.227]$ \\ 
    $m_b(m_b)$ & $[3.776,4.616]$ GeV & $A_\mathrm{\scriptscriptstyle{CKM}}$ & $[0.81,0.83]$ \\
    $m_c(m_c)$ & $[1.24,1.34]$ GeV & $\bar{\rho}_\mathrm{\scriptscriptstyle{CKM}}$ & $[0.165,0.185]$ \\
    $m_s(2 ~\mathrm{GeV})$ & $[88,98]$ MeV & $\bar{\eta}_\mathrm{\scriptscriptstyle{CKM}}$ & $[0.365,0.385]$\\
    $m_d(2 ~\mathrm{GeV})$ & $[4.4,5.0]$ MeV & & \\
    $m_u(2 ~\mathrm{GeV})$ & $[1.9,2.5]$ MeV & & \\
    \hline
  \end{tabular}
  \caption{Ranges of the SM-parameter uniform priors used in the fits. The values of the quark masses are given at the corresponding renormalization scale, as indicated in the table. The CKM parameters are defined in the Wolfenstein parametrisation~\cite{Wolfenstein:1983yz}.}
  \label{tab:smparameters}
\end{table}

\subsection{General setup for theoretical predictions}
\label{sec:theory-setup}

All the observables considered in the global fit are consistently calculated up to linear terms in the SMEFT Wilson coefficients\footnote{In the \texttt{HEPfit} framework this  corresponds to using the \texttt{NPSMEFTd6General} model setting the \texttt{QuadraticTerms} flag to false.}.
Establishing the validity of the SMEFT expansion at linear order is certainly important, and estimating it by including quadratic effects (i.e. terms of $O(1/\Lambda^4)$ originating from dimension-6 operators), although not sufficient to prove such validity, can provide partial information. Nevertheless, given the scope of our study, which presents for the first time a global fit of the SMEFT including EW, Drell-Yan, Higgs-boson, top-quark, and flavour observables, with SMEFT Wilson coefficients and SM parameters simultaneously floating and full RGE effects taken into account, we deem it appropriate to focus on a consistent implementation of the SMEFT at linear order and address the effect of quadratic terms in future studies.

Given a choice of input parameters $\{p\}$, all observables are expressed in terms of the input-parameter experimental values $\{\tilde{p}\}$ ($\tilde{p}=p+\delta p_\mathrm{SMEFT}$),
\begin{equation}
  \label{eq:obs-input-parameters}
  {\cal O}(\{\tilde{p}\})=\left[{\cal O}_{\mathrm{SM}}(\{p\})+\delta{\cal O}_{\mathrm{SMEFT}}(\{p\})\right]_{\{p\}\to \{\tilde{p}\}}\,,
\end{equation}
where SMEFT linear contributions come from the linear-order terms in both the $\delta{\cal O}_{\mathrm{SMEFT}}(\{p\})$ direct contribution and from expressing $p=p(\{\tilde{p}\})$ in terms of measured input parameters in ${\cal O}_{\mathrm{SM}}(\{p\})$. Choosing a given input scheme modifies the dependence of ${\cal O}(\{\tilde{p}\})$ on SMEFT coefficients. 
We note that for the purpose of illustration all the results presented in Section~\ref{sec:results} have been obtained in the $M_W$-scheme, which is commonly used for global fits that include a majority of collider observables.~\footnote{
The only exceptions are the predictions for LHC Drell-Yan observables which, as will be explained below, are computed with the {\tt HighPT} package~\cite{Allwicher:2022mcg}. These are given in the $\alpha$-scheme. Given that the difference between the $\alpha$ and $M_W$ schemes involves only operators that are well constrained by EW and Higgs precision observables, while the high-energy Drell-Yan measurements target four-fermion interactions, we expect this scheme mismatch to have a very small effect in the results.
}

Using Eq.~(\ref{eq:obs-input-parameters}), for all the observables considered in our study we have implemented state-of-the-art SM predictions (including QCD and EW corrections that could be analytically or numerically parametrised in an efficient way in the code) while SMEFT corrections, obtained from the interference between SMEFT and SM amplitudes, have been calculated at the lowest perturbative order. For several observables, e.g. the high-$p_T$ processes in top-quark physics, these are rescaled by the effect of higher-order SM corrections, assuming that their effects in the SM and the SMEFT are similar. When needed, further details will be given in the following sections where we discuss specific sets of observables.

\subsection{Electroweak and Drell-Yan observables}
\label{sec:ewpo}

We include in the fits the whole set of EW precision observables measured at the $Z$ pole at LEP/SLC.\footnote{We use the results obtained without assuming lepton universality.} A thorough description of all elements entering  the EW global fit,
including references to the state-of-the-art theoretical
parametrisations used in \texttt{HEPfit}, is given in refs.~\cite{deBlas:2021wap,deBlas:2022hdk}, to which we refer the reader interested in such details.

For the $W$ mass we consider the inputs from LEP2, ATLAS, CMS, LHCb and Tevatron, excluding the recent CDF measurement~\cite{CDF:2022hxs} which, in the light of the very recent CMS result~\cite{CMS:2024lrd}, requires additional investigation.\footnote{In ref.~\cite{deBlas:2022hdk} we discussed the impact of the CDF measurement on the EW precision fit, and introduced a \textit{conservative} scenario in which the uncertainty on the $M_W$ average including CDF was inflated to ensure compatibility among experimental measurements. This is less justified after the very recent CMS result, whose accuracy is comparable to the CDF one but is fully consistent with all existing measurements except for the CDF one.} 
We combine all the hadronic $W$-mass measurements obtained using the CT18 PDF set, assuming 
the same PDF correlation indicated in~\cite{LHC-TeVMWWorkingGroup:2023zkn}, which we take to be the same for ATLAS and CMS (not present in that reference). We also assume that QED systematics are 100\% correlated across the different measurements. The hadron-collider result is then combined with the LEP2 $W$-mass determination in an uncorrelated manner. This results in an average value $M_W=80.3635 \pm 0.0080$ GeV.
Similarly, we combine the measurements of the $W$ width results from LEP2, ATLAS and Tevatron.
There is some tension between the ATLAS and Tevatron determinations ($\chi^2/\textrm{d.o.f.}\approx 3$), which we take into account by scaling the uncertainty of the combination, before adding the LEP2 result.  
Finally, since the measurements of the $W$ width are not available for all experiments providing results for $M_W$, we neglect the correlation between the two observables. We obtain $\Gamma_W=2.151\pm 0.049$ GeV.
For the $Z$-boson mass, we also take into account the CDF determinations in ref.~\cite{CDF:2022hxs}, in good agreement with the LEP value. The Higgs-boson mass completes the list of SM inputs needed in the $W$-mass scheme, and here we use
the same value as in ref.~\cite{deBlas:2022hdk}. 
The numerical values used in the fit for those observables directly related to the SM inputs are reported in Table~\ref{tab:SM-exp}.\footnote{The Fermi constant $G_F$, known to a significantly higher precision, is fixed in the fits.}

\begin{table}[htbp]
  \centering
  \begin{tabular}{|c|c|c|c|}
    \hline
    Observable & Experimental value & Observable & Experimental value \\
    \hline
    $M_W$ (GeV) & $80.3635\pm0.0080$ & $M_Z$ (GeV) & $ 91.1880 \pm 0.0020$ \\
    $M_h$ (GeV) & $125.1 \pm 0.09$ & $\alpha_s(M_Z)$ & $0.11873 \pm 0.00056$ \\
    $m_t^\mathrm{pole}$ (GeV) & $172.61 \pm 0.59$ & $m_b(m_b)$ (GeV) & $4.196 \pm 0.014$ \\
    $m_c(m_c)$ (GeV) & $1.2917 \pm 0.0048$ & $m_s(2~\mathrm{GeV})$ (MeV) & $93.14 \pm 0.55$ \\
    $m_d(2~\mathrm{GeV})$ (MeV) & $4.70 \pm 0.04$ & $m_u(2~\mathrm{GeV})$ (MeV) & $2.20 \pm 0.06$ \\
    \hline
  \end{tabular}
  \caption{Observables relevant for the determination of the SM input parameters used in the fits (in the $W$ mass scheme). The values of the quark masses are taken from FLAG averages~\cite{FLAG:2021npn}. For $\alpha_s(M_Z)$ we use the recent lattice QCD result of ref.~\cite{Brida:2025gii}. See the text for details on the $W$ mass and width.
  }
  \label{tab:SM-exp}
\end{table}

Measurements of the $W$ branching ratios are also included, both from LEP2~\cite{ALEPH:2013dgf} and from the recent (and more precise) determination by CMS~\cite{CMS:2022mhs}. We also include the measurements available for ratios of $W$ and $Z$ decays into different lepton flavours from Tevatron~\cite{D0:1999rsi,CDF:2005bdv,D0:1999bqi,CDF:1991mse} and LHC~\cite{LHCb:2016zpq,ATLAS:2016nqi,ATLAS:2020xea,LHCb:2018ogb}, which serve as tests of lepton flavour universality. 

Two-to-two fermion processes above the $Z$ pole are also considered in our analysis. First, we include the measurements of total cross section and asymmetries for the $e^+ e^- \to \mu^+ \mu^-, \tau^+ \tau^-$ channels, the hadronic cross section, and the differential cross sections in $e^+ e^- \to e^+ e^-$ taken from LEP2~\cite{ALEPH:2013dgf}.~\footnote{On the hadronic side, we do not include the measurements of $R_{b}$ and $R_{c}$, since each is derived assuming the other is SM-like, which in general will not be true in the SMEFT.}
Di-lepton ($\ell^+ \ell^-$) and single-lepton ($\ell\nu$) processes from LHC Drell-Yan measurements are also included, implementing in {\tt HEPfit} the predictions from {\tt HighPT}~\cite{Allwicher:2022mcg}. These two-to-two fermion processes measured at high energies are largely sensitive to four-fermion operators, since their effects grow with the energy.

Finally, we also include in our study the constraints from $W^+W^-$ production at LEP2. For this observable we take into account the information available for the inclusive and differential cross sections measured at energies between 189 and 208~GeV, following the study in~\cite{Berthier:2016tkq}.

\subsection{Higgs-boson observables}
\label{sec:higgs}
We consider the full combination of Higgs-boson signal strengths at 8~TeV from ATLAS and CMS as presented in ref.~\cite{ATLAS:2016neq} as well as
the most recent combined measurements of Higgs-boson
production cross sections, decay branching ratios, and simplified template cross sections (STXS) as provided by the ATLAS and CMS
collaborations using 139 fb$^{-1}$ of collision data at
$\sqrt{s}=13$~TeV~\cite{ATLAS-CONF-2021-053,ATLAS:2022vkf,ATLAS:2024lyh,ATLAS:2024fkg,CMS:2022dwd,CMS-PAS-HIG-21-018}. All measurements are separated by production and decay modes. These
include measurements of Higgs-boson production
via gluon-gluon fusion (ggF), vector-boson fusion (VBF), and
associated production with gauge bosons ($ZH$ and $WH$) and top quarks
($t\bar{t}H$ and $tH$) as well as measurements of Higgs-boson decays
into $\gamma\gamma, ZZ^*, WW^*,\tau^+\tau^-,b\bar{b},\mu^+\mu^-$, and $Z\gamma$.  
For the STXS we adopt the most recent Stage 1.2 binning and compare with the experimental
results presented in~\cite{ATLAS:2024lyh,CMS-PAS-HIG-21-018}. In all cases we include the full correlation matrices provided by the experimental collaborations.

We parametrise both cross sections and decay rates at linear order in
the SMEFT coefficients using \texttt{MG5\_aMC@NLO} \cite{Alwall:2014hca} with a UFO file developed for this
study.\footnote{The UFO file used for this study has been generated with the \texttt{SMEFTci2} Mathematica in-house code. This is based on \texttt{SMEFTsim}~\cite{Brivio:2017btx,Brivio:2020onw} and has also been validated against \texttt{SmeftFR v3}~\cite{Dedes:2023zws}. 
} To obtain the parametrisation of some SM loop-induced processes, e.g. ggF, we also use the {\tt SMEFT@NLO} package~\cite{Degrande:2020evl}.

As mentioned earlier, we work in the $M_W$-scheme and use as default input parameters the experimental values of $G_F$, $M_Z$, and $M_W$. We adopt the
NNPDF4.0 NLO set of parton distribution functions with
$\alpha_s(M_Z)=0.118$ and fix factorisation and renormalization scales to the relevant scale for each process (e.g. $\mu=M_H$ for ggF, etc.). All chosen scales are very close to $\mu_W$, the scale at which we evaluate the RG-evolved SMEFT Wilson coefficients such that, for the purpose of this study, the effects of running from $\mu_W$ to the actual scale at which the Higgs-boson production processes have been calculated is negligible, specially given the experimental uncertainty on the high-$p_T$ bins.\footnote{As shown in Ref.~\cite{Battaglia:2021nys} (also recently revisited in Ref.~\cite{Maltoni:2024dpn}) these effects are not critical given the current experimental precision but may be relevant in the near future when data from HL-LHC becomes available.} Likewise, the theoretical error induced by the choice of PDF and their variation (at the few per-cent level for most state-of-the-art SM calculations) is also below the sensitivity of our study.

The parametrisation of the STXS has been
obtained at parton level. 
For the purpose of
the SMEFT global fit, working at parton level captures the bulk of the
relevant SMEFT effects. Indeed, we did not notice any substantial difference from
the parametrisations reported e.g.~in~\cite{ATLAS-CONF-2021-053} where
parton-shower effects have been fully simulated, except for a few bins that could not be obtained at parton level.
This includes for instance the case of ggF Higgs-boson production with 0 jets (where the STXS bin is defined by a cut on the Higgs-boson transverse momentum) or the case of Higgs-boson plus two jets final states (where the STXS bins are defined by a cut on the transverse momenta of the whole final-state system) since in both cases the transverse momentum used to define the STXS bin cannot be properly defined  without considering extra final-state partons as produced by the parton-shower. In these cases we have removed the corresponding bin.

\subsection{Top-quark observables}
\label{sec:top}

\begin{table*}[!ht]
\centering
\resizebox{\textwidth}{!}{  
\begin{tabular}{|l|c|c|c|c|c|}
\hline
Process & Observable & $\sqrt{s}$  & $\int \cal{L}$  & SM  & Ref.\\ \hline
\rule{0pt}{3.ex}\rule[-1.5ex]{0pt}{2.5ex} $p\bar{p} \rightarrow t\bar{t}$ & $ \rm{d}A^{t\bar{t}}_{\rm{FB}}/\rm{d}m_{t\bar{t}}$ & 1.96~TeV & 9.7~fb$^{-1}$ & \cite{Czakon:2014xsa,Czakon:2016ckf} & \cite{CDF:2017cvy} \\\hline
\rule{0pt}{3.ex} \multirow{6}{*}{$pp \rightarrow t\bar{t} $} & $\sigma_{t\bar{t}}^{13\rm{TeV}}/\sigma_{t\bar{t}}^{8\rm{TeV}}$  & 13 \& 8 TeV & 20 \& 36~fb$^{-1}$ & \multirow{6}{*}{\cite{Czakon:2011xx,Czakon:2013goa}} & \cite{ATLAS:2019hau} \\
 & $\sigma_{t\bar{t}}^{8\rm{TeV}}/\sigma_{t\bar{t}}^{7\rm{TeV}}$  & 8 \& 7 TeV & 20 \& 5~fb$^{-1}$ & & \cite{ATLAS:2022aof} \\
 & $\sigma_{t\bar{t}}$  & 13 TeV & 36/139~fb$^{-1}$ & & \cite{ATLAS:2019hau,ATLAS:2020aln} \\
 & $\rm{d}\sigma_{t\bar{t}}/\rm{d}m_{t\bar{t}}$  & 13 TeV & 36~fb$^{-1}$ &  & \cite{ATLAS:2019hxz} \\
 & $(\rm{d}\sigma_{t\bar{t}}/\rm{d}m_{t\bar{t}})/\sigma_{t\bar{t}}$  & 13 TeV & 36/137~fb$^{-1}$ &  & \cite{CMS:2018adi,CMS:2021vhb} \\
 \rule[-1.5ex]{0pt}{2.5ex} & $\rm{d}A_C/\rm{d}m_{t\bar{t}}$ & 13 TeV & 140~fb$^{-1}$ &  & \cite{ATLAS:2022waa} \\\hline
\rule{0pt}{3.ex}\rule[-1.5ex]{0pt}{2.5ex} $p p \rightarrow t \bar{t} Z$ &  $d\sigma/dp_T^Z$  & 13 TeV &  77.5/140~fb$^{-1}$  & 
\cite{Broggio:2019ewu} & \cite{CMS:2019too}/\cite{ATLAS:2023eld} \\ \hline
 \rule{0pt}{3.ex}\rule[-1.5ex]{0pt}{2.5ex} $p p \rightarrow t \bar{t} \gamma$ & $d\sigma/dp_T^\gamma$  & 13 TeV & 140~fb$^{-1}$ & 
\cite{Bevilacqua:2018woc,Bevilacqua:2018dny} & \cite{CMS:2021klw,CMS:2022lmh,ATLAS:2020yrp}  \\ \hline
 \rule{0pt}{1.ex} \multirow{2}{*}{$p p \rightarrow t \bar{t} W$} & $\sigma_{ttW^\pm}$ & \multirow{2}{*}{13 TeV} & \multirow{2}{*}{140~fb$^{-1}$}  & 
\multirow{2}{*}{\cite{Frederix:2017wme,Buonocore:2023ljm}} &  \multirow{2}{*}{\cite{ATLAS:2024moy,CMS:2022tkv}} \\
\rule[-1.5ex]{0pt}{0.5ex} &  $\sigma_{ttW^+}/\sigma_{ttW^-}$ &  &   & &  \\ \hline
\multirow{2}{*}{$t \rightarrow Wb $} & \multirow{2}{*}{$F_0$, $F_L$}  & 8~TeV & 20~fb$^{-1}$ & 
\multirow{2}{*}{\cite{PhysRevD.81.111503}}  & \cite{Aad:2020jvx} \\
 &   & 13~TeV & 140~fb$^{-1}$ &   & \cite{ATLAS:2022rms} \\ \hline
\multirow{3}{*}{$p p \rightarrow tW$} & \multirow{3}{*}{$\sigma$} & 7~TeV  & 4.6 \& 1.5 fb$^{-1}$  & 
 \cite{Kidonakis:2010ux} &  \cite{ATLAS:2019hhu}   \\
  &  & 8~TeV &  20~fb$^{-1}$  & 
 \cite{Kidonakis:2010ux} &  \cite{ATLAS:2019hhu}   \\ 
  &  & 13~TeV & 3.2/140~fb$^{-1}$  & 
 \cite{Kidonakis:2010ux} &  \cite{ATLAS:2016ofl}/\cite{CMS:2022ytw}   \\ \hline
 \multirow{2}{*}{$p p \rightarrow t\bar{b}$ (s-ch)} & \multirow{2}{*}{$\sigma$} & 8~TeV & 20~fb$^{-1}$ & 
\multirow{2}{*}{\cite{Aliev:2010zk,Kant:2014oha}} & \cite{ATLAS:2019hhu} \\ 
 &  & 13~TeV & 140~fb$^{-1}$ & & \cite{Muller:2723842,ATLAS:2022wfk} \\ \hline
\multirow{3}{*}{$p p \rightarrow tq$ (t-ch)} & \multirow{3}{*}{$\sigma$} & 7~TeV & 4.6 \& 1.5~fb$^{-1}$ & 
\multirow{3}{*}{\cite{Aliev:2010zk,Kant:2014oha}} & \cite{ATLAS:2019hhu} \\ 
 &  & 8~TeV & 20~fb$^{-1}$ &  & \cite{ATLAS:2019hhu} \\
 &  & 13~TeV &36/140~fb$^{-1}$ & & \cite{CMS:2018lgn}/\cite{ATLAS:2024ojr} \\ 
 \hline
\rule{0pt}{3.ex}\rule[-1.5ex]{0pt}{2.5ex} $p p \rightarrow t\gamma q$ & $\sigma$ & 13 TeV & 140/36~fb$^{-1}$  & 
 \cite{Alwall:2014hca} & \cite{ATLAS:2023qdu}/\cite{CMS:2018hfk} \\ \hline
\rule{0pt}{3.ex}\rule[-1.5ex]{0pt}{2.5ex} $p p \rightarrow tZq$ & $\sigma$ & 13 TeV & 140~fb$^{-1}$  & 
  \cite{Alwall:2014hca}  & \cite{CMS:2021ugv,ATLAS:2020bhu} \\ \hline
\rule{0pt}{3.ex}\rule[-1.5ex]{0pt}{2.5ex} $p p \rightarrow t\bar{t}b\bar{b}$ & $\sigma$ & 13~TeV & 36~fb$^{-1}$  & 
 \cite{Alwall:2014hca} &  \cite{CMS:2020grm}   \\ \hline
\rule{0pt}{3.ex}\rule[-1.5ex]{0pt}{2.5ex} $p p \rightarrow t\bar{t}t\bar{t}$ & $\sigma$ & 13~TeV & 140~fb$^{-1}$  & 
 \cite{vanBeekveld:2022hty} &  \cite{ATLAS:2021kqb,CMS:2023ftu,ATLAS:2023ajo}   \\ 
%
\hline
\end{tabular}
}
\caption{Measurements of top-quark observables included in the SMEFT fit. For each measurement we give: the process, the observable, the center-of-mass energy, and the integrated luminosity. The last two columns list the references for the SM predictions and measurements that are included in the fit. }
\label{tab:measurements}
\end{table*}

The top-quark interactions have been characterised with excellent precision by the Tevatron and the LHC. Our top-quark data set, summarised in Table~\ref{tab:measurements}, contains the most relevant and updated top-quark measurements. This data set consists of differential measurements of top-quark pair production, including also the $t\bar{t}$ charge asymmetry at LHC and the forward-backward asymmetry at Tevatron, differential and inclusive measurements of top-quark production in association with EW boson,\footnote{The associated production with Higgs bosons is listed in Section~\ref{sec:higgs} and, therefore, not listed here.} EW single top-quark production measurements, helicity measurements of top-quark decays, and four heavy quark production measurements including four top-quark production and two top-quark production in association with two bottom quarks.\footnote{A subset of these observables have been previously implemented in \texttt{HEPfit} to study specifically the top-quark sector \cite{Durieux:2019rbz,Miralles:2021dyw,Cornet-Gomez:2025jot}.}

In the top-quark sector, the theoretical uncertainties can be sizable and comparable to the experimental ones, particularly for differential measurements. To ensure a reliable analysis, we have taken both theoretical and experimental uncertainties into account. Furthermore, for the differential measurements, we have also included the correlations among the different bins, providing a complete and accurate treatment of the uncertainties. 

As in the case of Higgs-boson observables, top-quark observables are also parametrised at linear order in the SMEFT coefficients using \texttt{MG5\_aMC@NLO} with the same UFO file, PDF, and scale choices as described in Section~\ref{sec:higgs}. In this case, however, we have included differential distributions for the $ pp \to t\bar{t} $ process with invariant masses reaching up to a few~TeV. As shown in ref.~\cite{Czakon:2016dgf}, the most suitable dynamical scale for such differential observables is
$
H_T/4 = \frac{1}{4}\left(\sqrt{m_t^2 + p_{T,t}^2} + \sqrt{m_t^2 + p_{T,\bar{t}}^2}\right),
$
which is below one quarter of the invariant mass of the $ t\bar{t}$ system. Hence, even for the highest invariant mass bins, the optimal scale choice remains at a few hundred~GeV, comparable to the fixed scale $ \mu_W $ used for the other observables. This suggests that the deviation introduced by using $ \mu_W $ instead of the optimal dynamic scale should lie within the theoretical uncertainty of our predictions, thereby justifying the choice of keeping $ \mu_W $ also for this sector.

\subsection{Flavour observables}
\label{sec:flavour}

We now discuss in detail the procedure followed to study the constraints arising in the flavour sector. The floating SM parameters are the CKM matrix elements and the quark masses. The fit involves also a set of hadronic parameters reported in Table \ref{tab:flavour-parameters} that can be computed in lattice QCD without (or with negligible) effects from the SMEFT.\footnote{Results obtained fixing the lattice spacing using hadron spectroscopy are not affected by SMEFT contributions. In addition, there is usually good agreement (within current uncertainties) among results obtained fixing the lattice spacing with spectroscopy or with the weak decay constants, the latter procedure being potentially affected by SMEFT contributions. We conclude that within current uncertainties we can safely use all available lattice results.} 

In the computation of all flavour observables, the SM contribution is evaluated using all available higher-order corrections to achieve the best possible accuracy. The SMEFT Lagrangian is matched to the LEFT at the scale $\mu_W$ using the results of ref.~\cite{Jenkins:2017jig}. The LEFT coefficients are then evolved to the relevant energy scale using the LEFT renormalization group equations \cite{Jenkins:2017dyc}. 

\begin{table*}[!hb]
\centering
\begin{tabular}{|l|c|c|}
\hline
Parameter & Value & Ref.\\ \hline
$F_{B_s}$ (GeV)& $0.2301 \pm 0.0012$ & \cite{FLAG:2021npn} \\
$F_{B_s}/F_{B_d}$ & $1.208 \pm 0.005$ & \cite{FLAG:2021npn} \\
$B_{B_s}(4.2\, \mathrm{GeV})$ & $0.888 \pm 0.040$ & \cite{FLAG:2021npn} \\
$B_{B_s} / B_{B_d}$ & $1.015 \pm 0.021$ & \cite{FLAG:2021npn} \\
$B_{B_s,4}(4.2\, \mathrm{GeV})$ & $0.98 \pm 0.08$ & \cite{FLAG:2021npn} \\
$B_{B_s,5}(4.2\, \mathrm{GeV})$ & $1.66 \pm 0.13$ & \cite{FLAG:2021npn} \\
$B_{B_d,4}(4.2\, \mathrm{GeV})$ & $0.99 \pm 0.08$ & \cite{FLAG:2021npn} \\
$B_{B_d,5}(4.2\, \mathrm{GeV})$ & $1.58 \pm 0.18$ & \cite{FLAG:2021npn} \\
$B_{K}(2\, \mathrm{GeV})$ & $0.552 \pm 0.012$ & \cite{FLAG:2021npn} \\
$B_{K,4}(2\, \mathrm{GeV})$ & $0.904 \pm 0.053$ & \cite{FLAG:2021npn} \\
$B_{K,5}(2\, \mathrm{GeV})$ & $0.618 \pm 0.114$ & \cite{FLAG:2021npn} \\
$\phi_{\varepsilon_K} (^\circ)$ & $43.51 \pm 0.05$ & \cite{ParticleDataGroup:2024cfk}\\ 
$\overline{\kappa}_{\varepsilon_K}$ & $0.97 \pm 0.02$ & \cite{Ciuchini:2021zgf} \\
$(\Delta M_K)^{\rm SM}$ (ns$^{-1}$) & $8.8 \pm 3.6$ & \cite{Wang:2022lfq} \\
$B_{D,1}(3\, \mathrm{GeV})$ & $0.765 \pm 0.025$ & \cite{FLAG:2021npn} \\
$B_{D,4}(3\, \mathrm{GeV})$ & $0.98 \pm 0.06$ & \cite{FLAG:2021npn} \\
$B_{D,5}(3\, \mathrm{GeV})$ & $1.05 \pm 0.09$ & \cite{FLAG:2021npn} \\
$F_D$ (GeV) & $0.2120 \pm 0.0007$ & \cite{FLAG:2021npn} \\
$(\Delta M_D)^{\rm SM}$ (ps$^{-1}$) & $0.005 \pm 0.005$ & \cite{Betti:2024ldy} \\
$f_K$ (GeV) & $0.15611 \pm 0.00021$ & \cite{DiCarlo:2019thl} \\
$f_K / f_\pi$  & $1.1966 \pm 0.0018$ & \cite{DiCarlo:2019thl} \\
$\delta R_\pi^{\rm phys}$  & $0.0153 \pm 0.0019$ & \cite{DiCarlo:2019thl} \\
$\delta P_{c,u}$  & $0.04 \pm 0.02$ & \cite{Isidori:2005xm} \\
\hline
\end{tabular}
\caption{Key hadronic parameters floated in the theoretical prediction of flavour observables in our analysis. In particular, we report the values of bag parameters and meson decay constants, along with the long-distance contributions assumed for both $K$- and $D$-meson mixing and for $K \to \pi \nu \bar{\nu}$.}
\label{tab:flavour-parameters}
\end{table*}

\begin{table*}[!ht]
\centering
\begin{tabular}{|l|c|c|}
\hline
Observable & Value & Ref.\\ \hline
$\Delta m_{B_s}$ $(\mathrm{ps}^{-1})$& $17.765 \pm 0.006$ & \cite{HFLAV:2022esi,ParticleDataGroup:2024cfk} \\
$\phi_s$ & $-0.049 \pm 0.019$ & \cite{HFLAV:2022esi,ParticleDataGroup:2024cfk} \\
$A_{sl}^{s}$ & $-0.0006 \pm 0.00028$ & \cite{HFLAV:2022esi,ParticleDataGroup:2024cfk} \\
$\Delta m_{B_d}$ $(\mathrm{ps}^{-1})$ & $0.5069 \pm 0.0019$ & \cite{HFLAV:2022esi,ParticleDataGroup:2024cfk} \\
$S_{J/\psi K_S}$ & $0.692 \pm 0.016$ & \cite{HFLAV:2022esi,ParticleDataGroup:2024cfk,Ciuchini:2005mg} \\
$A_{sl}^{d}$ & $-0.0021 \pm 0.0017$ & \cite{HFLAV:2022esi,ParticleDataGroup:2024cfk} \\
$\Delta M_K$ $(\mathrm{ns}^{-1})$ & $5.293 \pm 0.009$ & \cite{ParticleDataGroup:2024cfk} \\
$\epsilon_K$ & $(2.228 \pm 0.011) \times 10^{-3}$ & \cite{ParticleDataGroup:2024cfk} \\
$\phi^{M}_{12}(^\circ)$ & $1.9 \pm 1.6$ & \cite{Betti:2024ldy} \\
${\rm BR}(B \to \tau \nu)  \times 10^{4}$ & $1.09 \pm 0.24$ & \cite{ParticleDataGroup:2024cfk} \\
${\rm BR}(D \to \tau \nu) \times 10^{4}$ & $9.9 \pm 1.2$ & \cite{ParticleDataGroup:2024cfk} \\
${\rm BR}(D \to \mu \nu) \times 10^4$ & $3.981 \pm 0.089$ & \cite{ParticleDataGroup:2024cfk} \\
${\rm BR}(D_s \to \tau \nu)  \times 10^{3} $& $5.31 \pm 0.11$ & \cite{ParticleDataGroup:2024cfk} \\
${\rm BR}(D_s \to \mu \nu)  \times 10^{2}$ & $5.37 \pm 0.10$ & \cite{ParticleDataGroup:2024cfk} \\
$\Gamma(K \to \mu \nu)/\Gamma(\pi \to \mu \nu)$ & $1.3367 \pm 0.0029$ & \cite{ParticleDataGroup:2024cfk} \\
${\rm BR}(\pi \to \mu \nu) \times 10^5$ & $3.8408 \pm 0.0007$ & \cite{ParticleDataGroup:2024cfk} \\
$d\Gamma(B \to D \ell \nu)/dw$ & $[\Delta \Gamma_i / \Delta w]_{10 \times 10} $ & \cite{Belle:2015pkj} \\
${\rm BR}(K^+ \to \pi^+ \nu \bar{\nu}) \times 10^{10}$ & $1.175 \pm 0.365$ & \cite{ParticleDataGroup:2024cfk} \\
${\rm BR}(B \to X_s \gamma) \times 10^{4}$ & $3.49 \pm 0.19$ & \cite{ParticleDataGroup:2024cfk} \\
$\overline{{\rm BR}}(B_s \to \mu^+ \mu^-) \times 10^{9}$ & $3.41 \pm 0.29$ & \cite{ParticleDataGroup:2024cfk} \\
\hline
\end{tabular}
\caption{Experimental measurements used for all the flavour observables considered in the present analysis, along with the corresponding references. The set includes both $\Delta F = 1$ and $\Delta F = 2$ observables, which play a crucial role in constraining the parameters of the Unitarity Triangle analysis \cite{UTfit:2022hsi}, testing the SM consistency, and probing potential contributions from NP \cite{Bona:2024bue}.
}
\label{tab:flavour-observables}
\end{table*}

\subsubsection{$\Delta F=2$ observables}
\label{sec:flavour-df2}

The most general $\Delta F=2$ effective Hamiltonian involves $5+3$ operators \cite{Gabbiani:1996hi}. However, in the context of the SMEFT, only $3+1$ of them are generated, namely, in the notation of ref.~\cite{Gabbiani:1996hi}:
\begin{eqnarray}
  Q_1^{q_iq_j} &=& (\bar{q}_i^\alpha \gamma_\mu P_L q_j^\alpha)(\bar{q}_i^\beta \gamma^\mu P_L q_j^\beta)\,, \nonumber\\
  Q_4^{q_iq_j} &=& (\bar{q}_i^\alpha P_L q_j^\alpha)(\bar{q}_i^\beta P_R q_j^\beta)\,, \nonumber\\
  Q_5^{q_iq_j} &=& (\bar{q}_i^\alpha P_L q_j^\beta)(\bar{q}_i^\beta P_L q_j^\alpha)\,, \nonumber\\
  \tilde{Q}_1^{q_iq_j} &=& (\bar{q}_i^\alpha \gamma_\mu P_R q_j^\alpha)(\bar{q}_i^\beta \gamma^\mu P_R q_j^\beta)\,,
  \label{eq:df2-operators}
\end{eqnarray}
where $q_i=u_i,d_i$ are up- and down-type quark fields, $P_{L,R}=(1\mp\gamma_5)/2$ are the left- and right-handed projectors, and $\alpha,\beta$ are colour indices. Furthermore, within the $U(2)^5$ flavour assumption, the operator $\tilde{Q}_1^{q_iq_j}$ is negligible. In terms of the LEFT operators of ref.~\cite{Jenkins:2017jig}, the Wilson coefficients of the above operators are given by
\begin{eqnarray}
  C_1^{q_iq_j} &=& -C_{qq}^{V,LL[ijij]}\,, \nonumber\\
  C_4^{q_iq_j} &=& C_{qq}^{V8,LR[ijij]}\,, \nonumber\\
  C_5^{q_iq_j} &=& 2C_{qq}^{V1,LR[ijij]}-C_{qq}^{V8,LR[ijij]}/3\,, \nonumber\\
  \tilde{C}_1^{q_iq_j} &=& -C_{qq}^{V,RR[ijij]}\,.
\end{eqnarray}
For the relevant matrix elements we use the lattice B-parameters reported in Table \ref{tab:flavour-parameters}. For $B_q - \bar{B}_q$ mixing ($q=d,s$), we use in the fit the mass difference, the time-dependent CP asymmetries, and the semileptonic asymmetries. For $K - \bar{K}$ mixing, we use the $\epsilon_K$ CP-violating parameter and the mass difference $\Delta M_K$. For the SM contribution to the latter, we use the lattice QCD estimate in ref.~\cite{Wang:2022lfq}. Finally, for the dispersive phase of $D-\bar{D}$ mixing $\phi^M_{12}$, we use the results of the global fit of ref.~\cite{Betti:2024ldy} and, lacking an estimate of the SM contribution to the real part of the amplitude, assume that the SM contribution can saturate the experimental value. All the experimental results used in this section are summarised in Table \ref{tab:flavour-observables}, while the expressions for the observables in terms of the operators in eq.~(\ref{eq:df2-operators}) can be found for example in refs.~\cite{UTfit:2006onp,UTfit:2007eik}.

\subsubsection{$\Delta F=1$ observables}
\label{sec:flavour-df1}

Several $\Delta F=1$ processes enter our analysis for the purpose of determining the CKM matrix elements and constraining the SMEFT coefficients. Let us consider first the charged current leptonic decays, which include $\pi \to \mu \nu$, $K \to \mu \nu$, $B \to \tau \nu$, $D_{(s)} \to \tau \nu$, and $D_{(s)} \to \mu \nu$. The relevant LEFT operators are given by
\begin{eqnarray}
  \label{eq:LEFT_DF1_tree}
  Q_{\nu e d u}^{VLL[kkij]} &=& (\bar{\nu}_k \gamma_\mu P_L e_k)(\bar{d}_i \gamma^\mu P_L u_j)\,, \nonumber\\
  Q_{\nu e d u}^{VLR[kkij]} &=& (\bar{\nu}_k \gamma_\mu P_L e_k)(\bar{d}_i \gamma^\mu P_R u_j)\,, \nonumber\\
  Q_{\nu e d u}^{SRR[kkij]} &=& (\bar{\nu}_k P_R e_k)(\bar{d}_i P_R u_j)\,, \nonumber\\
  Q_{\nu e d u}^{SRL[kkij]} &=& (\bar{\nu}_k P_R e_k)(\bar{d}_i P_L u_j)\,, \nonumber\\
  Q_{\nu e d u}^{TRR[kkij]} &=& (\bar{\nu}_k \sigma_{\mu\nu} P_R e_k)(\bar{d}_i \sigma^{\mu\nu} P_R u_j)\,.
\end{eqnarray}
After the RGE below the EW scale, the relevant matrix elements are computed using the lattice QCD results for the decay constants, including the QED corrections for light meson decays \cite{DiCarlo:2019thl}, as reported in Table \ref{tab:flavour-parameters}. The experimental results used in this section are all summarised in Table \ref{tab:flavour-observables}. 

For the process $B \to D \ell \nu $, the relevant matching conditions remain those corresponding to the operators shown in eq.~\eqref{eq:LEFT_DF1_tree}. The corresponding matrix elements are computed using the Boyd-Grinstein-Lebed parameterization, employing a $z$-expansion truncated at second order in the conformal variable $z$. In this analysis, we incorporate constraints from lattice QCD given in \cite{MILC:2015uhg} as well as unitarity bounds on the expansion coefficients. The experimental data used also include a $ 10 \times 10 $ correlation matrix ($[\Delta \Gamma_i / \Delta w]_{10 \times 10} $), see reference in Table~\ref{tab:flavour-observables}.  

We now turn to the second class of $\Delta F = 1$ decays, namely the flavour-changing neutral current (FCNC) processes. In our analysis, we include two key decays that are particularly sensitive to potential contributions from BSM physics: the inclusive radiative decay $B\rightarrow X_s\gamma$ and the short-distance dominated exclusive decay $B_s \to \mu^+ \mu^-$. SM predictions for these processes are computed to next-to-next-to-leading order accuracy as detailed in references \cite{Misiak:2020vlo,Bobeth:2013uxa}. The leading effects of NP for the inclusive radiative channel involve the following operators:
\begin{eqnarray}
  \label{eq:LEFT_DF1_bsg}
  Q_{d \gamma}^{[ij]} &=&  (\bar{d}_i \sigma_{\mu \nu} P_R d_j) \, F^{\mu \nu}\,, \nonumber\\
  Q_{d G}^{[ij]} &=&  (\bar{d}_i \sigma_{\mu \nu} T_A P_R d_j) \, G^{\mu \nu}_A \,, \nonumber\\
  Q_{ud}^{V1LL[ijkl]} &=& (\bar{u}_i \gamma_{\mu} P_L u_j) \, (\bar{u}_k \gamma^{\mu}  P_L d_l) \,, \nonumber\\
  Q_{ud}^{V8LL[ijkl]} &=& (\bar{u}_i \gamma_{\mu} T^A P_L u_j) \, (\bar{u}_k \gamma^{\mu} T^A P_L d_l) \,, \nonumber\\
  Q_{ud}^{V1RR[ijkl]} &=& (\bar{u}_i \gamma_{\mu} P_R u_j) \, (\bar{u}_k \gamma^{\mu}  P_R d_l) \,, \nonumber\\
  Q_{ud}^{V8RR[ijkl]} &=& (\bar{u}_i \gamma_{\mu} T^A P_R u_j) \, (\bar{u}_k \gamma^{\mu} T^A P_R d_l) \nonumber\,,
\end{eqnarray}
where the current-current operators also contribute to the decay at leading order via one-loop running effects below the EW scale. On the other hand, the leading-order contributions of new physics in $B_s \to \mu^+ \mu^-$ are not affected by RGE effects below the EW scale and, in particular, are given by the tree-level matching of the SMEFT onto: 
\begin{eqnarray}
  \label{eq:LEFT_DF1_Bsmumu}
  Q_{e d}^{VLL[kkij]} &=&  (\bar{e}_k \gamma_{\mu} P_L e_k) \, (\bar{d}_i \gamma^{\mu} P_L d_j) \,, \nonumber\\
  Q_{d e}^{VLR[ijkk]} &=&  (\bar{d}_i \gamma_{\mu} P_L d_j) \, (\bar{e}_k \gamma^{\mu} P_R e_k) \nonumber\,.
\end{eqnarray}

Finally, we consider the FCNC process $K^+ \to \pi^+ \nu \bar{\nu}$, which is also theoretically clean and highly suppressed in the SM \cite{Brod:2010hi}. Due to its strong sensitivity to short-distance physics and minimal hadronic uncertainties \cite{Brod:2021hsj}, it provides a powerful probe of potential new physics effects. Assuming lepton universality and taking into account the flavour symmetries adopted in our study, the only LEFT operator relevant for this channel is:
\begin{eqnarray}
  \label{eq:LEFT_DF1_kpinunu}
  Q_{\nu d}^{VLL[kkij]} &=&  (\bar{\nu}_k \gamma_{\mu} P_L \nu_k) \, (\bar{d}_i \gamma^{\mu} P_L d_j) \, \nonumber.
\end{eqnarray}


\section{Results}
\label{sec:results}

In this section, we present the results of the fits we performed with different assumptions on the SMEFT flavour symmetries, namely $U(3)^5$ and $U(2)^5$ in both the UP and DOWN bases defined in Section~\ref{sec:smeft}. In order to assess the impact of the RGE of the SMEFT Wilson coefficients, we give results for $\Lambda=10$ and $\Lambda=3$ TeV with RGE, and $\Lambda = 1$ TeV without RGE (labeled in the plots and tables as ``noRGE'').
Given the leading logarithmic accuracy of the RGE, to ensure the dominance of log-enhanced terms we require $\ln{(\mu_1^2/\mu_2^2)} \gtrsim 2$, and consequently we cannot resolve scale ratios smaller than $\sim 3$, which justifies the chosen values of $\Lambda$. We perform both global fits with all SMEFT Wilson coefficients floating and fits varying one Wilson coefficient at a time (called \textit{individual fits} in the following), while SM input parameters and hadronic matrix elements are always floated. For individual fits, we also perform fits omitting different sets of observables, to assess the impact of each set on the results. Numeric results for all the fits are reported in Appendix \ref{sec:app-results}, while here we illustrate them in a series of plots and discuss their main features and physical interpretation. 

\subsection{Results for the \texorpdfstring{$U(3)^5$}{U3-5} flavour symmetric SMEFT}
\label{sec:results-u3}

\begin{figure}[h!]
    \centering
    \includegraphics[width=1.\linewidth]{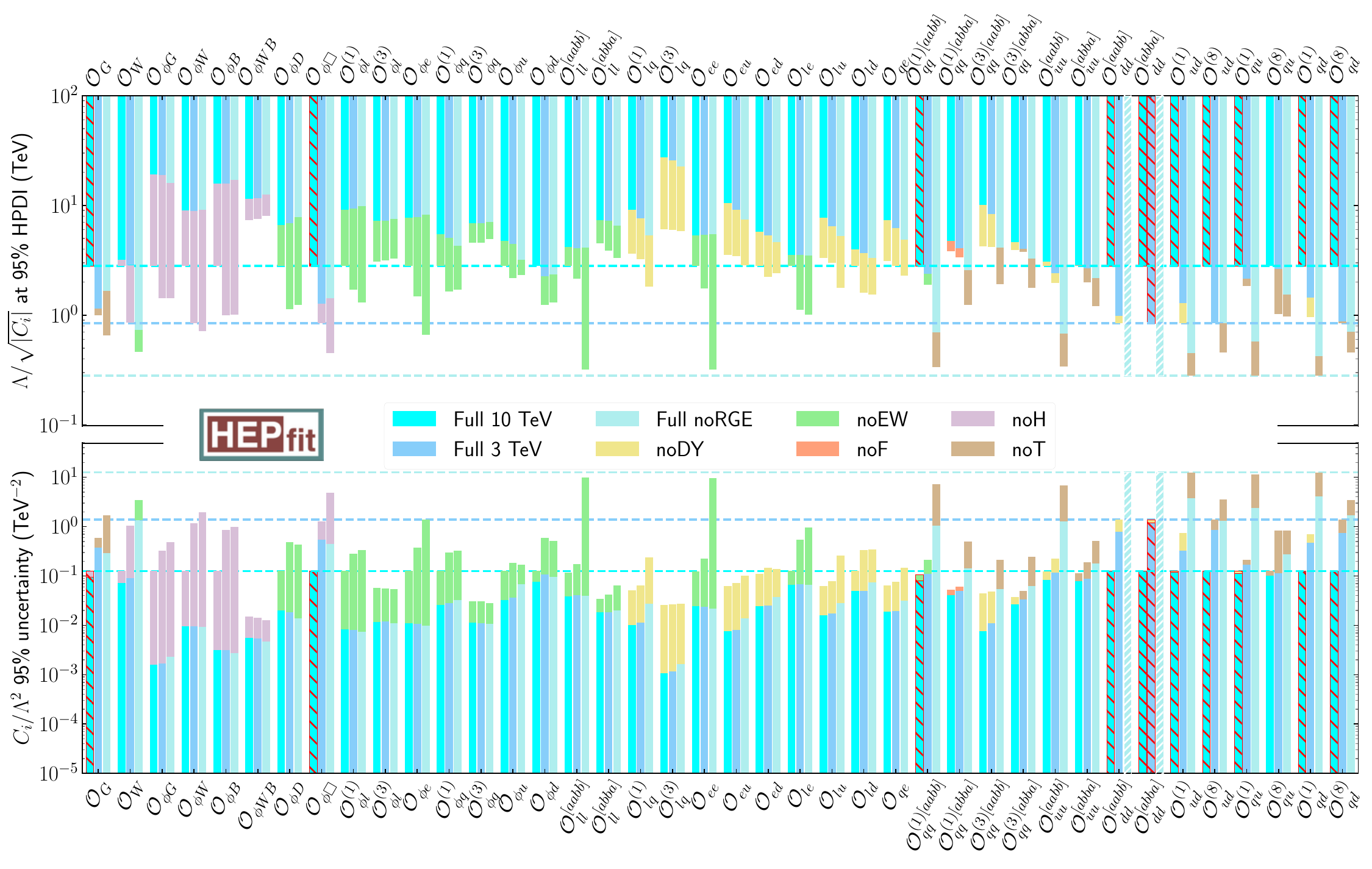}
    \caption{Results from individual fits in the $U(3)^5$ flavour symmetric SMEFT. For each coefficient $C_i$, the top panel shows the scale of NP allowed by the data at 95\% probability (normalized by the square root of the maximum of the 95\% HPDI interval for $|C_i|$). The bottom panel shows the width of the 95\% probability range divided by two. Both panels show results for the three cases 1) $\Lambda=10$~TeV with RGE, 2) $\Lambda=3$~TeV with RGE, and 3) $\Lambda=1$~TeV with no RGE. Furthermore, in each case we also include results obtained removing the most constraining data set for that particular coefficient (see Table~\ref{tab:u3-ind}). The color code is as explained in the legend. The horizontal lines indicate the maximum value allowed for each Wilson coefficient in the fit, corresponding to the perturbativity limit $4\pi$ (see Section~\ref{sec:fit-intro}), for the different values chosen for the NP scale $\Lambda$. The cases in which the $95\%$ HPDI interval touches the prior's edges, indicated in red in Table~\ref{tab:u3-ind}, are hatched with red diagonal lines. When the posterior distribution of a coefficient is completely flat the $95\%$ HPDI interval is hatched with diagonal white lines.
    }
    \label{fig:full_fit_ind_U3}
\end{figure}

Assuming that NP respects the $U(3)^5$ flavour symmetry, the SMEFT coefficients at the scale $\Lambda$ are given by eq.~(\ref{eq:u3}). Let us first consider individual fits. In Table 
\ref{tab:u3-ind} we report results for $\Lambda=10$ TeV and $\Lambda=3$ TeV including RGE, and for $\Lambda=1$ TeV without RGE. For each coefficient we give the 68\% HPDI interval and the bound on $\Lambda/\sqrt{|C_i|}$ obtained by taking the maximum of the 95\% HPDI interval for $|C_i|$. The cases in which the posterior for the relevant Wilson coefficient is not centered around zero so that the $68\%$ or $95\%$ HPDI (from which the bound on $\Lambda/\sqrt{\vert C_i\vert }$ is obtained) touches the edges of the prior's range (see Section~\ref{sec:fit-intro}) are indicated in red. The $68\%$ HPDI intervals for the $C_i$ roughly scale as $\Lambda^2$ up to RG effects, unless they hit the prior's edges. Conversely, the bounds on the effective scale $\Lambda/\sqrt{\vert C_i\vert }$ only change due to RG effects and possibly due to the prior's range. It happens for several less constrained coefficients that the sensitivity to experimental bounds gets too diluted at higher scales, so that the $95\%$ HPDI interval for $\vert C_i \vert$ touches the prior's edges, and consequently a bound on the scale cannot be obtained. RGE-induced effects are particularly relevant for coefficients such as $C_G$, which without RGE is mainly constrained by top-quark observables,\footnote{The $C_G$ coefficient can also be strongly constrained using multijet production \cite{Krauss:2016ely,Hirschi:2018etq} not present in our data set. As shown in those studies, the largest sensitivity to this operator comes from ${\cal O}(1/\Lambda^4)$ effects, while linear effects provide limited information.} or $C_W$, mainly constrained by EW observables, while including RGE both get stronger constraints from Higgs-boson observables. This is also illustrated in Fig.~\ref{fig:full_fit_ind_U3}, based on Table~\ref{tab:u3-ind}. In the lower panel, we report the half width of the $95\%$ HPDI interval for $\vert C_i \vert/\Lambda^2$ for the three cases 1) $\Lambda=10$~TeV with RGE, 2) $\Lambda=3$~TeV with RGE, and 3) $\Lambda=1$~TeV without RGE, both in the full fit (i.e. including all observables) and excluding the most constraining set of observables (using the color code explained in the figure). To guide the eye, we also draw the $4 \pi/\Lambda^2$ perturbativity bound as a horizontal dashed line. The cases in which the $95\%$ HPDI interval touches the prior's edges, indicated in red in Table \ref{tab:u3-ind}, are hatched with red diagonal lines. When the posterior is flat since the processes considered cannot bound the corresponding coefficient, the $95\%$ HPDI interval is hatched with diagonal white lines. Thus, ranges touching the perturbativity line but solid or hatched in red can be refined by improving the precision of the measurements already included in the fit, while ranges hatched with diagonal white lines can only be improved by including new observables. The results of the fits can also be translated into the allowed range at $95 \%$ probability for $\Lambda/\sqrt{\vert C_i \vert}$. Since no coefficient is driven away from zero at $95\%$ probability, we can only put a lower bound on the effective NP scale, as reported in the upper panel of Fig.~\ref{fig:full_fit_ind_U3}, with the same color code. Finally, we summarise the information on the lower bounds on the effective NP scale obtained for $\Lambda = 3$ TeV in Fig.~\ref{fig:bounds_U3}.

\begin{figure}[h!]
    \centering
    \includegraphics[width=1.\linewidth]{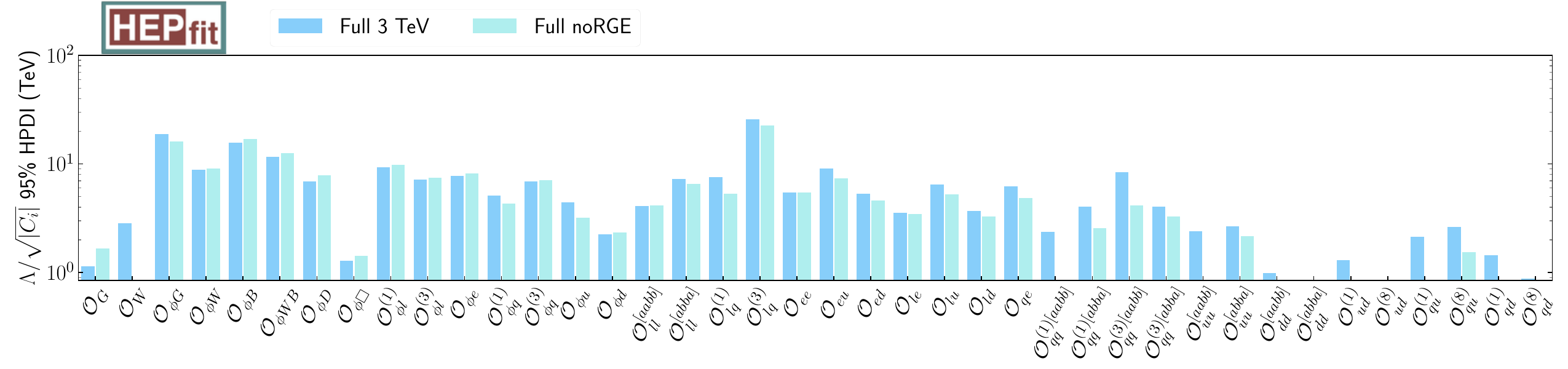}
    \caption{Summary of individual lower bounds on the effective NP scale in the $U(3)^5$ flavour symmetric SMEFT, obtained for $\Lambda = 3$ TeV. }
    \label{fig:bounds_U3}
\end{figure}

We now discuss the results of the global fit for the $U(3)^5$ assumption, whose results are presented in Table \ref{tab:u3-glob} and Figure~\ref{fig:global_fit_U3}. It is well known that, in the Warsaw basis, there are flat directions in the fit to EW precision observables in terms of SMEFT coefficients at the EW scale, see e.g. ref.~\cite{Falkowski:2014tna}. These flat directions are lifted by the inclusion of other observables, for example in the Higgs sector, but leave their footprint in terms of correlations among the coefficients. Furthermore, the flat directions at the EW scale get distorted by the RGE. While the full correlation matrix is too large to be presented here, we can mention a few examples of highly correlated coefficients. We observe a correlation of $0.95$ between $C_{\phi B}$ and $C_{\phi WB}$, of $-0.92$ between $C_{\phi WB}$ and $C_{\phi D}$, of $-0.89$ between $C_{\phi B}$ and $C_{\phi D}$, of $-0.85$ between $C_{\phi D}$ and $C_{\phi e}$, of $0.84$ between $C_{\phi l}^{(1)}$ and $C_{\phi e}$, and of $0.83$ between $C_{\phi W}$ and $C_{\phi WB}$. Those correlations are induced by EW and Higgs observables. Moreover, we observe a strong correlation of $-0.998$ between $C_{ll}^{[aabb]}$ and $C_{ee}$, of $0.83$ between $C_{ll}^{[aabb]}$ and $C_{ll}^{[abba]}$, and of $-0.86$ between $C_{ll}^{[abba]}$ and $C_{ee}$, all induced by EW observables, in particular, two-to-two fermion processes from LEP2. We also observe a correlation of $0.88$ between $C_{\phi G}$ and $C_{G}$, due to Higgs observables, and of $-0.83$ between $C_{qu}^{(1)}$ and $C_{qu}^{(8)}$, which is reasonably aligned with the expected RGE contributions from these two operators to ${\cal O}_{u\phi}^{[33]}$, whose coefficient modifies the top-quark Yukawa coupling. Finally, we mention a correlation of $-0.87$ between $C_{lq}^{(1)}$ and $C_{eu}$, due to DY observables. In spite of the dilution of the constraints due to interference between the coefficients, as can be seen from Figure~\ref{fig:global_fit_U3} and Table~\ref{tab:u3-glob}, we still obtain $\Lambda/\sqrt{\vert C_i \vert} \gtrsim 14$ TeV for $C_{\phi G}$, $\Lambda/\sqrt{\vert C_i \vert} \gtrsim 5$ TeV for $C_{lq}^{(3)}$, while several other coefficients give bounds of the order of $2$-$3$ TeV. 
Therefore, for a generic strongly-interacting $U(3)^5$ flavour symmetric theory, current data constrain NP to be well above the TeV scale, making the SMEFT approach fully consistent, while weakly-interacting $U(3)^5$ flavour symmetric theories are still allowed at the TeV scale, where however the SMEFT approach might not be fully consistent for high-$p_T$ observables, and direct searches could be more effective.

\begin{figure}[h!]
    \centering
    \includegraphics[width=1.\linewidth]{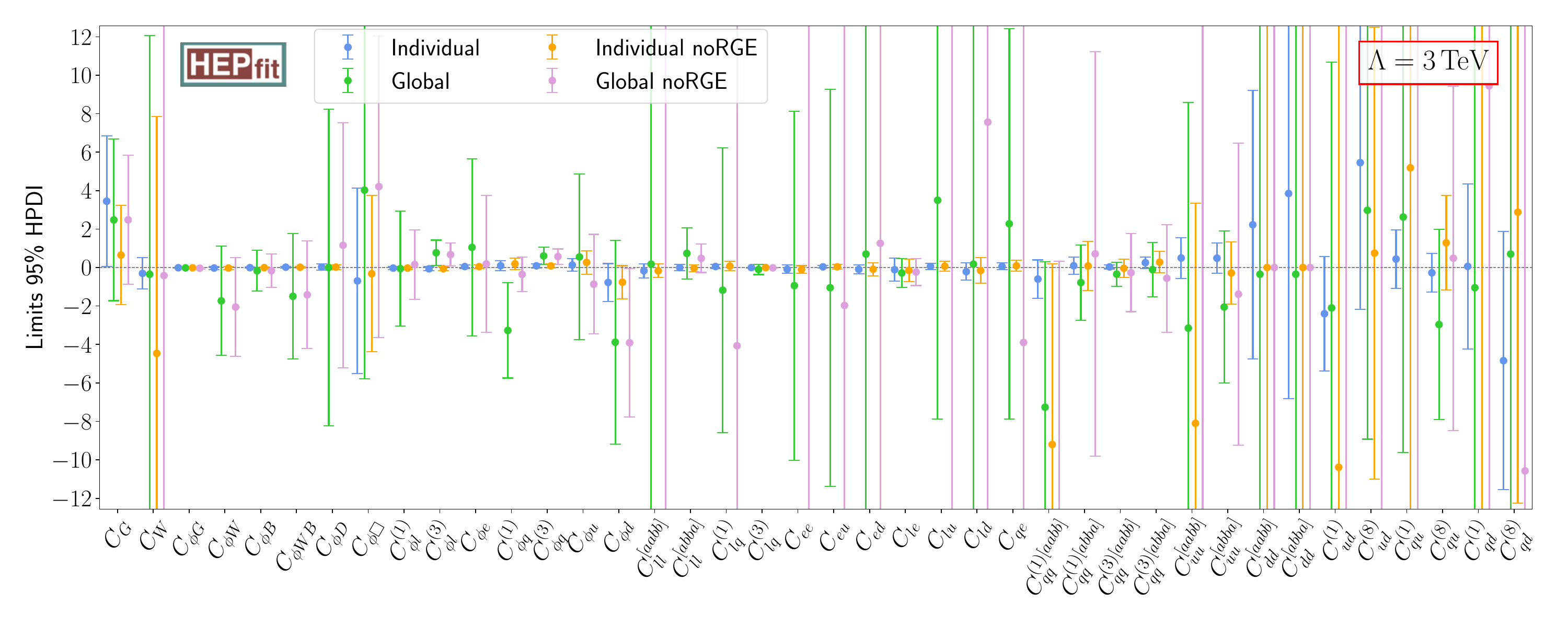}
    \caption{Comparison of individual and marginalised constraints from the global fit with $U(3)^5$ flavour symmetry assumption. The scale of NP has been set to $\Lambda=3$ TeV. The limits shown correspond to the 95\% HPDI. Results are presented for both individual and global fits, with and without the RGE effects (the latter also adjusted to a value of $\Lambda=3$ TeV for this comparison), following the colour scheme indicated in the legend.  }
    \label{fig:global_fit_U3}
\end{figure}

Finally, regarding the relaxation of the bounds in going from the individual fits to the global one due to mutual strong correlations, we would like to emphasise that this implies that a lower effective NP scale $\Lambda/\sqrt{\vert C_i \vert}$ with respect to the one indicated by individual fits is viable only for those UV completions that incorporate the phenomenologically required correlations. 
Conversely, we note that taking into account the theory prior imposing  perturbativity on the Wilson coefficients actually tightens many bounds.
Table~\ref{tab:u3-glob} shows that several limits on $\Lambda/\sqrt{\lvert C_i\rvert}$ at 10 TeV now exceed those at 3 TeV. In particular, although correlations with other operators substantially weaken the global-fit constraints on $C_{\phi B}$ and $C_{lq}^{(3)}$, restricting the allowed ranges of the correlated $C_i/\Lambda^2$ coefficients ensures that the bounds on $\Lambda/\sqrt{\lvert C_{\phi B}\rvert}$ and $\Lambda/\sqrt{\lvert C_{lq}^{(3)}\rvert}$ remain comparably strong at 10 TeV.

\subsection{Results for the \texorpdfstring{$U(2)^5$}{U2-5} flavour symmetric SMEFT}
\label{sec:results-u2}

\begin{figure}[h]
    \centering
    \includegraphics[width=1.\linewidth]{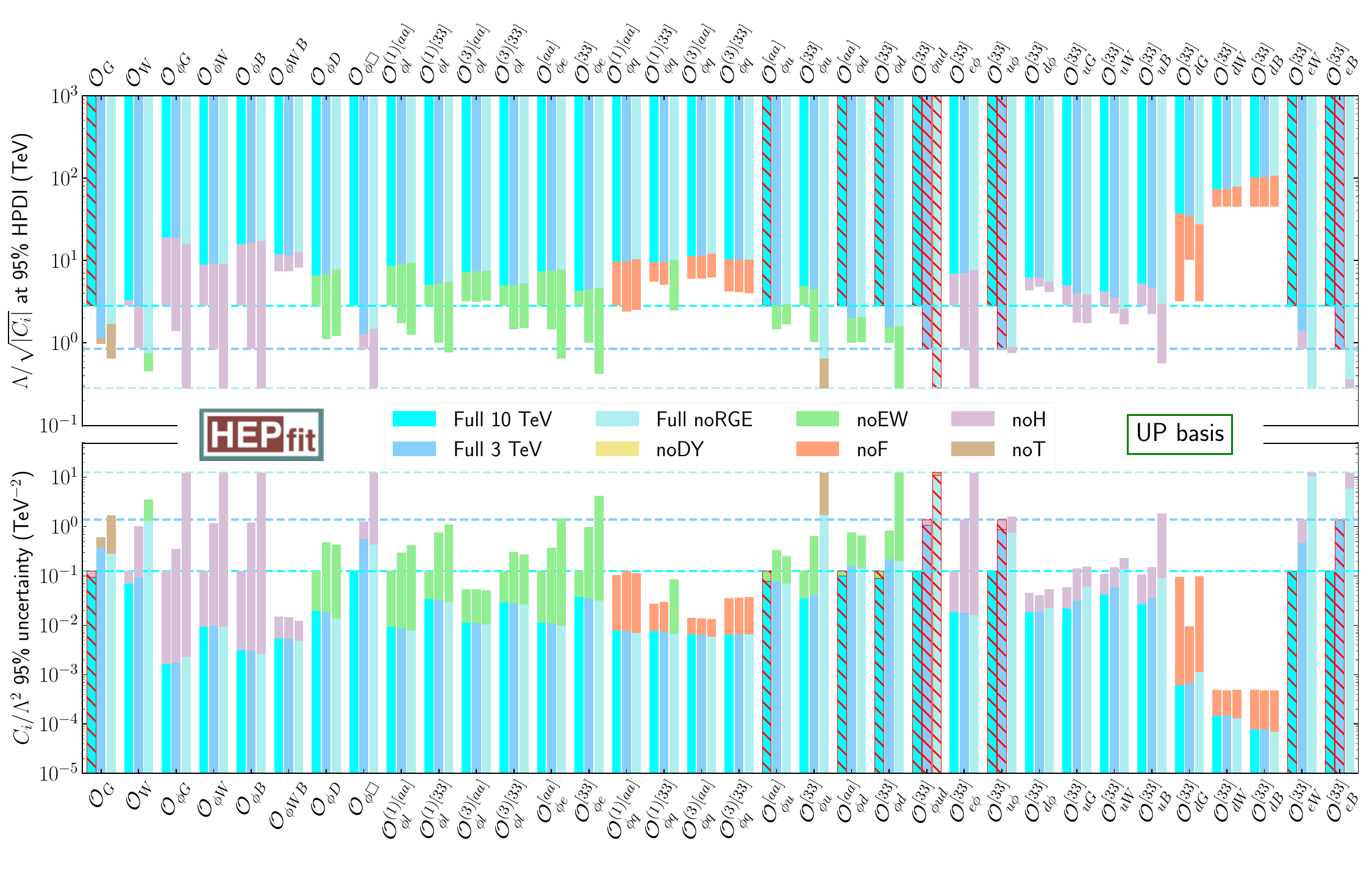}
    \caption{
     Same as Fig.~\ref{fig:full_fit_ind_U3} for bosonic and two-fermion operators in the $U(2)^5$ symmetric SMEFT (UP basis).
    }
    \label{fig:ind_bos_2F}
\end{figure}

\begin{figure}[h]
    \centering
    \includegraphics[width=1.\linewidth]{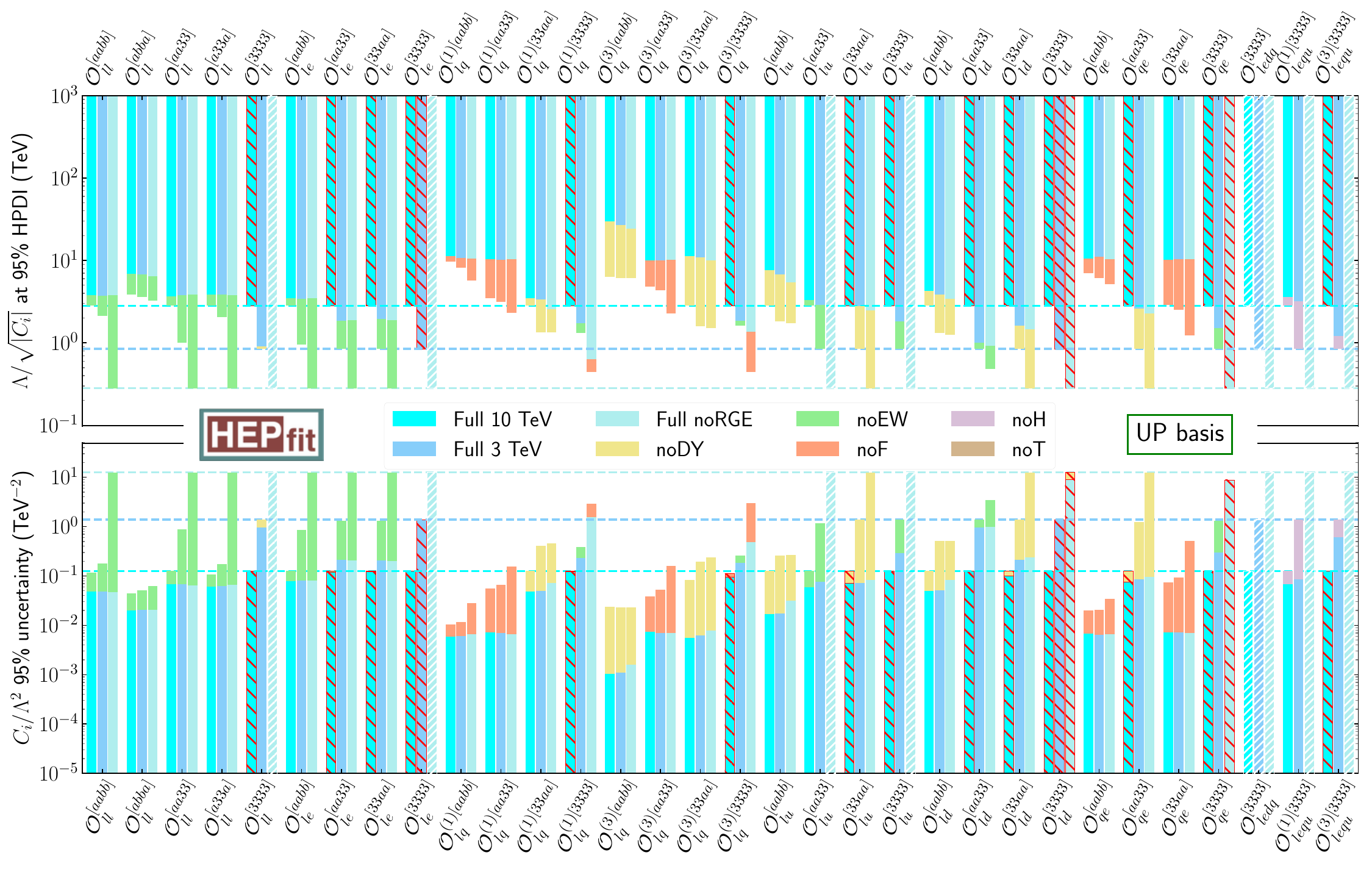}
    \caption{Same as Figure~\ref{fig:ind_bos_2F} but for the four-fermion operators that contain at least one leptonic field and one left-handed field.}
    \label{fig:ind_4FlepL}
\end{figure}

\begin{figure}[h]
    \centering
    \includegraphics[width=1.\linewidth]{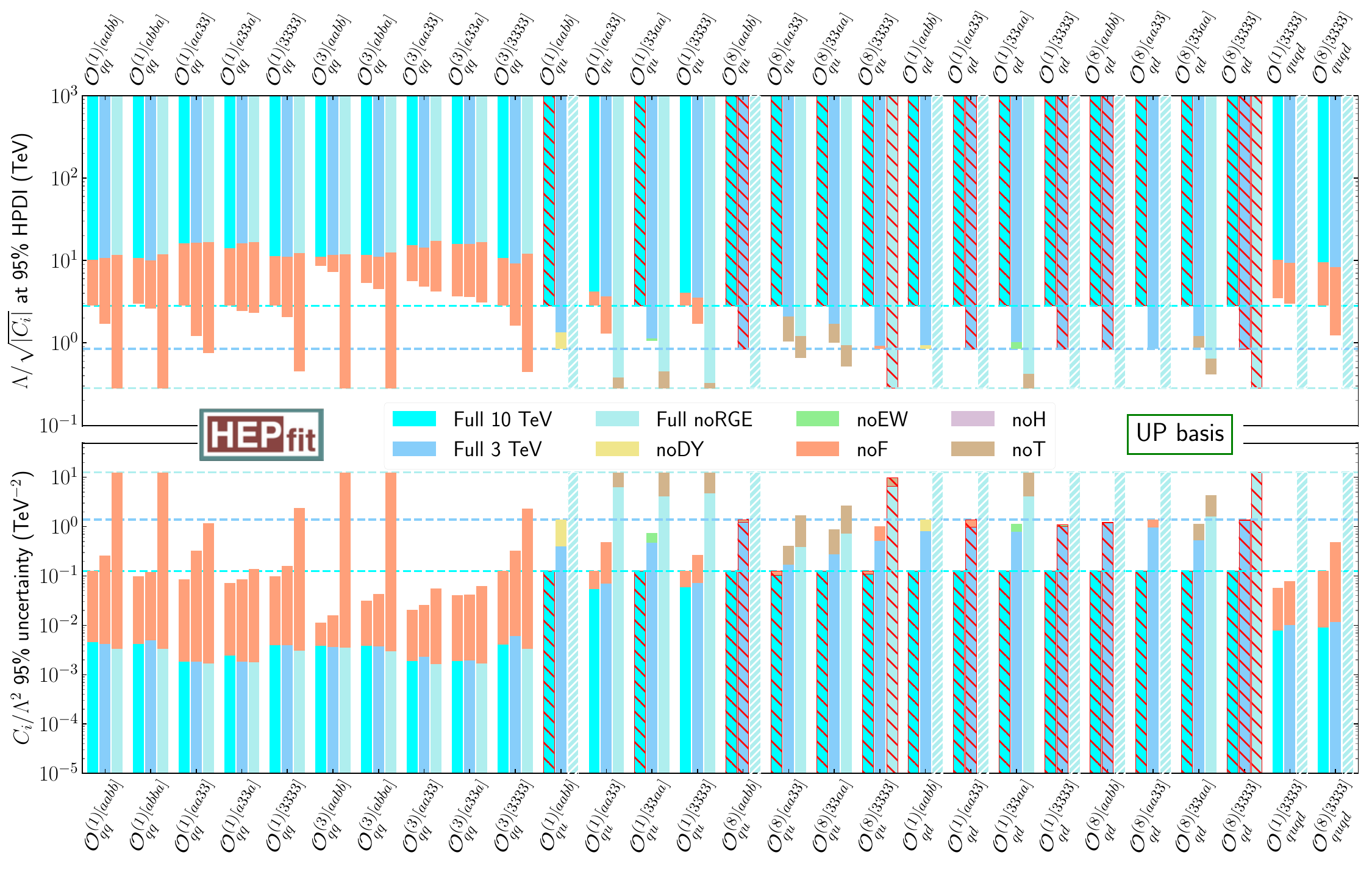}
    \caption{Same as Figure~\ref{fig:ind_bos_2F} but for the four-fermion operators that contain only quark fields and at least one left-handed field.}
    \label{fig:ind_4FquarkL}
\end{figure}

\begin{figure}[h]
    \centering
    \includegraphics[width=1.\linewidth]{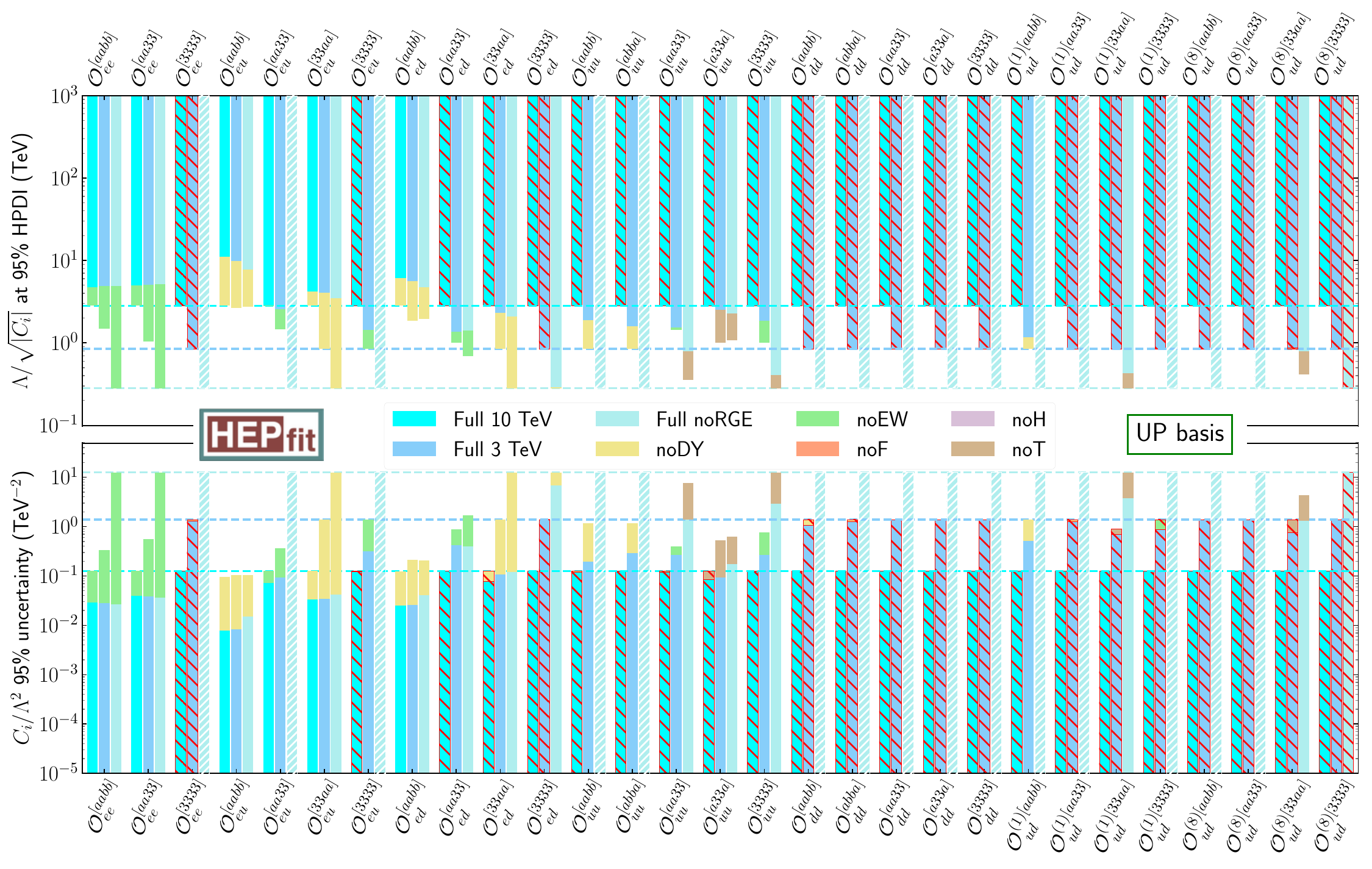}
    \caption{Same as Figure~\ref{fig:ind_bos_2F} but for the four-fermion operators that contain only right-handed fields.}
    \label{fig:ind_4FRR}
\end{figure}

The results of the individual fits for the $U(2)^5$ flavour symmetric SMEFT in the UP basis are reported in Table~\ref{tab:u2-ind} and in Figures \ref{fig:ind_bos_2F}-\ref{fig:ind_4FRR}, which have the same structure as Table~\ref{tab:u3-ind} and Figure~\ref{fig:full_fit_ind_U3} discussed in Section~\ref{sec:results-u3} for the $U(3)^5$ case, to which we refer. The limits on $\Lambda/\sqrt{|C_i|}$ are also summarised in Figure~\ref{fig:bounds_summary_U2}, where we only show the coefficients that can be constrained individually within the pertubative regime in the case of $\Lambda=3$ TeV.
Furthermore, a comparison between the case of the UP and DOWN bases is summarised in Table~\ref{tab:u2-up-down} and Figure~\ref{fig:bounds_UP_vs_DOWN} for the case of those coefficients that are affected by the choice of flavour basis. Finally, the global fit results are shown in Table \ref{tab:u2-glob} and Figures~\ref{fig:full_fit_bos_2f_2fqLL_Lam_10} and \ref{fig:full_fit_4f_Lam_10} where we show the results for the case of $\Lambda=10$ TeV.

The main points highlighted in the case of $U(3)^5$ still hold for $U(2)^5$, but there are several important new considerations that we discuss in the following. The first and most important one is that flavour observables play a crucial role in constraining many of the new coefficients appearing in the $U(2)^5$ case, consequently pushing the effective scale of NP to several tens of TeVs. This is the case for instance of $C_{dB}^{[33]}$, $C_{dW}^{[33]}$ and $C_{dG}^{[33]}$, which are very strongly constrained by $B \to X_s \gamma$ decays. Four-fermion operators are also strongly constrained by flavour observables, such as $C_{qq}^{(1)}$  and $C_{qq}^{(3)}$in all flavour combinations, which are constrained by meson-antimeson mixing, and $C_{qe}^{[33aa]}$ and $C_{qe}^{[aabb]}$ which are constrained by $B_s \to \mu^+ \mu^-$.

\begin{figure}
    \centering
    \includegraphics[width=1.\linewidth]{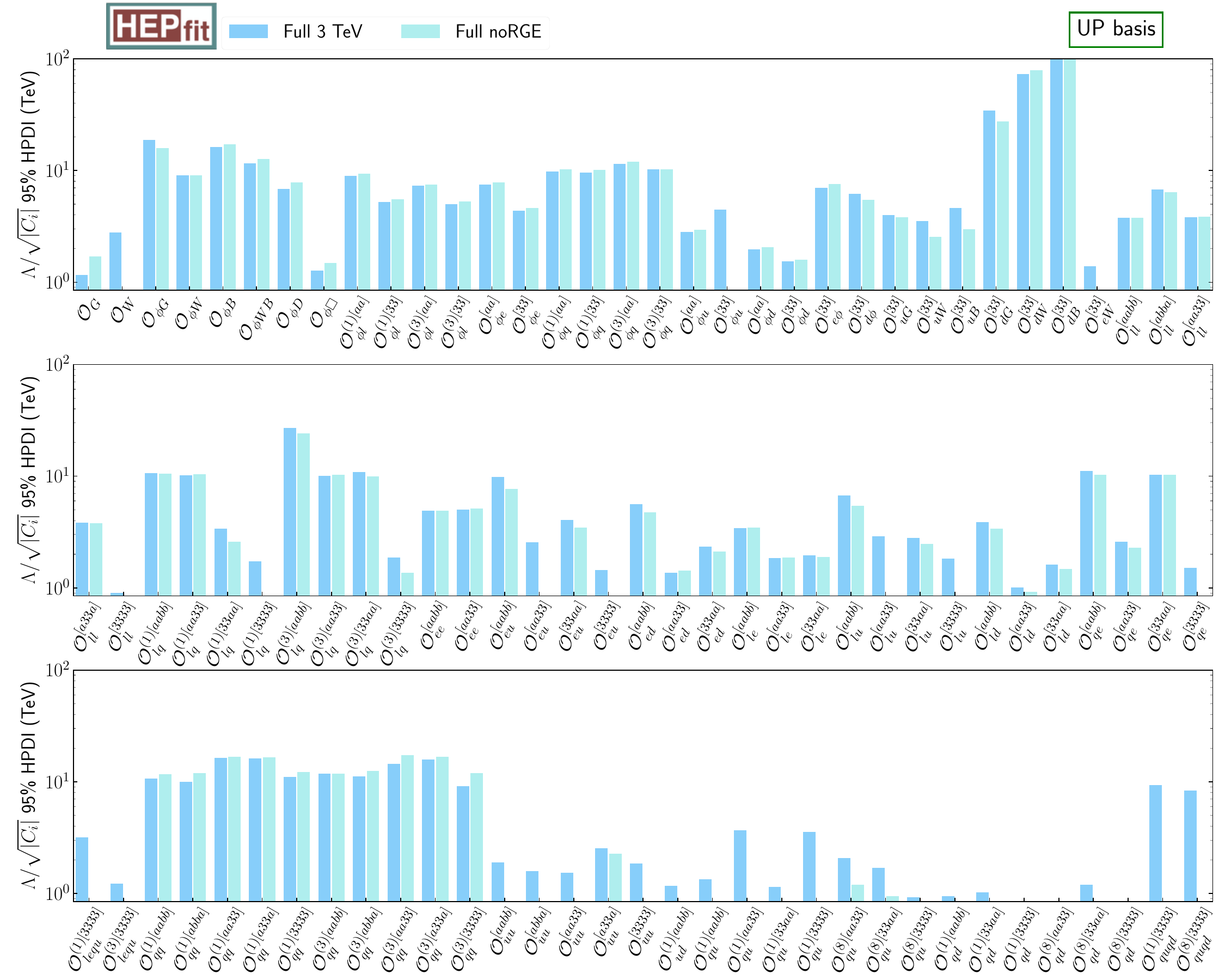}
    \caption{ Summary of individual lower bounds on the effective NP scale in the $U(2)^5$ flavour symmetric SMEFT obtained for $\Lambda = 3$ TeV in the UP basis. }
    \label{fig:bounds_summary_U2}
\end{figure}

Furthermore, as shown in Table~\ref{tab:u2-up-down} and Figure~\ref{fig:bounds_UP_vs_DOWN}, the choice of flavour basis has a pronounced impact on the mentioned $b$-quark dipole operators. Aligning to the DOWN basis, as expected, strongly suppresses contributions to $b$-quark flavour transitions, relaxing the bound on ${C}_{dB}^{[33]}$ by nearly two orders of magnitude. Even in this case, however, flavour observables continue to provide the most stringent constraints on these operators. This pattern extends, to varying degrees, to most of the operators shown in Figure~\ref{fig:bounds_UP_vs_DOWN}: bounds in the DOWN basis are systematically weaker due to flavour-space alignment and the precision of $B$-physics measurements.

\begin{figure}
    \centering
    \includegraphics[width=1.\linewidth]{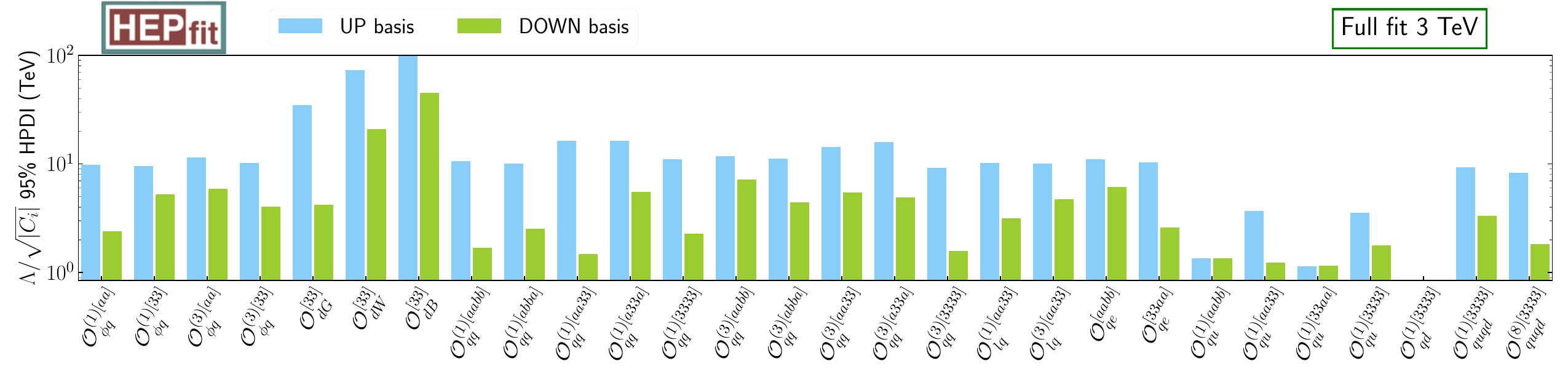}
    \caption{Comparison of the individual lower bounds on the effective NP scale in the $U(2)^5$ flavour symmetric SMEFT obtained for $\Lambda = 3$ TeV in the UP and DOWN basis. Only the operators constrained mostly from flavour observables are shown. The numerical values are also shown in Tab.~\ref{tab:u2-up-down}.
    }
    \label{fig:bounds_UP_vs_DOWN}
\end{figure}

We now discuss the results of the global fits for the $U(2)^5$ assumption, choosing as a representative example the results obtained with $\Lambda=10$ TeV in the UP basis shown in Table \ref{tab:u2-glob} and Figures~\ref{fig:full_fit_bos_2f_2fqLL_Lam_10} and \ref{fig:full_fit_4f_Lam_10}. We begin by highlighting the most important correlations between the different Wilson coefficients. Let us start with the coefficients mainly constrained by Higgs-boson and EW observables. For example, the coefficient $C_{\phi G}$ becomes degenerate with $C_{uG}^{[33]}$, with a correlation coefficient of $-0.91$. However, imposing the perturbativity prior on $C_{uG}^{[33]}$, one still gets a strong constraint on $C_{\phi G}$, with a $95\%$ HPDI interval of $[-0.30,0.85]$ corresponding to a bound on $\Lambda/\sqrt{\vert C_{\phi G}\vert }$ of $10.5$ TeV, to be compared with the bound of $19$ TeV obtained in the individual fit (see Table~\ref{tab:u2-ind}). A similar phenomenon occurs for other pairs of coefficients, such as $C_{e \phi}^{[33]}$ and $C_{lequ}^{(1)[3333]}$, with a correlation coefficient of $0.91$ and $C_{lequ}^{(1)[3333]}$ hitting the perturbativity limit; $C_{d \phi}^{[33]}$ and $C_{quqd}^{(1)[3333]}$, with a correlation coefficient of $-0.91$ and $C_{quqd}^{(1)[3333]}$ hitting the perturbativity limit; $C_{d B}^{[33]}$ and $C_{dW}^{[33]}$, with a correlation coefficient of $0.99$, and $C_{dW}^{[33]}$ hitting the perturbativity limit; $C_{ll}^{[aabb]}$ and $C_{ee}^{[aabb]}$, with a correlation coefficient of $-0.82$ and both coefficients hitting the perturbativity limit. Another pair of strongly correlated coefficients is given by $C_{\phi e}^{[33]}$ and $C_{\phi e}^{[aa]}$, with a correlation coefficient of $0.84$.
Concerning four fermion operators, one faces more complex interference patterns involving several coefficients. For example, the coefficient $C_{qq}^{(3)[aabb]}$ is strongly correlated with $C_{lq}^{(3)[aabb]}$ which is in turn strongly correlated with $C_{lq}^{(3)[33aa]}$, with correlation coefficients of $0.93$ and $0.85$, respectively. Therefore, a large set of four fermions operators cannot be constrained in the global fit with current data.

\begin{figure}[h]
    \centering
    \includegraphics[width=1.\linewidth]{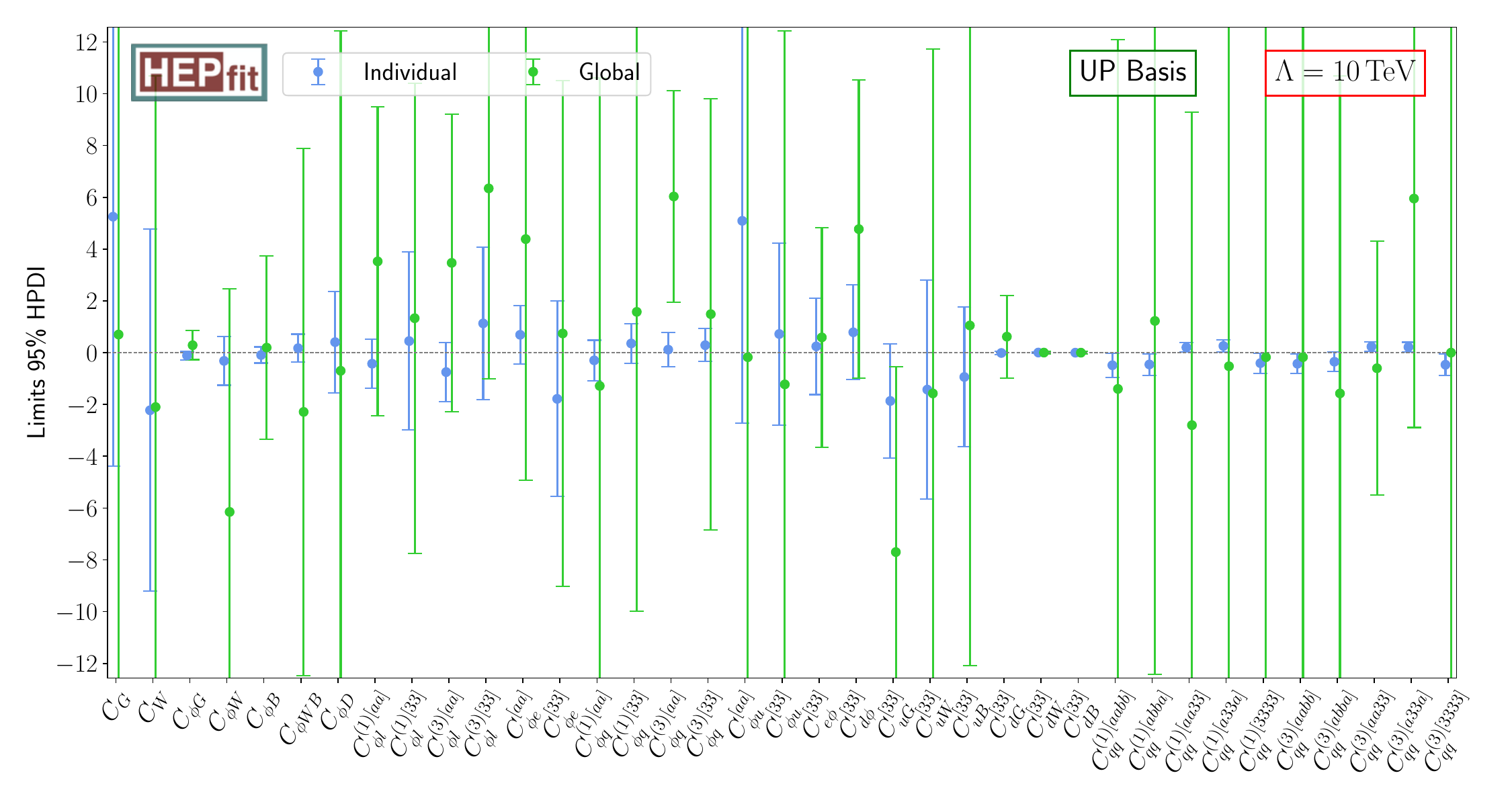}
    \caption{Comparison of individual and marginalised constraints from the global fit with $U(2)^5$ flavour symmetry assumption. The scale of NP is been set to $\Lambda=10$ TeV. The limits shown correspond to the 95\% HPDI. The limits are shown for the bosonic, two fermion, and $(\bar{L}L)(\bar{L}L)$ four quark operators. Only the operators that can be constrained at least in the individual fit are shown. }
    \label{fig:full_fit_bos_2f_2fqLL_Lam_10}
\end{figure}

\begin{figure}[h]
    \centering
    \includegraphics[width=1.\linewidth]{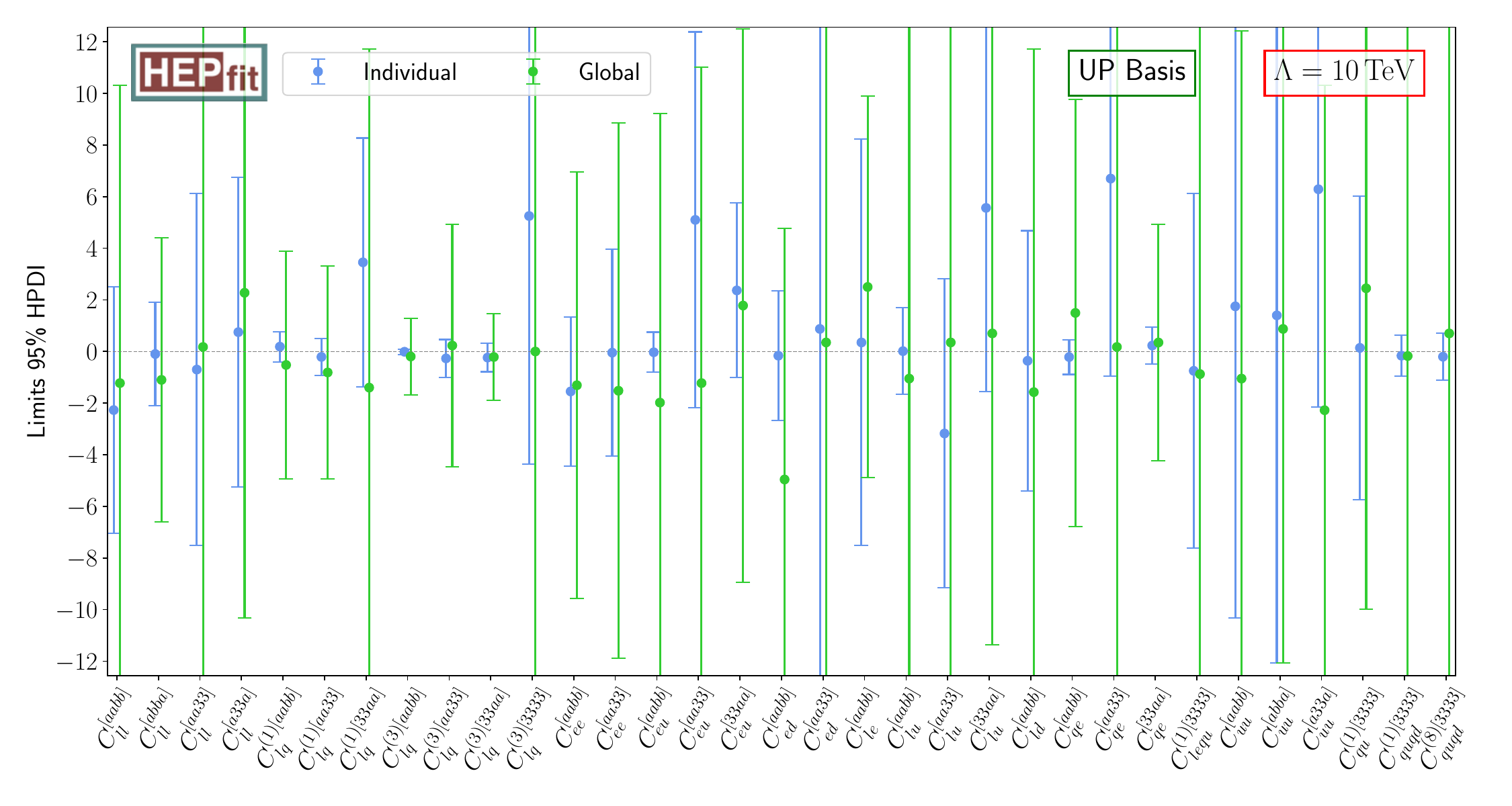}
    \caption{Same as Figure~\ref{fig:full_fit_bos_2f_2fqLL_Lam_10} but for the rest of four fermion operators.}
    \label{fig:full_fit_4f_Lam_10}
\end{figure}

A direct comparison between the global fit results in the $U(2)^5$ and $U(3)^5$ flavour-symmetric scenarios can be made by examining the last column of Tables~\ref{tab:u2-glob} and~\ref{tab:u3-glob}. We have already discussed how certain dipole interactions involving the bottom quark (${C}_{dB,dW,dG}^{[33]}$), which are absent under the $U(3)^5$ assumption, receive strong constraints from flavour observables in the $U(2)^5$ framework. On the other hand, when comparing the global bounds on operators such as ${C}_{\phi q}^{(1)}$ and ${C}_{\phi q}^{(3)}$, which modify the neutral and charged current couplings of left-handed quarks, we find that the $U(2)^5$ symmetry still offers significant protection against flavour constraints, yielding bounds comparable to those in the $U(3)^5$ scenario. However, this does not hold for individual fits. In particular, interactions involving the left-handed third-generation quarks are more strongly constrained by flavour observables, while those involving the light generations remain primarily constrained by electroweak data.\footnote{Interestingly, and as also pointed out in other studies, for the right-handed interactions with the top-quark, ${C}_{\phi u}^{[33]}$, the strongest individual bound comes from electroweak precision observables as a result of mixing between this operator and the custodial symmetry breaking operator
${C}_{\phi D}$~\cite{deBlas:2015aea} (see also~\cite{Stefanek:2024kds}).
A similar effect is induced by the left-handed counterpart ${C}_{\phi q}^{{(1})[33]}$, 
making electroweak bounds comparable to the flavour ones in this case (see Table~\ref{tab:u2-ind}).
}

Similar conclusions can be derived by looking at lepton-quark left-handed operators, 
where the separation of the third and first two families in the $U(2)^5$ also helps to identify the leading source of the bound in the $U(3)^5$-symmetric case. For both the ${C}_{lq}^{(1),(3)}$ operators, this comes from the entries involving only the first two quark families, where Drell-Yan at high energies still seems to be the leading constraint. Unfortunately, the lack of constraints for many of the interactions in the $U(2)^5$ global fit does not allow to derive further information from the comparison with the $U(3)^5$ case.

As a final point, we note from Figures~\ref{fig:full_fit_bos_2f_2fqLL_Lam_10} and~\ref{fig:full_fit_4f_Lam_10} that, similarly to the $U(3)^5$ case, the individual bounds are significantly weakened in the global case. Given the current precision of experimental measurements and theoretical predictions, constraining the most general $U(2)^5$-flavour-symmetric SMEFT becomes challenging for NP scales that ensure a reliable perturbative expansion, such as $\Lambda = 10$ TeV. Nevertheless, as previously noted, the presence of strong correlations between the coefficients does not prevent us from obtaining meaningful constraints on many of them when restricting to the perturbative regime.



\section{Conclusions}
\label{sec:conclusions}

The increasing accuracy reached in both experimental measurements and theoretical predictions for a large spectrum of observables over a broad range of scales allows to explore NP within the consistent and systematic framework of the SMEFT. This framework provides a powerful and model-independent approach to probe the reach of indirect NP searches at energies beyond the EW scale.

In this paper we have presented a comprehensive study aimed at
assessing our capability to indirectly probe NP beyond the EW scale via a state-of-the-art global analysis of the dimension-6 SMEFT that still assumes lepton and baryon number conservation and no additional CP-violation beyond the SM. We have considered a broad set of observables including EW, Drell-Yan, top-quark, Higgs-boson, and key flavour measurements adopted for the unitarity triangle analysis as well as FCNC channels sensitive to NP. We have fitted the SMEFT Wilson coefficients together with floating SM and hadronic parameters, by comparing the measurements of such observables to the corresponding predictions in the SMEFT at linear order and with LO RGE of the Wilson coefficients. We investigated two benchmark flavour symmetry scenarios: $U(3)^5$ and $U(2)^5$. Furthermore, in order to assess the extent to which current data constrain new physics at different energy scales, we have explored two representative benchmark values for the NP scale: $\Lambda = 3$, and $10$~TeV including full RGE, and also compared them with results obtained without RGE (using $\Lambda=1$~TeV). A summary of the individual bounds obtained in the case of $U(3)^5$ is given in Figure~\ref{fig:bounds_U3} and a corresponding  comparison of individual and marginalised constraints from the global fit can be found in Figure~\ref{fig:global_fit_U3}. The corresponding results for the case of $U(2)^5$ are illustrated in Figure~\ref{fig:bounds_summary_U2} and Figures~\ref{fig:full_fit_bos_2f_2fqLL_Lam_10}-\ref{fig:full_fit_4f_Lam_10}.

In the $U(3)^5$-symmetric case, the constraints on individual Wilson coefficients translate into lower bounds on the effective NP scale $\Lambda/\sqrt{|C_i|}$ reaching values up to $\sim 25$~TeV, particularly when RGE effects are included. We observe that the running of SMEFT operators plays a crucial role in tightening constraints for operators such as $\mathcal{O}_G$ and $\mathcal{O}_W$, which gain sensitivity through mixing into observables in the Higgs and EW sectors. In global fits -- where all relevant Wilson coefficients are varied simultaneously under the assumed flavour symmetry -- strong correlations among them tend to relax many of the individual bounds. For instance, large correlations arise among several bosonic and four-fermion operators. Nevertheless, meaningful bounds persist: $\Lambda/\sqrt{|C_{\phi G}|} \gtrsim 14$~TeV and $\Lambda/\sqrt{|C_{lq}^{(3)}|} \gtrsim 5$~TeV, while other coefficients yield bounds in the $2$--$3$~TeV range. These findings indicate that generic strongly-coupled $U(3)^5$-symmetric NP scenarios are pushed above the TeV scale, ensuring the internal consistency of the SMEFT expansion. Conversely, weakly-coupled UV completions may still reside near the TeV scale, though in such cases the SMEFT description must be applied with care, particularly for high-$p_T$ observables where EFT validity can break down.

In the $U(2)^5$ case, the orientation in flavour space becomes relevant, and the inclusion of flavour observables has a decisive impact on the analysis. In this study, we considered two complementary cases: one in which NP effects are introduced in the basis where the up-quark Yukawa matrix is diagonal -- the UP basis -- and another where NP effects are defined in the basis with diagonal down-quark Yukawas, the DOWN basis. 
In both cases, flavour misalignment stems from the left-handed quark sector, while the right-handed quarks can always be aligned with the direction of flavour-diagonal Yukawa couplings. In the lepton sector, the charged-lepton flavours can also always be rotated such that their Yukawa interactions are flavour diagonal.

Focusing on the UP basis, we find that several operators -- such as $C_{dB}^{[33]}$, $C_{dW}^{[33]}$, and $C_{dG}^{[33]}$ -- are strongly constrained by $B \to X_s \gamma$, with effective NP scales reaching several tens of TeV. Four-fermion operators, including various $C_{qq}^{(1),(3)}$ flavour configurations, are tightly constrained by meson mixing, and operators like $C_{qe}^{[33aa]}$ and $C_{qe}^{[aabb]}$ by rare decays such as $B_s \to \mu^+ \mu^-$. In global fits, the interplay of correlations once again leads to the saturation of perturbativity priors for certain coefficients; nevertheless, non-trivial bounds are retained for their correlated partners. For instance, the strong correlation between $C_{\phi G}$ and $C_{uG}^{[33]}$ still permits a meaningful constraint on $C_{\phi G}$, even when $C_{uG}^{[33]}$ approaches its prior limit.  A comprehensive picture of the individual lower bounds derived in the $U(2)^5$ case within the UP basis is summarised in Figure~\ref{fig:bounds_summary_U2} and a corresponding  comparison of individual and marginalised constraints from the global fit is highlighted in Figures~\ref{fig:full_fit_bos_2f_2fqLL_Lam_10} and~\ref{fig:full_fit_4f_Lam_10}. Finally, in Figure~\ref{fig:bounds_UP_vs_DOWN}, we compare the constraints on the set of operators that are sensitive to the choice of flavour basis -- UP versus DOWN -- for a reference NP scale of $\Lambda = 3$~TeV. The operators shown are those whose bounds are primarily driven by flavour observables and are therefore most affected by the basis-dependent treatment of third-generation quarks. 

Comparing the $U(3)^5$-symmetric case with the one of $U(2)^5$ in the UP basis, we note that several operators absent under $U(3)^5$ --- notably the dipole interactions involving the bottom quark, $\mathcal{O}_{dB}^{[33]}$, $\mathcal{O}_{dW}^{[33]}$, and $\mathcal{O}_{dG}^{[33]}$ --- receive strong constraints in the $U(2)^5$ scenario due to the inclusion of flavour observables. This highlights the increased sensitivity to third-generation-specific operators once the full flavour symmetry is relaxed.
On the other hand, for operators such as $\mathcal{O}_{\phi q}^{(1)}$ and $\mathcal{O}_{\phi q}^{(3)}$, which modify the neutral and charged currents of left-handed quarks, we find that the $U(2)^5$ symmetry still provides significant protection from flavour constraints. The resulting global bounds are comparable to those obtained in the $U(3)^5$ limit. However, the picture changes when examining individual coefficients: interactions involving third-generation left-handed quarks become more sensitive to flavour constraints, while those involving the light generations remain predominantly constrained by EW precision data.
A similar pattern emerges in the case of lepton-quark interactions, particularly for the $\mathcal{O}_{lq}^{(1)}$ and $\mathcal{O}_{lq}^{(3)}$ operators. In the $U(3)^5$ case, the leading bounds are driven by contributions involving only the first two quark generations, with high-energy Drell-Yan processes providing the dominant constraints. In the $U(2)^5$ scenario, the separation between the third and first two generations helps isolate the origin of these bounds. However, the limited sensitivity of the $U(2)^5$ global fit to many operators hinders a more granular comparison with the $U(3)^5$ scenario. This limitation underscores the need for an expanded set of precision observables --- particularly in the flavour sector --- as well as improved experimental measurements in relation to EW, Higgs and top-quark physics.

Finally, in Figure~\ref{fig:bounds_UP_vs_DOWN} we provide a direct inspection of flavour constraints in relation to the choice of flavour basis. For instance, in the DOWN basis, where the third generation is aligned with the down-quark Yukawa couplings, the contributions of dipole operators like $\mathcal{O}_{dB}^{[33]}$ to flavour-changing $b$-quark processes are suppressed, leading to significantly weaker bounds --- by nearly two orders of magnitude in some cases. Despite this suppression, flavour observables remain the primary source of constraints on these operators. This trend holds for most of the operators shown in Figure~\ref{fig:bounds_UP_vs_DOWN}, where bounds in the DOWN basis are systematically weaker than in the UP basis. Notable exceptions are the $\mathcal{O}_{\phi q}^{(1)}$ and $\mathcal{O}_{\phi q}^{(3)}$ operators, for which electroweak precision observables continue to impose strong and largely basis-independent limits.

In summary, this analysis represents a significant advance in quantifying the indirect sensitivity of current data to physics beyond the SM within the SMEFT framework. Overall, our results underscore the interplay and complementarity of different classes of observables in constraining simultaneously SM and SMEFT parameters, and highlight the importance of continued progress in both theoretical precision and experimental accuracy to probe NP up to multi-TeV scales.

To conclude, one of the key takeaways of this work is the crucial role played by flavour assumptions in the SMEFT framework. Relaxing the $U(3)^5$ ansatz to a $U(2)^5$ structure introduces a greater number of SMEFT Wilson coefficients, and we have shown to what extent this weakens the constraints achievable in a global analysis. Looking ahead, and with the prospect of more data from current and future collider and low-energy experiments, we aim to further expand the spectrum of observables and to explore scenarios with different flavour assumptions, such as minimal flavour violation.


\FloatBarrier

\subsection*{Acknowledgements}
V.M. is grateful to Marcel Vos for insightful discussions during their joint work. He also thanks Antonio Jaques Costa for valuable advice on several experimental measurements. We thank P. Stangl for useful comments on the manuscript. The work of J.B. has been partially funded by 
the FEDER/Junta de Andaluc\'ia project grant P18-FRJ-3735,  MICIU/AEI/\allowbreak10.13039/501100011033 and FEDER/UE (grant PID2022-139466NB-C21). The work of V.M. is supported by the MICIU through a Beatriz Galindo Junior grant (BG24/00038), the European Research Council (ERC) under the European Union’s Horizon 2020 research and innovation programme (Grant agreement No. 949451) and a Royal Society University Research Fellowship through grant URF/R1/201553.
The work of L.R. is supported in part by the U.S. Department of Energy under grant DE-SC0010102 and by
INFN through a Foreign Visiting Scientist Fellowship. This work was supported in part by the European Union - Next Generation
EU under italian MUR grant PRIN-2022-RXEZCJ. L.S. and M.V. acknowledge support from the project ``Exploring New Physics'' funded by INFN.

\newpage
\appendix

\section{Relations between SMEFT operators under different flavour assumptions}
\label{sec:app-SMEFT-flav}

In this appendix we present the defining relations of the $U(3)^5$ and $U(2)^5$ flavour-symmetric scenarios introduced in Section~\ref{sec:smeft}. 
Starting from the complete set of Wilson coefficients of the SMEFT, we specify the constraints imposed by
the assumptions of the NP respecting the two flavour symmetries at the scale where the UV theory is matched with the effective field theory.

\FloatBarrier

\subsection{\texorpdfstring{$U(3)^5$}{U3-5} flavour symmetric SMEFT}
\label{sec:app-SMEFT-U35}

In this scenario, NP is assumed to be flavour-blind. The set of SMEFT bosonic interactions is unaffected by this assumption. For the fermionic Wilson coefficients the following identities are satisfied, with the left-handed side of each expression referring to the general SMEFT case, and the right-handed side introduces the notation we follow for the $U(3)^5$-symmetric interactions:

\begin{align}
  \label{eq:u3}
  &C_{e\phi}^{[pr]} = C_{u\phi}^{[pr]} = C_{d\phi}^{[pr]} =
  C_{eX}^{[pr]} = C_{uX}^{[pr]} = C_{dX}^{[pr]} = C_{\phi ud}^{[pr]} =0\,, \\
  &C_{\phi l}^{(1),(3)[pr]} = C_{\phi l}^{(1),(3)} \delta_{pr}\,, \qquad
  C_{\phi q}^{(1),(3)[pr]} = C_{\phi q}^{(1),(3)} \delta_{pr}\,, \nonumber \\
  &C_{\phi e}^{[pr]} = C_{\phi e} \delta_{pr}\,, \qquad
  C_{\phi u}^{[pr]} = C_{\phi u} \delta_{pr}\,, \qquad
  C_{\phi d}^{[pr]} = C_{\phi d} \delta_{pr}\,, \nonumber \\
  &C_{ll}^{[prst]} = C_{ll}^{[aabb]} \delta_{pr} \delta_{st} + C_{ll}^{[abba]} \delta_{pt} \delta_{rs}\,, \nonumber \\
  &C_{qq}^{(1),(3)[prst]} = C_{qq}^{(1),(3)[aabb]} \delta_{pr} \delta_{st} + C_{qq}^{(1),(3)[abba]} \delta_{pt} \delta_{rs}\,, \nonumber \\
  &C_{lq}^{(1),(3)[prst]} = C_{lq}^{(1),(3)} \delta_{pr} \delta_{st} \,, \qquad
  C_{ee}^{[prst]} = C_{ee} \delta_{pr} \delta_{st}\,, \nonumber \\
  &C_{uu}^{[prst]} = C_{uu}^{[aabb]} \delta_{pr} \delta_{st} + C_{uu}^{[abba]} \delta_{pt} \delta_{rs}\,, \nonumber \\
  &C_{dd}^{[prst]} = C_{dd}^{[aabb]} \delta_{pr} \delta_{st} + C_{dd}^{[abba]} \delta_{pt} \delta_{rs}\,, \nonumber \\
  &C_{eu}^{[prst]} = C_{eu} \delta_{pr} \delta_{st}\,, \qquad
  C_{ed}^{[prst]} = C_{ed} \delta_{pr} \delta_{st}\,, \qquad
  C_{ud}^{(1),(8)[prst]} = C_{ud}^{(1),(8)} \delta_{pr} \delta_{st}\,, \nonumber \\
  &C_{le}^{[prst]} = C_{le} \delta_{pr} \delta_{st}\,, \qquad
  C_{lu}^{[prst]} = C_{lu} \delta_{pr} \delta_{st}\,, \qquad 
  C_{ld}^{[prst]} = C_{ld} \delta_{pr} \delta_{st}\,, \nonumber \\
  &C_{qe}^{[prst]} = C_{qe} \delta_{pr} \delta_{st}\,, \qquad
  C_{qu}^{(1),(8)[prst]} = C_{qu}^{(1),(8)} \delta_{pr} \delta_{st}\,, \qquad
  C_{qd}^{(1),(8)[prst]} = C_{qd}^{(1),(8)} \delta_{pr} \delta_{st}\,,\nonumber \\
  &C_{quqd}^{(1),(8)[prst]} = C_{lequ}^{(1),(3)[prst]} = C_{ledq}^{[prst]} = 0 \,. \nonumber
\end{align}

\vspace{1cm}

\subsection{\texorpdfstring{$U(2)^5$}{U2-5} flavour symmetric SMEFT}
\label{sec:app-SMEFT-U25}

In the $U(2)^5$-symmetric scenario there are a total of 124 independent Wilson coefficients. 
The relations between the general SMEFT fermionic Wilson coefficients and the $U(2)^5$-symmetric ones is given below in eq.~(\ref{eq:U25relations}). 
As in the previous case, 
the right-handed side of the expressions sets the notation we employ to refer to the independent degrees of freedom in this
scenario. 
Note that, as emphasised in Section~\ref{sec:smeft}, this $U(2)^5$ limit is only completely determined upon specifying the direction in flavour space that defines the third family.

\begin{align}
  C_{\psi_R\phi}^{[pr]} &= C_{\psi_R\phi}^{[33]} \delta_{p3} \delta_{r3} \qfor \psi_R={e,u,d}\,\nonumber, \\
  C_{\psi_R X}^{[pr]} &= C_{\psi_R X}^{[33]} \delta_{p3} \delta_{r3} \qfor \psi_R={e,u,d}\,, \nonumber \\
  C_{\phi \psi_L}^{(1),(3)[pr]} &= C_{\phi \psi_L}^{(1),(3)[33]} \delta_{p3} \delta_{r3} + C_{\phi \psi_L}^{(1),(3)[aa]} \delta_{pr} \qty(\delta_{p1}+\delta_{p2}) \qfor \psi_L={l,q}\,, \nonumber \\
  C_{\phi \psi_R}^{[pr]} &= C_{\phi \psi_R}^{[33]} \delta_{p3} \delta_{r3} + C_{\phi \psi_R}^{[aa]} \delta_{pr} \qty(\delta_{p1}+\delta_{p2}) \qfor \psi_R={e,u,d}\,, \nonumber \\
  C_{\phi u d}^{[pr]} &= C_{\phi u d}^{[33]} \delta_{p3} \delta_{r3} \,, \nonumber \\
  C_{ll}^{[prst]} &= C_{ll}^{[3333]} \delta_{p3} \delta_{r3} \delta_{s3} \delta_{t3} + 
    C_{ll}^{[aa33]} \delta_{pr} \qty(\delta_{p1}+\delta_{p2}) \delta_{s3} \delta_{t3} + 
    C_{ll}^{[a33a]} \delta_{pt} \qty(\delta_{p1}+\delta_{p2}) \delta_{r3} \delta_{s3} + \nonumber \\
  & C_{ll}^{[aabb]} \delta_{pr} \delta_{st} \qty(\delta_{p1}+\delta_{p2}) \qty(\delta_{s1}+\delta_{s2})  + 
    C_{ll}^{[abba]} \delta_{pt} \delta_{rs} \qty(\delta_{p1}+\delta_{p2}) \qty(\delta_{s1}+\delta_{s2})\,, \nonumber \\
  C_{qq}^{(1),(3)[prst]} &= C_{qq}^{(1),(3)[3333]} \delta_{p3} \delta_{r3} \delta_{s3} \delta_{t3} + 
    C_{qq}^{(1),(3)[aa33]} \delta_{pr} \qty(\delta_{p1}+\delta_{p2}) \delta_{s3} \delta_{t3} + \nonumber \\
    &  C_{qq}^{(1),(3)[a33a]} \delta_{pt} \qty(\delta_{p1}+\delta_{p2}) \delta_{r3} \delta_{s3} + 
    C_{qq}^{(1),(3)[aabb]} \delta_{pr} \delta_{st} \qty(\delta_{p1}+\delta_{p2}) \qty(\delta_{s1}+\delta_{s2})  +\nonumber \\
    &  C_{qq}^{(1),(3)[abba]} \delta_{pt} \delta_{rs} \qty(\delta_{p1}+\delta_{p2}) \qty(\delta_{s1}+\delta_{s2})\,, \nonumber \\
  C_{lq}^{(1),(3)[prst]} &= C_{lq}^{(1),(3)[3333]} \delta_{p3} \delta_{r3} \delta_{s3} \delta_{t3} +
    C_{lq}^{(1),(3)[aa33]} \delta_{pr} \qty(\delta_{p1}+\delta_{p2}) \delta_{s3} \delta_{t3} + \nonumber \\
    & C_{lq}^{(1),(3)[33aa]} \delta_{st} \qty(\delta_{s1}+\delta_{s2}) \delta_{p3} \delta_{r3} + 
    C_{lq}^{(1),(3)[aabb]} \delta_{pr} \delta_{st} \qty(\delta_{p1}+\delta_{p2}) \qty(\delta_{s1}+\delta_{s2})\,, \nonumber \\
  C_{ee}^{[prst]} &= C_{ee}^{[3333]} \delta_{p3} \delta_{r3} \delta_{s3} \delta_{t3} +
    C_{ee}^{[aa33]} \delta_{pr} \qty(\delta_{p1}+\delta_{p2}) \delta_{s3} \delta_{t3} + \nonumber \\
    & C_{ee}^{[aabb]} \delta_{pr} \delta_{st} \qty(\delta_{p1}+\delta_{p2}) \qty(\delta_{s1}+\delta_{s2})\,, \nonumber \\
    C_{uu}^{[prst]} &= C_{uu}^{[3333]} \delta_{p3} \delta_{r3} \delta_{s3} \delta_{t3} +
    C_{uu}^{[aa33]} \delta_{pr} \qty(\delta_{p1}+\delta_{p2}) \delta_{s3} \delta_{t3} + \nonumber \\
    & C_{uu}^{[a33a]} \delta_{pt} \qty(\delta_{p1}+\delta_{p2}) \delta_{r3} \delta_{s3} +
    C_{uu}^{[aabb]} \delta_{pr} \delta_{st} \qty(\delta_{p1}+\delta_{p2}) \qty(\delta_{s1}+\delta_{s2}) + \nonumber \\
    &C_{uu}^{[abba]} \delta_{pt} \delta_{rs} \qty(\delta_{p1}+\delta_{p2}) \qty(\delta_{s1}+\delta_{s2}) \,, \nonumber \\
  C_{dd}^{[prst]} &= C_{dd}^{[3333]} \delta_{p3} \delta_{r3} \delta_{s3} \delta_{t3} +
    C_{dd}^{[aa33]} \delta_{pr} \qty(\delta_{p1}+\delta_{p2}) \delta_{s3} \delta_{t3} + \nonumber \\
    & C_{dd}^{[a33a]} \delta_{pt} \qty(\delta_{p1}+\delta_{p2}) \delta_{r3} \delta_{s3} +
    C_{dd}^{[aabb]} \delta_{pr} \delta_{st} \qty(\delta_{p1}+\delta_{p2}) \qty(\delta_{s1}+\delta_{s2}) + \nonumber \\
    &C_{dd}^{[abba]} \delta_{pt} \delta_{rs} \qty(\delta_{p1}+\delta_{p2}) \qty(\delta_{s1}+\delta_{s2}) \,, \nonumber \\
    C_{eu}^{[prst]} &= C_{eu}^{[3333]} \delta_{p3} \delta_{r3} \delta_{s3} \delta_{t3} +
    C_{eu}^{[aa33]} \delta_{pr} \qty(\delta_{p1}+\delta_{p2}) \delta_{s3} \delta_{t3} + \label{eq:U25relations}\\
    & C_{eu}^{[33aa]} \delta_{st} \qty(\delta_{s1}+\delta_{s2}) \delta_{p3} \delta_{r3} + 
    C_{eu}^{[aabb]} \delta_{pr} \delta_{st} \qty(\delta_{p1}+\delta_{p2}) \qty(\delta_{s1}+\delta_{s2})\,, \nonumber \\
    C_{ed}^{[prst]} &= C_{ed}^{[3333]} \delta_{p3} \delta_{r3} \delta_{s3} \delta_{t3} +
    C_{ed}^{[aa33]} \delta_{pr} \qty(\delta_{p1}+\delta_{p2}) \delta_{s3} \delta_{t3} + \nonumber \\
    & C_{ed}^{[33aa]} \delta_{st} \qty(\delta_{s1}+\delta_{s2}) \delta_{p3} \delta_{r3} +
    C_{ed}^{[aabb]} \delta_{pr} \delta_{st} \qty(\delta_{p1}+\delta_{p2}) \qty(\delta_{s1}+\delta_{s2})\,, \nonumber \\
    C_{ud}^{(1),(8)[prst]} &= C_{ud}^{(1),(8)[3333]} \delta_{p3} \delta_{r3} \delta_{s3} \delta_{t3} +
    C_{ud}^{(1),(8)[aa33]} \delta_{pr} \qty(\delta_{p1}+\delta_{p2}) \delta_{s3} \delta_{t3} + \nonumber \\
    & C_{ud}^{(1),(8)[33aa]} \delta_{st} \qty(\delta_{s1}+\delta_{s2}) \delta_{p3} \delta_{r3} +
    C_{ud}^{(1),(8)[aabb]} \delta_{pr} \delta_{st} \qty(\delta_{p1}+\delta_{p2}) \qty(\delta_{s1}+\delta_{s2})\,, \nonumber \\
    C_{le}^{[prst]} &= C_{le}^{[3333]} \delta_{p3} \delta_{r3} \delta_{s3} \delta_{t3} +
    C_{le}^{[aa33]} \delta_{pr} \qty(\delta_{p1}+\delta_{p2}) \delta_{s3} \delta_{t3} + \nonumber \\
    & C_{le}^{[33aa]} \delta_{p3} \delta{r3} \qty(\delta_{s1}+\delta_{s2}) \delta_{st} + 
    C_{le}^{[aabb]} \delta_{pr} \delta_{st} \qty(\delta_{p1}+\delta_{p2}) \qty(\delta_{s1}+\delta_{s2})\,, \nonumber \\
    C_{lu}^{[prst]} &= C_{lu}^{[3333]} \delta_{p3} \delta_{r3} \delta_{s3} \delta_{t3} +
    C_{lu}^{[aa33]} \delta_{pr} \qty(\delta_{p1}+\delta_{p2}) \delta_{s3} \delta_{t3} + \nonumber \\
    & C_{lu}^{[33aa]} \delta_{p3} \delta{r3} \qty(\delta_{s1}+\delta_{s2}) \delta_{st} + 
    C_{lu}^{[aabb]} \delta_{pr} \delta_{st} \qty(\delta_{p1}+\delta_{p2}) \qty(\delta_{s1}+\delta_{s2})\,, \nonumber \\
    C_{ld}^{[prst]} &= C_{ld}^{[3333]} \delta_{p3} \delta_{r3} \delta_{s3} \delta_{t3} +
    C_{ld}^{[aa33]} \delta_{pr} \qty(\delta_{p1}+\delta_{p2}) \delta_{s3} \delta_{t3} + \nonumber \\
    & C_{ld}^{[33aa]} \delta_{p3} \delta{r3} \qty(\delta_{s1}+\delta_{s2}) \delta_{st} + 
    C_{ld}^{[aabb]} \delta_{pr} \delta_{st} \qty(\delta_{p1}+\delta_{p2}) \qty(\delta_{s1}+\delta_{s2})\,, \nonumber \\
    C_{qe}^{[prst]} &= C_{qe}^{[3333]} \delta_{p3} \delta_{r3} \delta_{s3} \delta_{t3} +
    C_{qe}^{[aa33]} \delta_{pr} \qty(\delta_{p1}+\delta_{p2}) \delta_{s3} \delta_{t3} + \nonumber \\
    & C_{qe}^{[33aa]} \delta_{p3} \delta{r3} \qty(\delta_{s1}+\delta_{s2}) \delta_{st} + 
    C_{qe}^{[aabb]} \delta_{pr} \delta_{st} \qty(\delta_{p1}+\delta_{p2}) \qty(\delta_{s1}+\delta_{s2})\,, \nonumber \\
    C_{qu}^{(1),(8)[prst]} &= C_{qu}^{(1),(8)[3333]} \delta_{p3} \delta_{r3} \delta_{s3} \delta_{t3} +
    C_{qu}^{(1),(8)[aa33]} \delta_{pr} \qty(\delta_{p1}+\delta_{p2}) \delta_{s3} \delta_{t3} + \nonumber \\
    & C_{qu}^{(1),(8)[33aa]} \delta_{st} \qty(\delta_{s1}+\delta_{s2}) \delta_{p3} \delta_{r3} +
    C_{qu}^{(1),(8)[aabb]} \delta_{pr} \delta_{st} \qty(\delta_{p1}+\delta_{p2}) \qty(\delta_{s1}+\delta_{s2})\,, \nonumber \\
    C_{qd}^{(1),(8)[prst]} &= C_{qd}^{(1),(8)[3333]} \delta_{p3} \delta_{r3} \delta_{s3} \delta_{t3} +
    C_{qd}^{(1),(8)[aa33]} \delta_{pr} \qty(\delta_{p1}+\delta_{p2}) \delta_{s3} \delta_{t3} + \nonumber \\
    & C_{qd}^{(1),(8)[33aa]} \delta_{st} \qty(\delta_{s1}+\delta_{s2}) \delta_{p3} \delta_{r3} +
    C_{qd}^{(1),(8)[aabb]} \delta_{pr} \delta_{st} \qty(\delta_{p1}+\delta_{p2}) \qty(\delta_{s1}+\delta_{s2})\,, \nonumber\\
    C_{quqd}^{(1),(8)[prst]} &= C_{quqd}^{(1),(8)[3333]} \delta_{p3} \delta_{r3} \delta_{s3} \delta_{t3}\,, \nonumber\\
    C_{lequ}^{(1),(3)[prst]} &= C_{lequ}^{(1),(3)[3333]} \delta_{p3} \delta_{r3} \delta_{s3} \delta_{t3}\,, \nonumber\\
    C_{ledq}^{[prst]} &= C_{ledq}^{[3333]} \delta_{p3} \delta_{r3} \delta_{s3} \delta_{t3} \,. \nonumber
\end{align}

\section{Tables of results}
\label{sec:app-results}

In this appendix we collect a series of tables with the numerical results presented in the figures discussed in the main text, as well as some supplementary results. We report both the $68\%$ HPDI interval for the Wilson coefficients at the different scales considered in the study, as well as the bound on $\Lambda/\sqrt{\vert C_i\vert }$ in units of TeV obtained from the maximum of the $95\%$ HPDI interval for $\vert C_i \vert$. As indicated in the main text, in some cases the $68\%$ or $95\%$ HPDI (from which the bound on $\Lambda/\sqrt{\vert C_i\vert }$ is obtained) touches the edges of the prior dictated by perturbativity. These cases are indicated in the tables in red colour. On the other hand, in those cases where the posterior distribution of a given coefficient does not allow to set a bound, we indicate that using a dash.


\FloatBarrier

\subsection{Results for the \texorpdfstring{$U(3)^5$}{U3-5} flavour symmetric SMEFT}

\begin{tiny}
\begin{longtable}{|l|c|c|c|c|c|c|c|c|c|} 
\caption[]{Results of the individual fits for the $U(3)^5$ flavour symmetric SMEFT. For each coefficient, we report the $68\%$ HPDI interval, the bound on $\Lambda/\sqrt{\vert C_i\vert }$ in units of TeV obtained taking the maximum of the $95\%$ HPDI interval for $\vert C_i \vert$, and the set of observables that give the strongest constraint (EW stands for Electroweak, F for Flavour, H for Higgs, T for top, and DY for Drell-Yan). Results are shown with RG evolution for $\Lambda = 10$ TeV, for $\Lambda = 3$ TeV and without RG evolution. A dash denotes that no bound can be set from that particular fit. The meaning of the colour code is discussed in the text.
\label{tab:u3-ind}  }\\
\hline
 & \multicolumn{3}{c|}{10 TeV}  & \multicolumn{3}{c|}{3 TeV}  & \multicolumn{3}{c|}{1 TeV - noRGE }  \\ \hline 
& $C_i$ & $\Lambda/\sqrt{|C_i|}$ & Obs & $C_i$ & $\Lambda/\sqrt{|C_i|}$ & Obs & $C_i$ & $\Lambda/\sqrt{|C_i|}$ & Obs \\ 
\hline \endhead 
 \hline \endlastfoot 
$C_{\phi  \Box}$                & --                                    & --                     & --            & $ -0.7 \pm 3.0 $         & $ 1.2 $                  & H         & $ -0.02 \pm 0.29 $        & $ 1.3 $       &  H    \\ 
$C_{\phi D}$                    & $ 0.3 \pm 1.1 $                       & $ 6.4 $                & EW            & $ 0.040 \pm 0.090 $      & $ 6.5 $                  & EW        & $ 0.0024 \pm 0.0072 $     & $ 7.7 $       & EW    \\ 
$C_{\phi G}$                    & $ -0.112 \pm 0.084 $                  & $ 19.0 $               & H             & $ -0.0102 \pm 0.0084 $   & $ 18.5 $                 & H         & $ -0.0017 \pm 0.0012 $    & $ 15.8 $      & H  \\ 
$C_{\phi B}$                    & $ -0.10 \pm 0.17 $                    & $ 15.4 $               & H             & $ -0.008 \pm 0.015 $     & $ 15.6 $                 & H         & $ -0.0008 \pm 0.0014 $    & $ 17.0 $      &  H \\ 
$C_{\phi W}$                    & $ -0.33 \pm 0.52 $                    & $ 8.7 $                & H             & $ -0.028 \pm 0.047 $     & $ 8.8 $                  & H         & $ -0.0029 \pm 0.0051 $    & $ 8.8 $       &  H \\ 
$C_{\phi WB}$                   & $ 0.19 \pm 0.29 $                     & $ 11.3 $               & H             & $ 0.018 \pm 0.025 $      & $ 11.6 $                 & H         & $ 0.0017 \pm 0.0025 $     & $ 12.2 $      &  H \\ 
$C_{G}$                         & $\textcolor{red}{10.3 \pm 4.7}$       &  \textcolor{red}{--}   & H/T           & $ 3.5 \pm 1.8 $          & $ 1.1 $                  & T         & $ 0.07 \pm 0.15 $         & $ 1.7 $       &  T \\ 
$C_{W}$                         & $ -2.2 \pm 3.8 $                      & $ 3.2 $                & H             & $ -0.25 \pm 0.43 $       & $ 2.9 $                  & H         & $ -0.50 \pm 0.70 $       & $ 0.7 $       & EW    \\ 
$C_{\phi l}^{(1)}$               & $ -0.33 \pm 0.45 $                   & $ 9.1 $               & EW            & $ -0.030 \pm 0.039 $     & $ 9.2 $                  & EW        & $ -0.0029 \pm 0.0039 $   & $ 9.8 $       & EW    \\ 
$C_{\phi l}^{(3)}$               & $ -0.67 \pm 0.64 $                   & $ 7.3 $               & EW            & $ -0.058 \pm 0.056 $     & $ 7.3 $                  & EW        & $ -0.0061 \pm 0.0058 $   & $ 7.6 $       & EW    \\ 
$C_{\phi e}$                     & $ 0.62 \pm 0.59 $                    & $ 7.6 $               & EW            & $ 0.052 \pm 0.051 $      & $ 7.7 $                  & EW        & $ 0.0053 \pm 0.0053 $    & $ 8.2 $       & EW    \\ 
$C_{\phi q}^{(1)}$               & $ 0.8 \pm 1.4 $                      & $ 5.4 $               & EW            & $ 0.10 \pm 0.13 $        & $ 5.0 $                  & EW        & $ 0.022 \pm 0.017 $      & $ 4.2 $       & EW    \\ 
$C_{\phi q}^{(3)}$               & $ 1.01 \pm 0.58 $                    & $ 6.7 $               & EW            & $ 0.086 \pm 0.052 $      & $ 6.8 $                  & EW        & $ 0.0090 \pm 0.0058 $    & $ 7.0 $       & EW    \\ 
$C_{\phi u}$                     & $ 1.3 \pm 1.7 $                      & $ 4.7 $               & EW            & $ 0.13 \pm 0.18 $        & $ 4.3 $                  & EW        & $ 0.033 \pm 0.035 $      & $ 3.1 $       & EW    \\ 
$C_{\phi d}$                     & $ \textcolor{red}{-10.3 \pm 4.4} $   & \textcolor{red}{--}   & EW            & $ -0.83 \pm 0.51 $       & $ 2.2 $                  & EW        & $ -0.087 \pm 0.051 $     & $ 2.3 $       & EW    \\ 
$C_{ll}^{[aabb]}$                & $ -1.7 \pm 2.1 $                     & $ 4.2 $               & EW            & $ -0.15 \pm 0.19 $       & $ 4.1 $                  & EW        & $ -0.019 \pm 0.021 $     & $ 4.0 $       & EW    \\ 
$C_{ll}^{[abba]}$                & $ -0.01 \pm 0.96 $                   & $ 7.3 $               & EW            & $ -0.005 \pm 0.091 $     & $ 6.9 $                  & EW        & $ -0.004 \pm 0.010 $     & $ 6.3 $       & EW    \\ 
$C_{qq}^{(1)[aabb]}$             & $ -5.8 \pm 4.6 $                     & $ 2.7 $               & EW            & $ -0.62 \pm 0.54 $       & $ 2.3 $                  & EW        & $ -1.00 \pm 0.55 $       & $ 0.7 $       & T     \\ 
$C_{qq}^{(1)[abba]}$             & $ 0.3 \pm 2.2 $                      & $ 4.7 $               & F/DY          & $ 0.09 \pm 0.23 $        & $ 4.0 $                  & F/DY      & $ 0.005 \pm 0.075 $      & $ 2.5 $       & T     \\ 
$C_{qq}^{(3)[aabb]}$             & $ 0.21 \pm 0.40 $                    & $ 9.9 $               & DY            & $ 0.034 \pm 0.050 $      & $ 8.3 $                  & DY        & $ -0.005 \pm 0.029 $     & $ 4.1 $       & T     \\ 
$C_{qq}^{(3)[abba]}$             & $ 2.0 \pm 1.4 $                      & $ 4.5 $               & DY/T          & $ 0.25 \pm 0.16 $        & $ 4.0 $                  & T/DY      & $ 0.031 \pm 0.034 $      & $ 3.3 $       & T     \\ 
$C_{lq}^{(1)}$                   & $ 0.23 \pm 0.51 $                    & $ 8.8 $               & DY            & $ 0.055 \pm 0.054 $      & $ 7.4 $                  & DY        & $ 0.007 \pm 0.014 $      & $ 5.3 $       & DY    \\ 
$C_{lq}^{(3)}$                   & $ -0.028 \pm 0.058 $                 & $ 26.7 $              & DY            & $ -0.0035 \pm 0.0055 $   & $ 24.9 $                 & DY        & $ -0.00034 \pm 0.00087 $ & $ 21.9 $      & DY    \\ 
$C_{ee}$                         & $ -1.0 \pm 1.3 $                     & $ 5.3 $               & EW            & $ -0.10 \pm 0.12 $       & $ 5.3 $                  & EW        & $ -0.011 \pm 0.012 $     & $ 5.4 $       & EW    \\
$C_{uu}^{[aabb]}$                & $ 2.2 \pm 4.4 $                      & $ 3.1 $               & DY            & $ 0.47 \pm 0.56 $        & $ 2.4 $                  & DY        & $ -0.87 \pm 0.65 $       & $ 0.7 $       & T     \\ 
$C_{uu}^{[abba]}$                & $ 4.2 \pm 4.1 $                      & $ 2.8 $               & T             & $ 0.51 \pm 0.40 $        & $ 2.6 $                  & T         & $ -0.032 \pm 0.092 $     & $ 2.1 $       & T     \\ 
$C_{dd}^{[aabb]}$                & --                                   & --                    &               & $ 2.4 \pm 3.7 $          & $ 1.0 $                  & DY        & --                        & --            & --    \\ 
$C_{dd}^{[abba]}$                & --                                   & --                    &               & $ 3.9 \pm 6.0 $          & \textcolor{red}{--}      & DY        & --                        & --            & --    \\ 
$C_{eu}$                         & $ 0.17 \pm 0.41 $                    & $ 10.2 $              & DY            & $ 0.036 \pm 0.038 $      & $ 9.0 $                  & DY        & $ 0.0046 \pm 0.0076 $    & $ 7.3 $       & DY    \\ 
$C_{ed}$                         & $ -0.5 \pm 1.3 $                     & $ 5.7 $               & DY            & $ -0.10 \pm 0.12 $       & $ 5.2 $                  & DY        & $ -0.010 \pm 0.020 $     & $ 4.6 $       & DY    \\ 
$C_{ud}^{(1)}$                   & $ \textcolor{red}{-8.2 \pm 6.8} $    & \textcolor{red}{--}   & EW            & $ -2.4 \pm 1.6 $         & $ 1.3 $                  & DY        & $ -1.3 \pm 2.0 $         & $ 0.4 $       & T     \\ 
$C_{ud}^{(8)}$                   & $ \textcolor{red}{6.3 \pm 8.8} $     & \textcolor{red}{--}   & T             & $ 5.5 \pm 4.1 $          & $ 0.8 $                  & T         & $ 0.05 \pm 0.70 $        & $ 0.8 $       & T     \\
$C_{le}$                         & $ -1.0 \pm 3.6 $                     & $ 3.5 $               & EW            & $ -0.11 \pm 0.31 $       & $ 3.5 $                  & EW        & $ -0.014 \pm 0.035 $     & $ 3.5 $       & EW    \\ 
$C_{lu}$                         & $ 0.12 \pm 0.85 $                    & $ 7.4 $               & DY            & $ 0.050 \pm 0.084 $      & $ 6.4 $                  & DY        & $ 0.008 \pm 0.015 $      & $ 5.2 $       & DY    \\ 
$C_{ld}$                         & $ -1.3 \pm 2.7 $                     & $ 3.9 $               & DY            & $ -0.21 \pm 0.24 $       & $ 3.7 $                  & DY        & $ -0.020 \pm 0.039 $     & $ 3.3 $       & DY    \\ 
$C_{qe}$                         & $ 0.0 \pm 1.0 $                      & $ 7.0 $               & DY            & $ 0.057 \pm 0.095 $      & $ 6.2 $                  & DY        & $ 0.008 \pm 0.017 $      & $ 4.8 $       & T     \\
$C_{qu}^{(1)}$                   & $ 1.1 \pm 6.3 $                      & $ 2.7 $               & DY/EW         & $ 0.40 \pm 0.80 $        & $ 2.1 $                  & DY/EW     & $ 0.5 \pm 1.3 $          & $ 0.6 $       & T     \\ 
$C_{qu}^{(8)}$                   & $ -2.6 \pm 5.4 $                     & $ 2.7 $               & T             & $ -0.27 \pm 0.53 $       & $ 2.6 $                  & T         & $ 0.15 \pm 0.14 $        & $ 1.5 $       & T     \\ 
$C_{qd}^{(1)}$                   & $ \textcolor{red}{5.5 \pm 9.5} $     & \textcolor{red}{--}   & EW            & $ -0.1 \pm 2.3 $         & $ 1.4 $                  & DY/EW     & $ -1.5 \pm 2.1 $         & $ 0.4 $       & T     \\ 
$C_{qd}^{(8)}$                   & $ \textcolor{red}{-6.8 \pm 8.2} $    & \textcolor{red}{--}   & T             & $ -4.8 \pm 3.5 $         & $ 0.9 $                  & T         & $ 0.28 \pm 0.88 $        & $ 0.7 $       & T

\end{longtable}
\end{tiny}

\begin{scriptsize}
\begin{longtable}{|l|c|c|c|c|c|c|c|c|}
\caption[]{ Results of the individual fits compared to those from the global fit for the $U(3)^5$ flavour symmetric SMEFT. For each coefficient, we report the $68\%$ HPDI interval and the bound on $\Lambda/\sqrt{\vert C_i\vert }$ in units of TeV obtained taking the maximum of the $95\%$ HPDI interval for $\vert C_i \vert$. Results are shown with RG evolution for $\Lambda = 3$ TeV and $\Lambda = 10$ TeV. \label{tab:u3-glob}  }\\
\hline
 & \multicolumn{2}{c|}{Individual 3 TeV}  & \multicolumn{2}{c|}{Global 3 TeV}  & \multicolumn{2}{c|}{Individual 10 TeV}  & \multicolumn{2}{c|}{Global 10 TeV}  \\ \hline 
& $C_i$ & $\Lambda/\sqrt{|C_i|}$ & $C_i$ & $\Lambda/\sqrt{|C_i|}$  & $C_i$ & $\Lambda/\sqrt{|C_i|}$ & $C_i$ & $\Lambda/\sqrt{|C_i|}$\\ 
\hline \endhead 
 \hline \endlastfoot 
$C_{\phi \Box}$                & $ -0.7 \pm 3.0 $         & $ 1.2 $                  & $  4.6 \pm 5.3 $                    & $0.8$                  & --                                    & --                     &  --    & --                             \\
$C_{\phi D}$                   & $ 0.040 \pm 0.090 $      & $ 6.5 $                  & $ -0.2 \pm 4.4 $                    & $ 1.0 $                & $ 0.3 \pm 1.1 $                       & $ 6.4 $                &  --    & --                             \\
$C_{\phi G}$                   & $ -0.0102 \pm 0.0084 $   & $ 18.5 $                 & $ 0.016 \pm 0.013 $                 & $ 14.1 $               & $ -0.112 \pm 0.084 $                  & $ 19.0 $               & $ -0.14 \pm 0.10 $ & $ 16.8 $           \\
$C_{\phi B}$                   & $ -0.008 \pm 0.015 $     & $ 15.6 $                 & $ -0.17 \pm 0.56 $                  & $ 2.7 $                & $ -0.10 \pm 0.17 $                    & $ 15.4 $               & $ -0.5 \pm 2.4 $ & $ 4.6 $              \\
$C_{\phi W}$                   & $ -0.028 \pm 0.047 $     & $ 8.8 $                  & $ -1.8 \pm 1.5 $                    & $ 1.4 $                & $ -0.33 \pm 0.52 $                    & $ 8.7 $                & $ \textcolor{red}{-12.0 \pm 2.8} $ & \textcolor{red}{--} \\
$C_{\phi WB}$                  & $ 0.018 \pm 0.025 $      & $ 11.6 $                 & $ -1.4 \pm 1.7 $                    & $ 1.4 $                & $ 0.19 \pm 0.29 $                     & $ 11.3 $               & $ -7.3 \pm 4.9 $  & \textcolor{red}{--}  \\
$C_{G}$                        & $ 3.5 \pm 1.8 $          & $ 1.1 $                  & $ 2.4 \pm 2.1 $                     & $ 1.2 $                & $\textcolor{red}{10.3 \pm 4.7}$       & \textcolor{red}{--}    & $\textcolor{red}{8.2 \pm 6.8}$       &  \textcolor{red}{--}  \\
$C_{W}$                        & $ -0.25 \pm 0.43 $       & $ 2.9 $                  & $ -1.1 \pm 7.0 $                    & $ 0.8 $                 & $ -2.2 \pm 3.8 $                      & $ 3.2 $                & -- & --  \\
$C_{\phi l}^{(1)}$            & $ -0.030 \pm 0.039 $     & $ 9.2 $                   & $ -0.2 \pm 1.5 $                    & $ 1.7 $                 & $ -0.33 \pm 0.45 $                   & $ 9.1 $                & $ 2.9 \pm 3.3 $ & $ 3.5 $           \\
$C_{\phi l}^{(3)}$            & $ -0.058 \pm 0.056 $     & $ 7.3 $                   & $ 0.80 \pm 0.35 $                   & $ 2.5 $                 & $ -0.67 \pm 0.64 $                   & $ 7.3 $                & $ 4.8 \pm 3.0 $ & $ 3.1 $   \\
$C_{\phi e}$                  & $ 0.052 \pm 0.051 $      & $ 7.7 $                   & $ 1.0 \pm 2.5 $                     & $ 1.3 $                 & $ 0.62 \pm 0.59 $                    & $ 7.6 $                & $ \textcolor{red}{9.8 \pm 5.1} $ & \textcolor{red}{--}          \\
$C_{\phi q}^{(1)}$            & $ 0.10 \pm 0.13 $        & $ 5.0 $                   & $ -3.5 \pm 1.3 $                    & $ 1.2 $                 & $ 0.8 \pm 1.4 $                      & $ 5.4 $                & $ -3.1 \pm 4.7 $ & $ 2.9 $     \\
$C_{\phi q}^{(3)}$            & $ 0.086 \pm 0.052 $      & $ 6.8 $                   & $ 0.61 \pm 0.25 $                   & $ 2.9 $                 & $ 1.01 \pm 0.58 $                    & $ 6.7 $                &  $ 6.4 \pm 2.1 $ & $ 3.1 $      \\
$C_{\phi u}$                  & $ 0.13 \pm 0.18 $        & $ 4.3 $                   & $ 1.0 \pm 2.3 $                     & $ 1.3 $                 & $ 1.3 \pm 1.7 $                      & $ 4.7 $                & $ \textcolor{red}{-6.8 \pm 7.2} $ &  \textcolor{red}{--}            \\
$C_{\phi d}$                  & $ -0.83 \pm 0.51 $       & $ 2.2 $                   & $ -3.9 \pm 3.0 $                    & $ 1.0 $                 & $ \textcolor{red}{-10.3 \pm 4.4} $   & \textcolor{red}{--}    & $ \textcolor{red}{-8.2 \pm 6.8} $ & \textcolor{red}{--}       \\
$C_{ll}^{[aabb]}$             & $ -0.15 \pm 0.19 $       & $ 4.1 $                   & --                                  & --                      & $ -1.7 \pm 2.1 $                     & $ 4.2 $                & -- & -- \\
$C_{ll}^{[abba]}$             & $ -0.005 \pm 0.091 $     & $ 6.9 $                   & $ 0.82 \pm 0.74 $                   & $ 2.1 $                 & $ -0.01 \pm 0.96 $                   & $ 7.3 $                & $ 2.1 \pm 3.4 $ & $ 3.6 $       \\
$C_{qq}^{(1)[aabb]}$          & $ -0.62 \pm 0.54 $       & $ 2.3 $                   & \textcolor{red}{$ -11.2 \pm 3.8 $}  & \textcolor{red}{--}     & $ -5.8 \pm 4.6 $                     & $ 2.7 $                & $ \textcolor{red}{-8.6 \pm 6.5 }$   & \textcolor{red}{--}    \\
$C_{qq}^{(1)[abba]}$          & $ 0.09 \pm 0.23 $        & $ 4.0 $                   & $ -0.8 \pm 1.0 $                    & $ 1.8 $                 & $ 0.3 \pm 2.2 $                      & $ 4.7 $                & $ 5.3 \pm 3.4 $ & $ 2.9 $     \\
$C_{qq}^{(3)[aabb]}$          & $ 0.034 \pm 0.050 $      & $ 8.3 $                   & $ -0.36 \pm 0.33 $                  & $ 3.0 $                 & $ 0.21 \pm 0.40 $                    & $ 9.9 $                &  $ -0.5 \pm 3.1 $ & $ 3.8 $      \\
$C_{qq}^{(3)[abba]}$          & $ 0.25 \pm 0.16 $        & $ 4.0 $                   & $ -0.24 \pm 0.74 $                  & $ 2.4 $                 & $ 2.0 \pm 1.4 $                      & $ 4.5 $                & $ 3.7 \pm 2.5 $ & $ 3.3 $      \\
$C_{lq}^{(1)}$                & $ 0.055 \pm 0.054 $      & $ 7.4 $                   & $ -2.5 \pm 4.3 $                    & $ 1.0 $                 & $ 0.23 \pm 0.51 $                    & $ 8.8 $                & $ -3.7 \pm 4.9 $ & $ 2.7 $     \\
$C_{lq}^{(3)}$                & $ -0.0035 \pm 0.0055 $   & $ 24.9 $                  & $ -0.10 \pm 0.14 $                  & $ 4.9 $                 & $ -0.028 \pm 0.058 $                 & $ 26.7 $               & $ -0.03 \pm 0.51 $ & $ 9.7 $   \\
$C_{ee}$                      & $ -0.10 \pm 0.12 $       & $ 5.3 $                   & $ -4.6 \pm 5.4 $                    & $ 0.9 $                 & $ -1.0 \pm 1.3 $                     & $ 5.3 $                & $ -2.9 \pm 6.8 $ &  $ 2.7 $         \\
$C_{uu}^{[aabb]}$             & $ 0.47 \pm 0.56 $        & $ 2.4 $                   & $ -9.5 \pm 5.6 $                    & \textcolor{red}{--}     & $ 2.2 \pm 4.4 $                      & $ 3.1 $                & --   & --   \\
$C_{uu}^{[abba]}$             & $ 0.51 \pm 0.40 $        & $ 2.6 $                   & $ -2.3 \pm 2.0 $                    & $ 1.2 $                 & $ 4.2 \pm 4.1 $                      & $ 2.8 $                & $ \textcolor{red}{-8.9 \pm 6.1} $ & \textcolor{red}{--}       \\
$C_{dd}^{[aabb]}$             & $ 2.4 \pm 3.7 $          & $ 1.0 $                   & --                                  & --                      & --                                   & --                     & -- & --   \\
$C_{dd}^{[abba]}$             & $ 3.9 \pm 6.0 $          & \textcolor{red}{--}       & --                                  & --                      & --                                   & --                     & -- & -- \\
$C_{eu}$                      & $ 0.036 \pm 0.038 $      & $ 9.0 $                   & $ 0.9 \pm 6.1 $                     & $ 0.9 $                 & $ 0.17 \pm 0.41 $                    & $ 10.2 $               & $ 2.6 \pm 8.2.5 $ & \textcolor{red}{--}       \\
$C_{ed}$                      & $ -0.10 \pm 0.12 $       & $ 5.2 $                   & $\textcolor{red}{6.5 \pm 7.2} $     & \textcolor{red}{--}     & $ -0.5 \pm 1.3 $                     & $ 5.7 $                & -- & --  \\
$C_{ud}^{(1)}$                & $ -2.4 \pm 1.6 $         & $ 1.3 $                   & $\textcolor{red}{ -7.9 \pm 7.2} $   & \textcolor{red}{--}     & $ \textcolor{red}{-8.2 \pm 6.8} $    & \textcolor{red}{--}    & -- & --  \\
$C_{ud}^{(8)}$                & $ 5.5 \pm 4.1 $          & $ 0.8 $                   & $\textcolor{red}{8.4 \pm 6.6} $     & \textcolor{red}{--}     & $ \textcolor{red}{6.3 \pm 8.8} $     & \textcolor{red}{--}    & -- & -- \\
$C_{le}$                      & $ -0.11 \pm 0.31 $       & $ 3.5 $                   &  $ -0.28 \pm 0.39 $                 & $ 2.9 $                 & $ -1.0 \pm 3.6 $                     & $ 3.5 $                &  $ -1.7 \pm 3.6 $ & $ 3.3 $         \\
$C_{lu}$                      & $ 0.050 \pm 0.084 $      & $ 6.4 $                   & $ 5.3 \pm 6.7 $                     & $ 0.8 $                 & $ 0.12 \pm 0.85 $                    & $ 7.4 $                & --   & -- \\
$C_{ld}$                      & $ -0.21 \pm 0.24 $       & $ 3.7 $                   & --                                  & --                      & $ -1.3 \pm 2.7 $                     & $ 3.9 $                &  -- & --    \\
$C_{qe}$                      & $ 0.057 \pm 0.095 $      & $ 6.2 $                   & $ 1.6 \pm 5.8 $                     & $ 0.8 $                 & $ 0.0 \pm 1.0 $                      & $ 7.0 $                & -- & --      \\
$C_{qu}^{(1)}$                & $ 0.40 \pm 0.80 $        & $ 2.1 $                   & $ 6.8 \pm 7.2 $                     & \textcolor{red}{--}     & $ 1.1 \pm 6.3 $                      & $ 2.7 $                & --   &  --    \\
$C_{qu}^{(8)}$                & $ -0.27 \pm 0.53 $       & $ 2.6 $                   & $ -3.7 \pm 2.6 $                    & $ 1.0 $                 & $ -2.6 \pm 5.4 $                     & $ 2.7 $                & $ -1.8 \pm 7.0 $   & $ 2.7 $       \\
$C_{qd}^{(1)}$                & $ -0.1 \pm 2.3 $         & $ 1.4 $                   & --                                  & --                      & $ \textcolor{red}{5.5 \pm 9.5} $     & \textcolor{red}{--}    &  --   & --   \\
$C_{qd}^{(8)}$                & $ -4.8 \pm 3.5 $         & $ 0.9 $                   & --                                  & --                      & $ \textcolor{red}{-6.8 \pm 8.2} $    & \textcolor{red}{--}    & -- & --

\end{longtable}
\end{scriptsize}

\subsection{Results for the \texorpdfstring{$U(2)^5$}{U2-5} flavour symmetric SMEFT}

\begin{tiny}
\begin{longtable}{|l|c|c|c|c|c|c|c|c|c|} 
\caption[]{ Same as Table~\ref{tab:u3-ind}, presenting the results of the individual fits for the $U(2)^5$ flavour symmetric SMEFT in the UP basis. 
\label{tab:u2-ind}  }\\
\hline
 & \multicolumn{3}{c|}{10 TeV}  & \multicolumn{3}{c|}{3 TeV}  & \multicolumn{3}{c|}{1 TeV - noRGE }  \\ \hline 
& $C_i$ & $\Lambda/\sqrt{|C_i|}$ & Obs & $C_i$ & $\Lambda/\sqrt{|C_i|}$ & Obs & $C_i$ & $\Lambda/\sqrt{|C_i|}$ & Obs  \\ 
\hline \endhead 
 \hline \endlastfoot 
$C_{\phi  \Box}$& -- & --  & -- & $ -0.7 \pm 3.0 $ & $ 1.2 $  & H & $ -0.02 \pm 0.29 $ & $ 1.3 $  &  H \\ 
$C_{\phi D}$& $ 0.3 \pm 1.1 $ & $ 6.4 $  & EW & $ 0.040 \pm 0.090 $ & $ 6.5 $  & EW & $ 0.0024 \pm 0.0072 $ & $ 7.7 $  & EW \\ 
$C_{\phi G}$& $ -0.112 \pm 0.084 $ & $ 19.0 $  & H & $ -0.0102 \pm 0.0084 $ & $ 18.5 $  & H & $ -0.0017 \pm 0.0012 $ & $ 15.8 $  & H  \\ 
$C_{\phi B}$& $ -0.10 \pm 0.17 $ & $ 15.4 $  & H & $ -0.008 \pm 0.015 $ & $ 15.6 $  & H & $ -0.0008 \pm 0.0014 $ & $ 17.0 $  &  H \\ 
$C_{\phi W}$& $ -0.33 \pm 0.52 $ & $ 8.7 $  & H & $ -0.028 \pm 0.047 $ & $ 8.8 $  & H & $ -0.0029 \pm 0.0051 $ & $ 8.8 $  &  H \\ 
$C_{\phi WB}$& $ 0.19 \pm 0.29 $ & $ 11.3 $  & H & $ 0.018 \pm 0.025 $ & $ 11.6 $  & H & $ 0.0017 \pm 0.0025 $ & $ 12.2 $  &  H \\ 
$C_{G}$& $\textcolor{red}{10.3 \pm 4.7}$ &  \textcolor{red}{--}   & H/T & $ 3.5 \pm 1.8 $ & $ 1.1 $  & T & $ 0.07 \pm 0.15 $ & $ 1.7 $  &  T \\ 
$C_{W}$& $ -2.2 \pm 3.8 $ & $ 3.2 $  & H & $ -0.25 \pm 0.43 $ & $ 2.9 $  & H & $ -0.50 \pm 0.70 $ & $ 0.7 $ & EW\\ 
$C_{\phi l}^{(1)[33]}$& $ 0.4 \pm 1.8 $ & $ 5.0 $  & EW & $ 0.03 \pm 0.16 $ & $ 5.1 $  & EW & $ 0.002 \pm 0.015 $ & $ 5.4 $  &  EW \\ 
$C_{\phi l}^{(1)[aa]}$& $ -0.43 \pm 0.47 $ & $ 8.5 $  & EW & $ -0.040 \pm 0.043 $ & $ 8.6 $  & EW & $ -0.0034 \pm 0.0040 $ & $ 9.2 $  &  EW \\ 
$C_{\phi l}^{(3)[33]}$& $ 1.2 \pm 1.5 $ & $ 4.9 $  & EW & $ 0.10 \pm 0.14 $ & $ 4.9 $  & EW & $ 0.010 \pm 0.014 $ & $ 5.1 $  &  EW \\ 
$C_{\phi l}^{(3)[aa]}$& $ -0.82 \pm 0.61 $ & $ 7.1 $  & EW & $ -0.071 \pm 0.054 $ & $ 7.2 $  & EW & $ -0.0072 \pm 0.0058 $ & $ 7.4 $  & EW \\ 
$C_{\phi e}^{[33]}$& $ -1.7 \pm 2.0 $ & $ 4.3 $  & EW & $ -0.15 \pm 0.17 $ & $ 4.3 $  & EW & $ -0.015 \pm 0.017 $ & $ 4.6 $  &  EW \\ 
$C_{\phi e}^{[aa]}$& $ 0.76 \pm 0.62 $ & $ 7.2 $  & EW & $ 0.063 \pm 0.053 $ & $ 7.3 $  & EW & $ 0.0067 \pm 0.0054 $ & $ 7.6 $  &  EW \\ 
$C_{\phi q}^{(1)[33]}$& $ 0.32 \pm 0.40 $ & $ 9.4 $ & F & $ 0.029 \pm 0.036 $ & $ 9.5 $  & F & $ 0.0029 \pm 0.0037 $ & $ 9.9 $  &  F\\ 
$C_{\phi q}^{(1)[aa]}$&  $ -0.28 \pm 0.41 $ & $ 9.5 $   & F &$ -0.022 \pm 0.037 $ & $ 9.5 $  & F & $ -0.0024 \pm 0.0036 $ & $ 10.2 $   &  F \\ 
$C_{\phi q}^{(3)[33]}$& $ 0.27 \pm 0.33 $ & $ 10.2 $ & F & $ 0.026 \pm 0.030 $ & $ 10.1 $ & F &  $ 0.0031 \pm 0.0037 $ & $ 9.9 $  &  F \\ 
$C_{\phi q}^{(3)[aa]}$& $ 0.15 \pm 0.35 $ & $ 11.0 $ & F & $ 0.010 \pm 0.030 $ & $ 11.1 $  & F & $ 0.0012 \pm 0.0032 $ & $ 11.9 $ & F \\ 
$C_{\phi u}^{[33]}$& $ 0.8 \pm 1.9 $ & $ 4.7 $  & EW & $ 0.10 \pm 0.20 $ & $ 4.4 $  & EW & $ 0.586 \pm 0.946 $ &  $0.64$ & T \\ 
$C_{\phi u}^{[aa]}$& $ 4.8 \pm 4.3 $ & $ 2.8 $  & EW & $ 0.43 \pm 0.37 $ & $ 2.8 $  & EW & $ 0.043 \pm 0.036 $ & $ 3.0 $  &  EW \\ 
$C_{\phi d}^{[33]}$& $ \textcolor{red}{-10.9 \pm 4.2} $ & \textcolor{red}{--}  & EW & $ -2.0 \pm 1.1 $ & $ 1.5 $  & EW & $ -0.21 \pm 0.10 $ & $ 1.6 $  & EW \\ 
$C_{\phi d}^{[aa]}$& $ \textcolor{red}{-8.8 \pm 6.0} $  &  \textcolor{red}{--} & EW & $ -0.86 \pm 0.78 $ & $ 1.9 $  & EW & $ -0.080 \pm 0.074 $ & $ 2.1 $  &  EW \\ 
$C_{\phi ud}^{[33]}$& -- & --  & & $ 4.6 \pm 5.6 $ & $ 0.8 $  & H & $ \textcolor{red}{-8.9 \pm 6.1} $  & \textcolor{red}{--}  & T \\ 
$C_{e\phi }^{[33]}$& $ 0.20 \pm 0.98 $ & $ 6.7 $  & H & $ 0.020 \pm 0.086 $ & $ 6.7 $  & H & $ 0.0021 \pm 0.0085 $ & $ 7.3 $  & H \\ 
$C_{u\phi }^{[33]}$& -- & --  & & $ 5.3 \pm 4.3 $ & $ 0.8 $  & H & $ 0.52 \pm 0.41 $ & $ 0.9 $  &  H \\ 
$C_{d\phi }^{[33]}$& $ 0.7 \pm 1.0 $ & $ 6.1 $  & H & $ 0.080 \pm 0.093 $ & $ 6.0 $  & H & $ 0.009 \pm 0.012 $ & $ 5.5 $  &  H \\ 
$C_{eB}^{[33]}$& -- & --  & & $ \textcolor{red}{6.3 \pm 8.1} $ & \textcolor{red}{--}   & H & $ 1.6 \pm 3.2 $ & $ 0.4 $  &  H \\ 
$C_{uB}^{[33]}$& $ -0.7 \pm 1.4 $ & $ 5.4 $  & H & $ -0.10 \pm 0.17 $ & $ 4.6 $  & H & $ 0.017 \pm 0.048 $ & $ 2.9 $  &  H \\ 
$C_{dB}^{[33]}$& $ -0.0015 \pm 0.0041 $ & $ 100.2 $ & F & $ -0.00012 \pm 0.00037 $ & $ 101.0 $ & F & $ -0.000016 \pm 0.000036 $ & $ 106.2 $   &  F  \\ 
$C_{eW}^{[33]}$& -- & --  & & $ 0.4 \pm 2.2 $ & $ 1.4 $  & H & $ -3.3 \pm 4.9 $ & \textcolor{red}{--} & H \\ 
$C_{uW}^{[33]}$& $ -1.2 \pm 2.2 $ & $ 4.2 $  & H & $ -0.12 \pm 0.26 $ & $ 3.8 $  & H & $ -0.007 \pm 0.073 $ & $ 2.6 $  & H \\ 
$C_{dW}^{[33]}$& $ 0.0032 \pm 0.0078 $ & $ 72.4 $  & F & $ 0.00028 \pm 0.00070 $ & $ 72.3 $  & F & $ 0.000027 \pm 0.000068 $ & $ 76.8 $  &  F \\ 
$C_{uG}^{[33]}$& $ -1.9 \pm 1.2 $ & $ 4.9 $   & H & $ -0.27 \pm 0.15 $ & $ 4.0 $ & H & $ 0.009 \pm 0.033 $ & $ 3.7 $  &  H \\ 
$C_{dG}^{[33]}$& $ -0.014 \pm 0.032 $ & $ 36.9 $  & F & $ -0.0014 \pm 0.0031 $ & $ 33.6 $   & F & $ -0.00022 \pm 0.00060 $ & $ 26.5 $ &  F \\ 
$C_{ll}^{[3333]}$& --  & --  & & $ 2.8 \pm 4.6 $ & $ 0.9 $  & DY & -- & --  &  \\ 
$C_{ll}^{[aa33]}$& $ -0.1 \pm 3.7 $ & $ 3.7 $  & EW & $ 0.00 \pm 0.32 $ & $ 3.8 $  & EW & $ -0.002 \pm 0.034 $ & $ 3.8 $  &  EW \\ 
$C_{ll}^{[a33a]}$& $ 0.8 \pm 3.2 $ & $ 3.7 $  & EW & $ 0.06 \pm 0.29 $ & $ 3.8 $  & EW & $ -0.003 \pm 0.035 $ & $ 3.8 $  &  EW \\ 
$C_{ll}^{[aabb]}$& $ -2.5 \pm 2.6 $ & $ 3.7 $  & EW & $ -0.21 \pm 0.23 $ & $ 3.7 $  & EW & $ -0.026 \pm 0.024 $ & $ 3.6 $  &  EW \\ 
$C_{ll}^{[abba]}$& $ -0.1 \pm 1.0 $ & $ 6.8 $  & EW & $ -0.005 \pm 0.099 $ & $ 6.7 $  & EW & $ -0.003 \pm 0.011 $ & $ 6.3 $  &  EW \\ 
$C_{qq}^{(1)[3333]}$& $ -0.45 \pm 0.23 $ & $ 10.7 $  & F & $ -0.043 \pm 0.020 $ & $ 10.7 $  & F & $ -0.0036 \pm 0.0018 $ & $ 11.9 $  &  F \\ 
$C_{qq}^{(1)[aa33]}$& $ 0.235 \pm 0.099 $ & $ 15.1 $  & F & $ 0.0192 \pm 0.0096 $ & $ 15.4 $  & F & $ 0.00182 \pm 0.00088 $ & $ 16.8 $  & F \\ 
$C_{qq}^{(1)[a33a]}$& $ 0.214 \pm 0.096 $ & $ 15.5 $  & F & $ 0.0188 \pm 0.0096 $ & $ 15.3 $  & F & $ 0.00178 \pm 0.00086 $ & $ 17.0 $  &  F \\ 
$C_{qq}^{(1)[aabb]}$& $ -0.43 \pm 0.22 $ & $ 10.2 $  & F & $ -0.036 \pm 0.018 $ & $ 11.2 $  & F & $ -0.0035 \pm 0.0016 $ & $ 12.2 $  &  F \\ 
$C_{qq}^{(1)[abba]}$& $ -0.47 \pm 0.22 $ & $ 10.7 $  & F & $ -0.036 \pm 0.021 $ & $ 10.1 $  & F & $ -0.0039 \pm 0.0018 $ & $ 11.8 $  &  F \\ 
$C_{qq}^{(3)[3333]}$& $ -0.56 \pm 0.23 $ & $ 10.2 $   & F & $ -0.040 \pm 0.020 $ & $ 10.4 $  & F & $ -0.0038 \pm 0.0018 $ & $ 11.9 $  & F \\ 
$C_{qq}^{(3)[aa33]}$& $ 0.22 \pm 0.13 $ & $ 13.5 $  & F & $ 0.0207 \pm 0.0096 $ & $ 15.1 $  & F & $ 0.00192 \pm 0.00090 $ & $ 16.3 $  & F \\ 
$C_{qq}^{(3)[a33a]}$& $ 0.18 \pm 0.11 $ & $ 15.7 $  & F & $ 0.0194 \pm 0.0094 $ & $ 15.8 $  & F & $ 0.00192 \pm 0.00096 $ & $ 16.7 $  & F \\ 
$C_{qq}^{(3)[aabb]}$& $ -0.36 \pm 0.21 $ & $ 11.5 $  & F & $ -0.032 \pm 0.019 $ & $ 11.2 $  & F & $ -0.0038 \pm 0.0018 $ & $ 12.0 $  &  F \\ 
$C_{qq}^{(3)[abba]}$& $ -0.40 \pm 0.21 $ & $ 10.6 $  & F & $ -0.036 \pm 0.018 $ & $ 11.1 $   & F & $ -0.0039 \pm 0.0019 $ & $ 11.3 $  &  F \\ 
$C_{lq}^{(1)[3333]}$& $ \textcolor{red}{-7.1 \pm 7.1 }$ & \textcolor{red}{--}  & EW & $ -0.9 \pm 1.1 $ & $ 1.7 $  & EW & $ -0.21 \pm 0.24 $  & $ 1.2 $  &  F \\ 
$C_{lq}^{(1)[aa33]}$& $ -0.21 \pm 0.38 $ & $ 10.1 $  & F & $ -0.019 \pm 0.035 $ & $ 10.1 $   & F & $ -0.0021 \pm 0.0036 $ & $ 10.3 $   &  F \\ 
$C_{lq}^{(1)[33aa]}$& $ 3.6 \pm 2.5 $ & $ 3.4 $  & DY & $ 0.33 \pm 0.24 $ & $ 3.3 $  & DY & $ 0.076 \pm 0.038 $ & $ 2.6 $  &  DY \\ 
$C_{lq}^{(1)[aabb]}$& $ 0.17 \pm 0.31 $ & $ 11.3 $  & F & $ 0.024 \pm 0.030 $ & $ 10.4 $  & F & $ 0.0022 \pm 0.0035 $ & $ 10.2 $   & F \\ 
$C_{lq}^{(3)[3333]}$& $ 8.2 \pm 5.8 $ & \textcolor{red}{--}  & EW/DY & $ 0.93 \pm 0.91 $ & $ 1.8 $  & EW & $ 0.15 \pm 0.24 $ & $ 1.3 $  &  F \\ 
$C_{lq}^{(3)[aa33]}$& $ -0.25 \pm 0.39 $ & $ 9.7 $   & F & $ -0.022 \pm 0.034 $ & $ 10.0 $  & F & $ -0.0023 \pm 0.0038 $ & $ 10.0 $   &  F \\ 
$C_{lq}^{(3)[33aa]}$& $ -0.21 \pm 0.29 $ & $ 11.1 $  & DY & $ -0.018 \pm 0.031 $ & $ 10.6 $  & DY & $ -0.0022 \pm 0.0042 $ & $ 9.8 $  & DY \\ 
$C_{lq}^{(3)[aabb]}$& $ -0.011 \pm 0.056 $ & $ 29.2 $  & DY & $ -0.0016 \pm 0.0053 $ & $ 27.1 $ & DY & $ -0.00007 \pm 0.00084 $ & $ 24.3 $  & DY \\ 
$C_{ee}^{[3333]}$& -- & -- & & $ 7.2 \pm 7.2 $ & \textcolor{red}{--}   & EW/DY & -- & -- &  \\ 
$C_{ee}^{[aa33]}$& $ 0.0 \pm 2.1 $ & $ 4.9 $  & EW & $ 0.01 \pm 0.18 $ & $ 4.9 $  & EW & $ -0.003 \pm 0.019 $ & $ 4.9 $  &  EW \\ 
$C_{ee}^{[aabb]}$& $ -1.4 \pm 1.5 $ & $ 4.8 $  & EW & $ -0.13 \pm 0.13 $ & $ 4.8 $  & EW & $ -0.015 \pm 0.014 $ & $ 4.8 $  &  EW \\ 
$C_{uu}^{[3333]}$& $2.5 \pm 8.1$ & \textcolor{red}{--}  & EW & $ 0.2 \pm 1.3 $ & $ 1.8 $  & EW & $ -2.912 \pm 1.56 $ & 0.41  & T \\ 
$C_{uu}^{[aa33]}$& $ \textcolor{red}{6.8 \pm 7.2} $  & \textcolor{red}{--} & T & $ 1.3 \pm 1.3 $ & $ 1.5 $  & EW & $ -0.17 \pm 0.56 $ & $ 1.0 $  &  T\\ 
$C_{uu}^{[a33a]}$& $ 6.3 \pm 4.7 $ & $ 2.6 $  & T & $ 0.55 \pm 0.44 $ & $ 2.5 $  & T & $ 0.014 \pm 0.092 $ & $ 2.2 $  & T \\ 
$C_{uu}^{[aabb]}$& $ 1.6 \pm 6.8 $ & $ 2.7 $  & DY & $ 0.72 \pm 0.91 $ & $ 1.9 $  & DY & -- & --  &  \\ 
$C_{uu}^{[abba]}$& $ 2.5 \pm 8.4 $ & \textcolor{red}{--}  & DY & $ 1.0 \pm 1.3 $ & $ 1.6 $  & DY & -- & --  &  \\ 
$C_{dd}^{[3333]}$& -- & -- & & -- & --   & & -- & -- &  \\ 
$C_{dd}^{[aa33]}$& -- & -- & & $ 6.0 \pm 8.1 $ & \textcolor{red}{--}   & EW/DY & --& --  &  \\ 
$C_{dd}^{[a33a]}$& -- & -- & & -- & --  & & -- & -- &  \\ 
$C_{dd}^{[aabb]}$& -- & -- & & $ 3.9 \pm 5.3 $ & $ 0.8 $  & DY & -- & --  &  \\ 
$C_{dd}^{[abba]}$& -- & -- & & $ 4.2 \pm 6.7 $ & \textcolor{red}{--}  & DY & -- & -- &  \\
$C_{eu}^{[3333]}$& $ \textcolor{red}{-8.2 \pm 6.8} $ & \textcolor{red}{--}  & EW & $ -1.3 \pm 1.5 $ & $ 1.4 $  & EW & -- & --  &  \\ 
$C_{eu}^{[aa33]}$& $ 5.4 \pm 4.0 $ & $ 2.8 $  & EW & $ 0.56 \pm 0.44 $ & $ 2.5 $  & EW &-- &--  &  \\ 
$C_{eu}^{[33aa]}$& $ 2.7 \pm 1.8 $ & $ 4.0 $  & DY & $ 0.24 \pm 0.17 $ & $ 4.1 $  & DY & $ 0.041 \pm 0.022 $ & $ 3.4 $  & DY \\ 
$C_{eu}^{[aabb]}$& $ -0.02 \pm 0.41 $ & $ 11.0 $  & DY & $ 0.021 \pm 0.039 $ & $ 9.5 $  & DY & $ -0.0013 \pm 0.0077 $ & $ 7.9 $  &  DY \\ 
$C_{ed}^{[3333]}$& -- & --  & & -- & -- & & $ -4.8 \pm 3.7 $ & $ 0.3 $  &  DY \\ 
$C_{ed}^{[aa33]}$& $\textcolor{red}{5.6 \pm 9.5}$ & \textcolor{red}{--}  & EW & $ 1.1 \pm 2.0 $ & $ 1.3 $  & EW & $ 0.08 \pm 0.22 $ & $ 1.4 $  &  EW \\ 
$C_{ed}^{[33aa]}$& $ \textcolor{red}{-7.9 \pm 4.7}$ & \textcolor{red}{--}  & DY & $ -0.73 \pm 0.50 $ & $ 2.3 $  & DY & $ -0.110 \pm 0.061 $ & $ 2.1 $  &  DY \\ 
$C_{ed}^{[aabb]}$& $ 0.0 \pm 1.4 $ & $ 5.9 $  & DY & $ -0.06 \pm 0.12 $ & $ 5.4 $  & DY & $ 0.003 \pm 0.021 $ & $ 4.8 $  &  DY \\ 
$C_{ud}^{(1)[3333]}$& $\textcolor{red}{-6.5 \pm 8.3}$ & \textcolor{red}{--} & EW & $ \textcolor{red}{-10.9 \pm 4.2} $ & \textcolor{red}{--}  & EW & -- & --  &  \\ 
$C_{ud}^{(1)[aa33]}$& -- & -- & & $ -5.1 \pm 6.8 $ & \textcolor{red}{--}  & DY & -- & --  &  \\ 
$C_{ud}^{(1)[33aa]}$& $ \textcolor{red}{-7.4 \pm 7.7} $ & \textcolor{red}{--}  & EW/T & $ -10.3 \pm 3.7 $ & \textcolor{red}{--}  & EW & $ -1.5 \pm 1.9 $ & $ 0.4 $  &  T \\ 
$C_{ud}^{(1)[aabb]}$& $ \textcolor{red}{-6.0 \pm 8.8} $  & \textcolor{red}{--}  & DY & $ -1.9 \pm 2.6 $ & $ 1.1 $  & DY & -- & -- &  \\ 
$C_{ud}^{(8)[3333]}$& -- & --  & & -- & -- & & $ \textcolor{red}{5.5 \pm 9.5} $  & \textcolor{red}{--} & T \\ 
$C_{ud}^{(8)[aa33]}$& -- & --  & & -- & --  & & -- & --  &  \\ 
$C_{ud}^{(8)[33aa]}$& $ \textcolor{red}{6.8 \pm 8.3} $  & \textcolor{red}{--}  & T & $ 5.2 \pm 3.6 $ & $ 0.9 $  & T & $ 0.33 \pm 0.71 $ & $ 0.8 $  & T \\ 
$C_{ud}^{(8)[aabb]}$& -- & --  & &  -- & -- & & -- & --  &  \\ 
$C_{le}^{[3333]}$& -- & --  & & $ \textcolor{red}{8.1 \pm 6.9} $  & \textcolor{red}{--}  & DY & -- & --  &  \\ 
$C_{le}^{[aa33]}$& $ -4.2\pm 7.1$ & \textcolor{red}{--}  & EW & $ -0.65 \pm 0.98 $ & $ 1.9 $  & EW & $ -0.08 \pm 0.11 $ & $ 1.8 $  &  EW \\ 
$C_{le}^{[33aa]}$& $ \textcolor{red}{-6.7 \pm 7.7} $ & \textcolor{red}{--}  & EW & $ -0.55 \pm 0.99 $ & $ 1.9 $  & EW & $ -0.08 \pm 0.11 $ & $ 1.8 $  &  EW \\ 
$C_{le}^{[aabb]}$& $ 0.4 \pm 4.2 $ & $ 3.3 $  & EW & $ 0.04 \pm 0.38 $ & $ 3.4 $  & EW & $ 0.000 \pm 0.041 $ & $ 3.4 $  &  EW \\ 
$C_{lu}^{[3333]}$& $ 2.5 \pm 8.8 $ & \textcolor{red}{--} & EW & $ 0.3 \pm 1.4 $ & $ 1.7 $  & EW & -- & --  &  \\ 
$C_{lu}^{[aa33]}$& $ -2.8 \pm 3.3 $ & $ 3.3 $  & EW & $ -0.32 \pm 0.37 $ & $ 2.9 $  & EW & -- & --  &  \\ 
$C_{lu}^{[33aa]}$& $ 5.7 \pm 3.8 $ & $ 2.8 $  & DY & $ 0.49 \pm 0.35 $ & $ 2.8 $  & DY & $ 0.083 \pm 0.045 $ & $ 2.4 $  &  DY \\ 
$C_{lu}^{[aabb]}$& $ -0.03 \pm 0.89 $ & $ 7.4 $  & DY & $ 0.042 \pm 0.084 $ & $ 6.6 $  & DY & $ -0.002 \pm 0.016 $ & $ 5.6 $  & DY \\ 
$C_{ld}^{[3333]}$& -- & --  & & -- & --  & & $ \textcolor{red}{-9.5 \pm 5.3} $ & \textcolor{red}{--}  & DY  \\ 
$C_{ld}^{[aa33]}$& -- & --  & & $ 0.4 \pm 4.6 $ & $ 1.0 $  & EW & $ 0.15 \pm 0.53 $ & $ 0.9 $  &  EW \\ 
$C_{ld}^{[33aa]}$& $ \textcolor{red}{-9.8 \pm 5.3} $ & \textcolor{red}{--}  & DY & $ -1.4 \pm 1.1 $ & $ 1.6 $  & DY & $ -0.23 \pm 0.13 $ & $ 1.4 $  &  DY \\ 
$C_{ld}^{[aabb]}$& $ -0.4 \pm 2.7 $ & $ 4.1 $  & DY & $ -0.14 \pm 0.24 $ & $ 3.8 $  & DY & $ 0.000 \pm 0.043 $ & $ 3.4 $  & DY \\ 
$C_{qe}^{[3333]}$& $ \textcolor{red}{7.7 \pm 7.4} $ & \textcolor{red}{--}  & EW & $ 1.1 \pm 1.5 $ & $ 1.5 $  & EW & $ \textcolor{red}{8.2 \pm 5.4}$ & \textcolor{red}{--}  &  DY \\ 
$C_{qe}^{[aa33]}$& $ 6.9 \pm 4.4 $ & $ 2.6 $  & DY & $ 0.62 \pm 0.40 $ & $ 2.6 $  & DY & $ 0.091 \pm 0.049 $ & $ 2.3 $  &  DY \\ 
$C_{qe}^{[33aa]}$& $ 0.22 \pm 0.38 $ & $ 10.0 $  & F & $ 0.020 \pm 0.035 $ & $ 9.9 $  & F & $ 0.0024 \pm 0.0036 $ & $ 10.1 $  & F \\ 
$C_{qe}^{[aabb]}$& $ -0.23 \pm 0.36 $ & $ 10.4 $  & F & $ -0.015 \pm 0.032 $ & $ 10.8 $  & F & $ -0.0021 \pm 0.0035 $ & $ 10.3 $  & F \\ 
$C_{qu}^{(1)[3333]}$& $ 0.3 \pm 3.1 $ & $ 3.9 $  & F & $ 0.08 \pm 0.36 $ & $ 3.4 $ & F &  $ 4.77 \pm 2.53 $ & $ 0.32 $ &  T\\ 
$C_{qu}^{(1)[aa33]}$& $ 0.1 \pm 2.9 $ & $ 4.2 $   & F & $ 0.02 \pm 0.34 $ & $ 3.6 $  & F & $ -1.1 \pm 3.3 $ & $ 0.4 $ &  T\\ 
$C_{qu}^{(1)[33aa]}$& $\textcolor{red}{ -7.0 \pm 6.9 }$  & \textcolor{red}{--} & F & $ -2.5 \pm 2.3 $ & $ 1.1 $   & F & \textcolor{red}{--} & \textcolor{red}{--} & \\ 
$C_{qu}^{(1)[aabb]}$& $ 3.6 \pm 9.1 $  & \textcolor{red}{--}  & DY & $ 1.5 \pm 1.9 $ & $ 1.3 $ & DY & -- & --  &  \\ 
$C_{qu}^{(8)[3333]}$& $ \textcolor{red}{-10.3 \pm 4.7} $ & \textcolor{red}{--} & F & $ -6.2 \pm 2.5 $ & $ 0.9 $  & F & $\textcolor{red}{ -7.6 \pm 2.5} $ & \textcolor{red}{--}  & T \\ 
$C_{qu}^{(8)[aa33]}$& $ \textcolor{red}{6.8 \pm 6.1} $ & \textcolor{red}{--}  & F/T& $ 0.50 \pm 0.78 $ & $ 2.1 $  & T & $ 0.32 \pm 0.21 $ & $ 1.2 $  & T \\ 
$C_{qu}^{(8)[33aa]}$& $ \textcolor{red}{-5.8 \pm 8.2} $ & \textcolor{red}{--}  & T & $ -0.6 \pm 1.3 $ & $ 1.7 $  & T & $ 0.43 \pm 0.36 $ & \textcolor{red}{--} &  T \\ 
$C_{qu}^{(8)[aabb]}$& -- & --  & & $ -8.1 \pm 6.7 $ & \textcolor{red}{--}  & T & -- & --  &  \\ 
$C_{qd}^{(1)[3333]}$& -- & --  & & $ 3.6 \pm 4.4 $ & $ 0.9 $  & F & -- & --  &  \\ 
$C_{qd}^{(1)[aa33]}$& -- & --  & &  -- & --  & & -- & --  &  \\ 
$C_{qd}^{(1)[33aa]}$& -- & --  & & $ -1.1 \pm 3.6 $ & $ 1.1 $  & EW & $-1.65 \pm 2.15 $ & 0.41  & T  \\ 
$C_{qd}^{(1)[aabb]}$& -- & --  & & $ -0.9 \pm 3.9 $ & $ 1.0 $  & DY & --& --  &  \\ 
$C_{qd}^{(8)[3333]}$& -- & --  & & $ -1.2 \pm 6.1 $ & $ 0.9 $  & F & $\textcolor{red}{6.0 \pm 9.0} $  & \textcolor{red}{--}  & T \\ 
$C_{qd}^{(8)[aa33]}$& -- & --  & &  -- & --  & & -- &  --  &  \\ 
$C_{qd}^{(8)[33aa]}$& -- & --  & & $ -1.1 \pm 2.7 $ & $ 1.1 $  & T & $ 0.85 \pm 0.87 $ & $ 0.6 $  &  T \\ 
$C_{qd}^{(8)[aabb]}$& -- & -- & & $ -3.6 \pm 4.4 $ & $ 0.8 $  & F & -- & --  &  \\ 
$C_{ledq}^{[3333]}$& -- & --  & & -- & --  & & -- & -- &  \\ 
$C_{quqd}^{(1)[3333]}$& $ -0.10 \pm 0.40 $ & $ 10.2 $  & F & $ -0.014 \pm 0.047 $ & $ 9.2 $  & F & -- & --  &  \\ 
$C_{quqd}^{(8)[3333]}$& $ -0.16 \pm 0.48 $ & $ 9.2 $  & F & $ -0.020 \pm 0.054 $ & $ 8.1 $  & F & -- & -- &  \\ 
$C_{lequ}^{(1)[3333]}$& $ -1.2 \pm 3.7 $ & $ 3.5 $  & H & $ -0.10 \pm 0.40 $ & $ 3.1 $  & H & -- & --  &  \\ 
$C_{lequ}^{(3)[3333]}$& -- & --  & & $ 0.5 \pm 2.9 $ & $ 1.2 $  & H & -- & --  &  

\end{longtable}
\end{tiny}


\begin{tiny}
\begin{longtable}{|l|c|c|c|c||c|c|c|c|}
\caption[]{ Results of the individual fits in the UP basis (first two columns) and in the DOWN basis (second two columns) for the $U(2)^5$ flavour symmetric SMEFT, limited to those coefficients that are mainly constrained by flavour observables. Results are presented for $\Lambda = 3$ TeV and under the same assumptions on HPDI used in previous tables for both coefficient and scale bounds. 
\label{tab:u2-up-down}}\\
\hline
  & \multicolumn{2}{c|}{Full Fit UP}  & \multicolumn{2}{c||}{no Flavour UP}  & \multicolumn{2}{c|}{Full Fit DOWN}  & \multicolumn{2}{c|}{no Flavour DOWN}  \\ \hline 
& $C_i$ & $\Lambda/\sqrt{|C_i|}$ & $C_i$ & $\Lambda/\sqrt{|C_i|}$ & $C_i$ & $\Lambda/\sqrt{|C_i|}$ & $C_i$ & $\Lambda/\sqrt{|C_i|}$  \\ 
\hline \endhead 
 \hline \endlastfoot 
$C_{\phi q}^{(1)[33]}$& $ 0.029 \pm 0.036 $ & $ 9.5 $  & $ 0.08 \pm 0.14 $ & $ 5.0 $  & $ 0.06 \pm 0.14 $ & $ 5.1 $  & $ 0.08 \pm 0.14 $ & $ 5.0 $   \\ 
$C_{\phi q}^{(1)[aa]}$& $ -0.022 \pm 0.037 $ & $ 9.5 $  & $ 0.46 \pm 0.60 $ & $ 2.4 $  & $ 0.48 \pm 0.56 $ & $ 2.4 $  & $ 0.38 \pm 0.60 $ & $ 2.4 $   \\ 
$C_{\phi q}^{(3)[33]}$& $ 0.026 \pm 0.030 $ & $ 10.1 $  & $ 0.21 \pm 0.18 $ & $ 4.0 $  & $ 0.22 \pm 0.17 $ & $ 4.0 $  & $ 0.20 \pm 0.18 $ & $ 4.0 $   \\ 
$C_{\phi q}^{(3)[aa]}$& $ 0.010 \pm 0.030 $ & $ 11.1 $  & $ 0.124 \pm 0.066 $ & $ 6.0 $  & $ 0.134 \pm 0.068 $ & $ 5.8 $  & $ 0.122 \pm 0.069 $ & $ 5.9 $   \\ 
$C_{dB}^{[33]}$& $ -0.00012 \pm 0.00037 $ & $ 101.0 $  & $ 3.7 \pm 5.1 $ & $ 0.8 $  & $ 1.3 \pm 1.7 $  & $ 1.3 $  & $ 4.4 \pm 5.1 $ & $ 0.8 $ \\ 
$C_{dW}^{[33]}$& $ 0.00028 \pm 0.00070 $ & $ 72.3 $  & $ 0.44 \pm 0.57 $ & $ 2.4 $ & $ -0.0033 \pm 0.0091 $ & $ 20.2 $  & $ 0.42 \pm 0.56 $ & $ 2.4 $ \\ 
$C_{dG}^{[33]}$& $ -0.0014 \pm 0.0031 $ & $ 33.6 $  & $ -2.3 \pm 4.0 $ & $ 0.9 $  & $ 0.08 \pm 0.22 $ & $ 4.2 $  & $ -2.3 \pm 3.9 $ & $ 1.0 $    \\ 
$C_{qq}^{(1)[3333]}$& $ -0.043 \pm 0.020 $ & $ 10.7 $ & $ -0.65 \pm 0.74 $ & $ 2.0 $ & $ -0.40 \pm 0.70 $ & $ 2.2 $ & $ -0.66 \pm 0.76 $ & $ 2.1 $  \\ 
$C_{qq}^{(1)[aa33]}$& $ 0.0192 \pm 0.0096 $ & $ 15.4 $ & $ -3.1 \pm 1.6 $ & $ 1.2 $ & $ -2.2 \pm 1.1 $ & $ 1.4 $ & $ -3.1 \pm 1.5 $ & $ 1.2 $  \\ 
$C_{qq}^{(1)[a33a]}$& $ 0.0188 \pm 0.0096 $ & $ 15.3 $ & $ 0.72 \pm 0.41 $ & $ 2.4 $ & $ 0.14 \pm 0.28 $ & $ 3.5 $ & $ 0.73 \pm 0.42 $ & $ 2.4 $  \\ 
$C_{qq}^{(1)[aabb]}$& $ -0.036 \pm 0.018 $ & $ 11.2 $ & $ 0.7 \pm 1.2 $ & $ 1.7 $ & $ 1.0 \pm 1.1 $ & $ 1.7 $ & $ 0.6 \pm 1.2 $ & $ 1.7 $  \\ 
$C_{qq}^{(1)[abba]}$& $ -0.036 \pm 0.021 $ & $ 10.1 $ & $ 0.24 \pm 0.59 $ & $ 2.6 $ & $ 0.38 \pm 0.58 $ & $ 2.5 $ & $ 0.23 \pm 0.59 $ & $ 2.5 $  \\ 
$C_{qq}^{(3)[3333]}$& $ -0.040 \pm 0.020 $ & $ 10.4 $ & $ 1.0 \pm 1.4 $ & $ 1.5 $ & $ 1.4 \pm 1.2 $ & $ 1.5 $ & $ 0.6 \pm 1.4 $ & $ 1.5 $  \\ 
$C_{qq}^{(3)[aa33]}$& $ 0.0207 \pm 0.0096 $ & $ 15.1 $ & $ 0.13 \pm 0.13 $ & $ 4.9 $ & $ 0.11 \pm 0.12 $ & $ 5.2 $ & $ 0.15 \pm 0.13 $ & $ 4.8 $  \\ 
$C_{qq}^{(3)[a33a]}$& $ 0.0194 \pm 0.0094 $ & $ 15.8 $ & $ 0.33 \pm 0.20 $ & $ 3.5 $ & $ 0.28 \pm 0.17 $ & $ 3.8 $ & $ 0.32 \pm 0.20 $ & $ 3.5 $  \\ 
$C_{qq}^{(3)[aabb]}$& $ -0.032 \pm 0.019 $ & $ 11.2 $ & $ 0.032 \pm 0.078 $ & $ 6.9 $ & $ 0.031 \pm 0.079 $ & $ 7.0 $ & $ 0.031 \pm 0.079 $ & $ 7.0 $  \\ 
$C_{qq}^{(3)[abba]}$& $ -0.036 \pm 0.018 $ & $ 11.1 $ & $ 0.06 \pm 0.20 $ & $ 4.5 $ & $ 0.06 \pm 0.20 $ & $ 4.4 $ & $ 0.06 \pm 0.20 $ & $ 4.4 $  \\ 
$C_{lq}^{(1)[aa33]}$& $ -0.019 \pm 0.035 $ & $ 10.1 $  & $ 0.28 \pm 0.32 $ & $ 3.1 $  & $ 0.35 \pm 0.30 $ & $ 3.2 $  & $ 0.33 \pm 0.33 $ & $ 3.1 $     \\ 
$C_{lq}^{(1)[aabb]}$& $ 0.024 \pm 0.030 $ & $ 10.4 $  & $ 0.035 \pm 0.056 $ & $ 8.0 $  & $ 0.028 \pm 0.055 $ & $ 8.0 $  & $ 0.030 \pm 0.056 $ & $ 8.1 $    \\ 
$C_{lq}^{(3)[aa33]}$& $ -0.022 \pm 0.034 $ & $ 10.0 $  & $ 0.00 \pm 0.25 $ & $ 4.2 $  & $ 0.02 \pm 0.22 $ & $ 4.7 $  & $ -0.01 \pm 0.25 $ & $ 4.2 $     \\ 
$C_{qe}^{[33aa]}$& $ 0.020 \pm 0.035 $ & $ 9.9 $ & $ -0.58 \pm 0.45 $ & $ 2.5 $ & $ -0.59 \pm 0.40 $ & $ 2.6 $ & $ -0.55 \pm 0.46 $ & $ 2.5 $  \\ 
$C_{qe}^{[aabb]}$& $ -0.015 \pm 0.032 $ & $ 10.8 $ & $ 0.055 \pm 0.099 $ & $ 6.0 $ & $ 0.06 \pm 0.10 $ & $ 6.0 $ & $ 0.05 \pm 0.10 $ & $ 6.0 $  \\ 
$C_{qu}^{(1)[3333]}$& $ 0.08 \pm 0.36 $ & $ 3.4 $  & $ 0.7 \pm 1.3 $ & $ 1.7 $  & $ 0.6 \pm 1.2 $ & $ 1.7 $  & $ 0.7 \pm 1.3 $ & $ 1.7 $    \\ 
$C_{qu}^{(1)[aa33]}$& $ 0.02 \pm 0.34 $ & $ 3.6 $  & $ 1.0 \pm 2.2 $ & $ 1.3 $  & $ 1.5 \pm 2.3 $ & $ 1.2 $  & $ 1.1 \pm 0.0 $ & $ 1.3 $    \\ 
$C_{qd}^{(1)[3333]}$& $ 3.6 \pm 4.4 $ & $ 0.9 $ & $ 11.2 \pm 3.9 $ & $ 0.8 $ & $ 11.2 \pm 3.9 $ & $ 0.8 $ & $ 10.9 \pm 4.2 $ & $ 0.8 $  \\ 
$C_{quqd}^{(1)[3333]}$& $ -0.014 \pm 0.047 $ & $ 9.2 $  & $ 0.31 \pm 0.37 $ & $ 3.0 $  & $ 0.24 \pm 0.29 $ & $ 3.3 $  & $ 0.31 \pm 0.37 $ & $ 3.0 $    \\ 
$C_{quqd}^{(8)[3333]}$& $ -0.020 \pm 0.054 $ & $ 8.1 $  & $ 1.9 \pm 2.3 $ & $ 1.2 $  & $ 0.7 \pm 1.0 $ & $ 1.8 $  & $ 1.9 \pm 2.3 $ & $ 1.2 $   

\end{longtable}
\end{tiny}

\begin{small}
\begin{longtable}{|l|c|c|c|c|}
\caption[]{ Results of the individual fits (first two columns) and of the global fit (second two columns) for the $U(2)^5$ flavour symmetric SMEFT. For each coefficient, we report the $68\%$ HPDI interval and the bound on $\Lambda/\sqrt{\vert C_i\vert }$ in units of TeV obtained taking the maximum of the $95\%$ HPDI interval for $\vert C_i \vert$. Results are shown with RG evolution for $\Lambda = 10$ TeV. All results are obtained in the UP basis.  
\label{tab:u2-glob}}\\
\hline
 & \multicolumn{2}{c|}{10 TeV Individual}  & \multicolumn{2}{c|}{10 TeV Global}  \\ \hline 
& $C_i$ & $\Lambda/\sqrt{|C_i|}$ & $C_i$ & $\Lambda/\sqrt{|C_i|}$ \\ 
\hline \endhead 
 \hline \endlastfoot 
$C_{\phi D}$& $ 0.3 \pm 1.1 $ & $ 6.4 $  & -- & --  \\
$C_{\phi G}$& $ -0.112 \pm 0.084 $ & $ 19.0 $ & $ 0.31 \pm 0.30 $ & $ 10.5 $  \\
$C_{\phi B}$& $ -0.10 \pm 0.17 $ & $ 15.4 $  & $ 0.3 \pm 2.0 $ & $ 5.1 $  \\
$C_{\phi W}$& $ -0.33 \pm 0.52 $ & $ 8.7 $  & $ -6.5 \pm 4.7 $ & \textcolor{red}{--}  \\
$C_{\phi WB}$& $ 0.19 \pm 0.29 $ & $ 11.3 $ & $ -3.7 \pm 4.8 $ & $ 2.7 $   \\
$C_{G}$& $ \textcolor{red}{10.3 \pm 4.7} $ & \textcolor{red}{--}  & -- &--   \\
$C_{W}$& $ -2.2 \pm 3.8 $ & $ 3.2 $  & -- & --  \\
$C_{\phi l}^{(1)[33]}$& $ 0.4 \pm 1.8 $ & $ 5.0 $  & $ 1.2 \pm 5.1 $ & $ 3.1 $  \\
$C_{\phi l}^{(1)[aa]}$& $ -0.43 \pm 0.47 $ & $ 8.5 $ & $ 4.0 \pm 3.3 $ & $ 3.1 $  \\
$C_{\phi l}^{(3)[33]}$& $ 1.2 \pm 1.5 $ & $ 4.9 $ & $ 3.8 \pm 3.4 $ & $ 2.7 $    \\
$C_{\phi l}^{(3)[aa]}$& $ -0.82 \pm 0.61 $ & $ 7.1 $ & $ 3.2 \pm 3.2 $ & $ 3.2 $   \\
$C_{\phi e}^{[33]}$& $ -1.7 \pm 2.0 $ & $ 4.3 $  & $ 3.0 \pm 5.1 $ & $ 3.1 $ \\
$C_{\phi e}^{[aa]}$& $ 0.76 \pm 0.62 $ & $ 7.2 $  & $ 6.4 \pm 5.3 $ & $2.7$  \\
$C_{\phi q}^{(1)[33]}$& $ 0.32 \pm 0.40 $ & $ 9.4 $  & $ 1.6 \pm 8.3 $ & $ 2.7 $   \\
$C_{\phi q}^{(1)[aa]}$& $ -0.28 \pm 0.41 $ & $ 9.5 $ & $ -3.0 \pm 7.6 $ & $ 2.7 $  \\
$C_{\phi q}^{(3)[33]}$& $ 0.27 \pm 0.33 $ & $ 10.2 $ & $ 8.6 \pm 6.4 $ & \textcolor{red}{--}   \\
$C_{\phi q}^{(3)[aa]}$& $ 0.15 \pm 0.35 $ & $ 11.0 $ & $ 6.2 \pm 2.6 $ & $ 3.1 $   \\
$C_{\phi u}^{[33]}$& $ 0.8 \pm 1.9 $ & $ 4.7 $ & -- & --    \\
$C_{\phi u}^{[aa]}$& $ 4.8 \pm 4.3 $ & $ 2.8 $ & $\textcolor{red}{ -8 \pm 7 }$ & \textcolor{red}{--}   \\
$C_{\phi d}^{[33]}$& $ \textcolor{red}{-10.9 \pm 4.2} $ & \textcolor{red}{--} & $ \textcolor{red}{-8.1 \pm 6.7 }$ & \textcolor{red}{--}  \\
$C_{\phi d}^{[aa]}$& $ \textcolor{red}{-8.8 \pm 6.0} $  &  \textcolor{red}{--} & -- & --  \\
$C_{e\phi }^{[33]}$& $ 0.20 \pm 0.98 $ & $ 6.7 $ & $ 0.3 \pm 2.4 $ & $ 4.5 $   \\
$C_{d\phi }^{[33]}$& $ 0.7 \pm 1.0 $ & $ 6.1 $ & $ 5.5 \pm 3.0 $ & $ 3.1 $  \\
$C_{uB}^{[33]}$& $ -0.7 \pm 1.4 $ & $ 5.4 $  & -- & -- \\
$C_{dB}^{[33]}$& $ -0.0015 \pm 0.0041 $ & $ 100.2 $ & $ 0 \pm 4.3 $ & $ 3.5 $ \\
$C_{uW}^{[33]}$& $ -1.2 \pm 2.2 $ & $ 4.2 $ & -- & --  \\
$C_{dW}^{[33]}$& $ 0.0032 \pm 0.0078 $ & $ 72.4 $  & -- & -- \\
$C_{uG}^{[33]}$& $ -1.9 \pm 1.2 $ & $ 4.9 $  & \textcolor{red}{--} & \textcolor{red}{--} \\
$C_{dG}^{[33]}$& $ -0.014 \pm 0.032 $ & $ 36.9 $  & $ 1.1 \pm 0.9 $ & $ 6.6 $ \\
$C_{ll}^{[aa33]}$& $ -0.1 \pm 3.7 $ & $ 3.7 $  & -- & --  \\
$C_{ll}^{[a33a]}$& $ 0.8 \pm 3.2 $ & $ 3.7 $  & -- & -- \\
$C_{ll}^{[aabb]}$& $ -2.5 \pm 2.6 $ & $ 3.7 $  & -- & --  \\
$C_{ll}^{[abba]}$& $ -0.1 \pm 1.0 $ & $ 6.8 $  & $ -0.6 \pm 3.3 $ & $ 3.8 $  \\
$C_{qq}^{(1)[3333]}$& $ -0.45 \pm 0.23 $ & $ 10.7 $  & -- & --  \\
$C_{qq}^{(1)[aa33]}$& $ 0.235 \pm 0.099 $ & $ 15.1 $  & -- & -- \\
$C_{qq}^{(1)[a33a]}$& $ 0.214 \pm 0.096 $ & $ 15.5 $  & -- & -- \\
$C_{qq}^{(1)[aabb]}$& $ -0.43 \pm 0.22 $ & $ 10.2 $  & -- & -- \\
$C_{qq}^{(1)[abba]}$& $ -0.47 \pm 0.22 $ & $ 10.7 $  & -- & --  \\
$C_{qq}^{(3)[3333]}$& $ -0.56 \pm 0.23 $ & $ 10.2 $   & -- & --  \\
$C_{qq}^{(3)[aa33]}$& $ 0.22 \pm 0.13 $ & $ 13.5 $  & $-0.5 \pm 3.2 $ & $4.2$ \\
$C_{qq}^{(3)[a33a]}$& $ 0.18 \pm 0.11 $ & $ 15.7 $  & $3.8 \pm 4.2 $ & $2.8$  \\
$C_{qq}^{(3)[aabb]}$& $ -0.36 \pm 0.21 $ & $ 11.5 $  & -- & --  \\
$C_{qq}^{(3)[abba]}$& $ -0.40 \pm 0.21 $ & $ 10.6 $  & -- & --  \\
$C_{lq}^{(1)[3333]}$& $ \textcolor{red}{-7.1 \pm 7.1 }$ & \textcolor{red}{--}  & -- & --  \\
$C_{lq}^{(1)[aa33]}$& $ -0.21 \pm 0.38 $ & $ 10.1 $  & $ \textcolor{red}{-7.0 \pm 8.1} $ & \textcolor{red}{--}  \\
$C_{lq}^{(1)[33aa]}$& $ 3.6 \pm 2.5 $ & $ 3.4 $  & $ -2.6 \pm 8.2 $ & --  \\
$C_{lq}^{(1)[aabb]}$& $ 0.17 \pm 0.31 $ & $ 11.3 $  &  $ -3.9 \pm 4.6 $ & $ 2.8 $  \\
$C_{lq}^{(3)[3333]}$& $ 8.2 \pm 5.8 $ & \textcolor{red}{--}  &  -- & -- \\
$C_{lq}^{(3)[aa33]}$& $ -0.25 \pm 0.39 $ & $ 9.7 $  & $ 5.6 \pm 7.0 $ & \textcolor{red}{--}  \\
$C_{lq}^{(3)[33aa]}$& $ -0.21 \pm 0.29 $ & $ 11.1 $  & $ -0.2 \pm 1.2 $ & $ 7.3 $  \\
$C_{lq}^{(3)[aabb]}$& $ -0.011 \pm 0.056 $ & $ 29.2 $  & $ -0.1 \pm 1.1 $ & $ 7.7 $ \\
$C_{ee}^{[aa33]}$& $ 0.0 \pm 2.1 $ & $ 4.9 $  & $ -2.0 \pm 5.2 $ & $ 2.8 $  \\
$C_{ee}^{[aabb]}$& $ -1.4 \pm 1.5 $ & $ 4.8 $  & $ -2.4 \pm 5.1 $ & $ 3.2 $ \\
$C_{uu}^{[3333]}$& $ 2.5 \pm 8.1 $ & \textcolor{red}{--}  & -- & --  \\
$C_{uu}^{[aa33]}$& $ \textcolor{red}{6.8 \pm 7.2} $  & \textcolor{red}{--} & --  & --  \\
$C_{uu}^{[a33a]}$& $ 6.3 \pm 4.7 $ & $ 2.6 $  & $ -1.9 \pm 5.8 $ & $ 3.0 $  \\
$C_{uu}^{[aabb]}$& $ 1.6 \pm 6.8 $ & $ 2.7 $  & -- & --  \\
$C_{uu}^{[abba]}$& $ 2.5 \pm 8.4 $ & \textcolor{red}{--} & -- & -- \\
$C_{eu}^{[3333]}$& $ \textcolor{red}{-8.2 \pm 6.8} $ & \textcolor{red}{--}  & -- & -- \\
$C_{eu}^{[aa33]}$& $ 5.4 \pm 4.0 $ & $ 2.8 $  & $0.9 \pm 6.9$ & $2.7$ \\
$C_{eu}^{[33aa]}$& $ 2.7 \pm 1.8 $ & $ 4.0 $  & $ 2.1 \pm 5.4 $ & $ 2.8 $  \\
$C_{eu}^{[aabb]}$& $ -0.02 \pm 0.41 $ & $ 11.0 $  & $ -1.5 \pm 4.5 $ & $2.7$ \\
$C_{ed}^{[aa33]}$& $\textcolor{red}{5.6 \pm 9.5}$ & \textcolor{red}{--}   & -- & --  \\
$C_{ed}^{[33aa]}$& $ \textcolor{red}{-7.9 \pm 4.7} $ & \textcolor{red}{--}  & -- & --   \\
$C_{ed}^{[aabb]}$& $ 0.0 \pm 1.4 $ & $ 5.9 $  & \textcolor{red}{--} & \textcolor{red}{--}   \\
$C_{ud}^{(1)[3333]}$& $\textcolor{red}{-6.5 \pm 8.3}$ & \textcolor{red}{--}  & -- & --   \\
$C_{ud}^{(1)[33aa]}$& $ \textcolor{red}{-7.4 \pm 7.7} $ & \textcolor{red}{--}  & -- & --   \\
$C_{ud}^{(1)[aabb]}$& $ \textcolor{red}{-6.0 \pm 8.8} $  & \textcolor{red}{--}  & --& --  \\
$C_{ud}^{(8)[33aa]}$& $ \textcolor{red}{6.8 \pm 8.3} $  & \textcolor{red}{--}  & -- & --  \\
$C_{le}^{[aa33]}$& $ -4.2\pm 7.1$ & \textcolor{red}{--}  &  \textcolor{red}{--}  & \textcolor{red}{--}    \\
$C_{le}^{[33aa]}$& $ \textcolor{red}{-6.7 \pm 7.7} $ & \textcolor{red}{--} &  \textcolor{red}{--} & \textcolor{red}{--}    \\
$C_{le}^{[aabb]}$& $ 0.4 \pm 4.2 $ & $ 3.3 $  & $ 2.1 \pm 4.8 $ & $ 3.2 $ \\
$C_{lu}^{[3333]}$& $ 2.5 \pm 8.8 $ & \textcolor{red}{--} & -- & -- \\
$C_{lu}^{[aa33]}$& $ -2.8 \pm 3.3 $ & $ 3.3 $  & -- & --   \\
$C_{lu}^{[33aa]}$& $ 5.7 \pm 3.8 $ & $ 2.8 $  & -- & --   \\
$C_{lu}^{[aabb]}$& $ -0.03 \pm 0.89 $ & $ 7.4 $  & -- & --   \\
$C_{ld}^{[33aa]}$& $ \textcolor{red}{-9.8 \pm 5.3} $ & \textcolor{red}{--}  & -- & --  \\
$C_{ld}^{[aabb]}$& $ -0.4 \pm 2.7 $ & $ 4.1 $  & -- & --   \\
$C_{qe}^{[3333]}$& $ \textcolor{red}{7.7 \pm 7.4} $ & \textcolor{red}{--}  & -- & --   \\
$C_{qe}^{[aa33]}$& $ 6.9 \pm 4.4 $ & $ 2.6 $  & -- & --   \\
$C_{qe}^{[33aa]}$& $ 0.22 \pm 0.38 $ & $ 10.0 $  & -- & -- \\
$C_{qe}^{[aabb]}$& $ -0.23 \pm 0.36 $ & $ 10.4 $  & -- & --  \\
$C_{qu}^{(1)[3333]}$& $ 0.3 \pm 3.1 $ & $ 3.9 $  & $ \textcolor{red}{7.5 \pm 7.5} $ & \textcolor{red}{--}  \\
$C_{qu}^{(1)[aa33]}$& $ 0.1 \pm 2.9 $ & $ 4.2 $  &--&--  \\
$C_{qu}^{(1)[33aa]}$& $\textcolor{red}{ -7.0 \pm 6.9 }$ & \textcolor{red}{--}  & --&--\\
$C_{qu}^{(1)[aabb]}$& $ 3.6 \pm 9.1 $  & \textcolor{red}{--}  & --&-- \\
$C_{qu}^{(8)[3333]}$& $ \textcolor{red}{-10.3 \pm 4.7} $ & \textcolor{red}{--} & --&--  \\
$C_{qu}^{(8)[aa33]}$& $ \textcolor{red}{6.8 \pm 6.1} $ & \textcolor{red}{--}  & --&--  \\
$C_{qu}^{(8)[33aa]}$& $ \textcolor{red}{-5.8 \pm 8.2} $ & \textcolor{red}{--}  & --&--  \\
$C_{quqd}^{(1)[3333]}$& $ -0.10 \pm 0.40 $ & $ 10.2 $  & --&-- \\
$C_{quqd}^{(8)[3333]}$& $ -0.16 \pm 0.48 $ & $ 9.2 $  & --&--  \\
$C_{lequ}^{(1)[3333]}$& $ -1.2 \pm 3.7 $ & $ 3.5 $  & --&--

\end{longtable}
\end{small}

\FloatBarrier

\bibliographystyle{JHEP}
\bibliography{hepfit}

\end{document}